\documentclass[aps,
               notitlepage,mathrsfs,
               amssymb,amsmath,amsfonts,
               superscriptaddress,twocolumn,rmp,
               longbibliography,compress,
               nofootinbib,floatfix]{revtex4-2}
\usepackage{times}
\usepackage[normalem]{ulem}
\usepackage{graphicx}
\usepackage[linktocpage,breaklinks]{hyperref}
\usepackage[capitalize]{cleveref}
\usepackage[usenames,dvipsnames]{xcolor}
\hypersetup{colorlinks=true,
            citecolor=NavyBlue,
            linkcolor=magenta,
            urlcolor=magenta}
\usepackage{multirow,array}
\DeclareMathAlphabet{\pazocal}{OMS}{zplm}{m}{n}

\usepackage{journals}

\newcommand{\tGW}{\tau_{\rm GW}}
\newcommand{\MA}{M_{A}}
\newcommand{\MB}{M_{B}}
\newcommand{\RA}{R_{A}}
\newcommand{\RB}{R_{B}}

\newcommand{\bx}{\boldsymbol{x}}
\newcommand{\Oorb}{\Omega_{\rm orb}}
\newcommand{\bOs}{\boldsymbol{\Omega}_{s}}
\newcommand{\bxi}{\boldsymbol{\xi}}

\begin{document}
\title{Premerger phenomena in neutron-star binary coalescences}

\date{\today}

\author{Arthur G. Suvorov}
\email{arthur.suvorov@ua.es}
\affiliation{Departament de F{\'i}sica Aplicada, Universitat d'Alacant, Ap. Correus 99, E-03080 Alacant, Spain}
\affiliation{Theoretical Astrophysics (IAAT), University of T{\"u}bingen, T{\"u}bingen, D-72076, Germany}

\author{Hao-Jui Kuan}
\affiliation{Max Planck Institute for Gravitational Physics (Albert Einstein Institute), 14476 Potsdam, Germany}

\author{Kostas D. Kokkotas}
\affiliation{Theoretical Astrophysics (IAAT), University of T{\"u}bingen, T{\"u}bingen, D-72076, Germany}

\begin{abstract}
A variety of high-energy events can take place in the seconds leading up to a binary neutron-star merger. Mechanisms involving tidal resonances, electrodynamic interactions, or shocks in mass-loaded wakes have been proposed as instigators of these \emph{precursors}. With a view of gravitational-wave and multimessenger astrophysics more broadly, premerger observations and theory are reviewed emphasising how gamma-ray precursors and dynamical tides can constrain the neutron-star equation of state, thermodynamic microphysics, and evolutionary pathways. Connections to post-merger phenomena, notably gamma-ray bursts, are discussed together with how magnetic fields, spin and misalignment, crustal elasticity, and stratification gradients impact observables.
\end{abstract}
\maketitle

\tableofcontents

\section{Introduction}
\label{sec:intro}

Massive stars that have exhausted their fuel reservoir eventually collapse, as the outward pressures no longer resist the inward gravitational pull. The brilliant supernovae that accompany these collapses rip away most of the outer layers of material, though the inner layers become more and more compressed. Subatomic fermions, {both individually through their Pauli degeneracy pressures and collectively through many-body interactions}, provide a last bastion of resistance against this compression. The strongest such pressure is from neutrons, and thus \emph{neutron stars} represent some of the most compact objects in the universe, behind only black holes and the hypothetical quark stars. Neutron-star core densities likely reach supranuclear levels \cite[$\sim 10^{15} \text{ g/cm}^3$;][] {of16,2017RvMP...89a5007O,2021ARNPS..71..433L} because their stellar radii are of order $\sim 12$~km. Studying their properties thus enables, for instance, tests of the ``low-energy'' limit of quantum chromodynamics with baryonic degrees of freedom and general relativity (GR). The exact chemical makeup of a star defines its equation of state (EOS), determining the particulars of which constitutes one of {the} open problems in high-energy astrophysics.

Neutron-star mergers are some of the brightest events in the universe. A crescendo of gravitational waves (GWs) are released during the final moments of inspiral culminating in a ``chirp'' and subsequent ringdown of the hyperactive remnant left behind at the crash site, being either a black hole or another, heavier neutron star \cite[e.g.][]{shib00,shib03}. In the latter case the object will likely be short-lived: it may maintain a state of metastability despite having a mass larger than the maximum (i.e. Tolman-Oppenheimer-Volkoff limit) applying to an ``ordinary'' neutron star if supported by centrifugal \cite{fr14}, thermal \cite{kap14}, or magnetic \cite{suvg22} pressures. 

Much of the fireworks come shortly after the actual coalescence in the form of gamma-ray bursts (GRBs), primarily of the short variety with prompt durations less than two seconds though there is evidence that some mergers can also fuel long bursts \cite[e.g.][]{troja22,lev24}. Through either a fireball-like (neutrino-anti-neutrino) or Poynting-like process a relativistic jet is collimated by the remnant, providing the impetus for a wealth of emissions that are both luminous (isotropic energies often reach $\sim 10^{53}$ erg) and broadband (spanning $\sim 20$ magnitudes in frequency space via long-lived afterglows). The multimessenger event GW170817/GRB170817A \cite{LIGOScientific17,gold17} has not only firmly established the merger-GRB connection but has been used to place strong constraints on the nature of matter in extreme environments and hypothetical departures from GR in the ultraviolet \cite[e.g.][]{abb19,chat20,diet21}.

A range of luminous and sometimes multiband events can also take place in the moments \emph{leading up to} a binary coalescence. One of the main subjects of this Review pertains to such \emph{precursors}: a fraction \cite[$\lesssim 5\%$;][]{wl21} of merger-driven GRBs show statistically-significant gamma-ray flashes before the main GRB. While other types of precursors may be released premerger (covered in Sec.~\ref{sec:multimessenger}), we typically write \emph{precursor} to specifically refer to these first-round gamma flares. Indeed, many reviews are devoted to the topic of merger phenomena, involving numerical simulations and observations of inspiral, remnant dynamics, and the subsequent forming of relativistic jets \cite[see][for instance]{faber12,br17,shib19,burns20,radi18,cio18,sarinlasky21,kiuc24}. Comparatively little attention has been given towards a complete description of what can be (and has been!) learned \emph{just prior to merger} where such precursors may be expected. Our goal is to fill this gap, motivated in part by recent observational campaigns devoted to searching for precursor flares \cite[e.g.][]{Coppin20,Wang20,deng24} and the ushering in of the LIGO-Virgo-KAGRA (LVK) collaboration's O4 science run. In the spirit of the burgeoning field of multimessenger astrophysics, we also cover the theory of tides and premerger GWs in detail in the hope that a reader not interested in gamma-ray precursors may still find value.

Precursor flashes give promise for a plethora of information about fundamental physics, naturally complementing that associated with merger and postmerger phenomena. A premerger precursor could be used, for instance, to improve sky localization for an impending collision if nothing else \cite[see, e.g.][]{cooper23}. One key difference though is that premerger objects are likely \emph{cold}, and thus arguably give a better handle on the EOS as the impact of $\gg$~MeV temperatures and a littered environment (e.g. from dynamical ejecta) do not require disentangling. Precursors are observationally diverse. They have waiting times --- relative to the main GRB (not merger!) --- spanning $\sim 30$~ms to $\sim $20~s. Their durations and luminosities span a commensurate number of magnitudes, with spectra varying from being almost a perfect blackbody to being highly non-thermal. Such an assortment of characteristics makes a strong case that there may be subpopulations of such precursors fuelled by different means \cite[e.g.][]{wl21}.  By dividing our discussion on precursors into observational (Sec.~\ref{sec:precursors}) and theoretical (Sec.~\ref{sec:prectheory}) elements, we hope to build a full picture of these rich systems. 

Premerger precursors are primarily thought to be caused by one of two means. The first involves direct, electromagnetic interactions between the two stars \cite[e.g.][]{hl01,palen13,wang18,wade20}. For instance, if the inspiralling constituents have dipole moments which are anti-aligned with respect to each other, ample reconnection can take place as the field lines get more and more entwined as the orbit decays. To get large luminosities one would expect the flares to be emitted very close to merger as the dipole fields die off like the cube of distance. One may anticipate relatively short \emph{waiting times} (see Sec.~\ref{sec:precursors}) in this case therefore, unless magnetars are involved in the merger. Spectral and other considerations of certain precursors however give reason to suspect that some mergers do indeed contain a magnetar \cite[e.g.][]{tsang23,suv24,xiao24}, despite the fact that the GW inspiral time exceeds the characteristic Ohmic-decay timescale by orders of magnitude (see Sec.~\ref{sec:magneticfields}). 

The second ignition mechanism involves tidal fields. Tides can not only deform the star geometrically (``equilibrium tide'') but also excite internal fluid motions (``dynamical tides''). These internal motions are characterised through a set of \emph{quasi-normal modes} (QNMs). These QNMs, which come in a few families (e.g. $g$-modes) that we properly introduce in Secs.~\ref{sec:dyntides} and \ref{sec:resfamilies} though sprinkle information about throughout, can be driven to large amplitudes when coming into resonance with the dynamical tidal field. This tidal driving can manifest gravitationally, through a dephasing of the gravitational waveform, and electromagnetically, by liberating potential energy a star whose crustal layers succumb to the high-amplitude pulsations. Ruptures and subsequent energy release models are akin to some of those considered for magnetar flares \cite[see, e.g.][]{td93,td95}, though in the premerger context the cause of \emph{crustal failure} is QNM resonances rather than the gradual build-up of mechanical stress from the secular evolution of a superstrong field.

Galactic neutron stars exhibit a variety of multiwavelength phenomena, the varying characteristics of which has invited rather detailed classification schemes over the years, with stars being grouped into categories such as recycled, radio, or X-ray pulsars \cite[e.g.][]{teru19}. We describe how precursors, and inspiral lead-up more generally, can inform us about each of these pathways. This Review is thus roughly divided into two halves: that devoted to GWs and that to electromagnetic precursors, described in more detail below. 

\subsection{Structure and purpose of this review}
\label{sec:intro2}

Premerger phenomena can be either gravitational or electromagnetic, both of which we cover in later sections. Before doing so however, it is important to give a sense of background regarding both neutron star macro- (Sec.~\ref{sec:macrostructure}) and micro-structure (Sec.~\ref{sec:microstructure}), as these aspects are that which we hope to learn about from multimessenger channels. We aim to make this Review self-contained therefore by exploring theoretical and observational elements of neutron-star properties. 

\begin{figure*}
\centering
\includegraphics[scale=0.95]{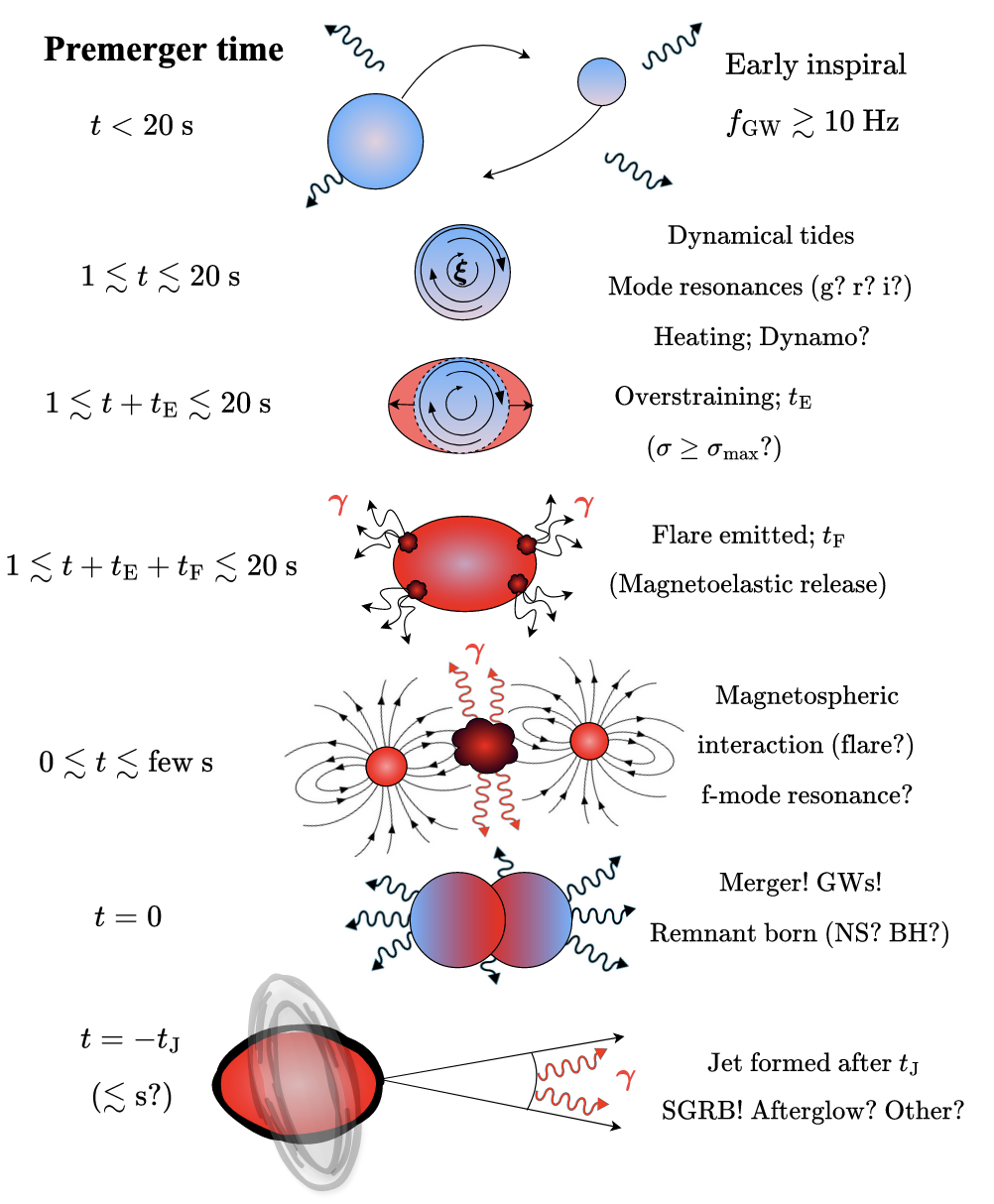}
	\caption{Cartoon depiction of premerger events culminating in a GRB with at least one precursor flare. A binary of cold neutron stars inspiral, and the equilibrium tidal imprint on the gravitational waveform can be used to deduce something about stellar structure. Once the orbital frequency matches that of some mode, dynamical tides start to induce a dephasing in the waveform through resonances; heating occurs meanwhile. Modes exert stress on the crust such that it may overstrain after time $t_{E}$ for favourable eigenfunctions. Magnetoelastic and/or thermal energy is released following a flare development timescale $t_{F}$, possibly modulated by quasi-periodic oscillations (QPOs; as in GRB 211211A). Once the separation is sufficiently small, magnetospheric interactions may also be powerful enough to produce bright gamma-ray or broadband emissions. In the final stages, the fundamental mode may also become resonant. The stars then coalesce, at the peak of the GW signal, the information of which can be further used to decipher the nature of the inspiralling stars and the remnant (as for GW 170817). The remnant, whether a neutron star of some meta-stable variety or a black hole, likely surrounded by a temporary accretion disk in any case, can then launch a GRB jet that successfully drills out after some delay timescale, $t_{J}$.}
	\label{fig:dia1}
\end{figure*}

The main purpose of this work is to detail some aspects that we feel have received little attention. For example, the theory of dynamical tides is rich and varied in the literature, with several different notations being used and so on. Similarly, precursors seem to have escaped attention despite their, as we argue at least, propensity for educating us about neutron-star physics. Figure~\ref{fig:dia1} depicts the various elements we describe throughout together a rough timeline of anticipated events.

We begin by introducing details about neutron-star structure generally (macrophysics in Sec.~\ref{sec:macrostructure} and microphysics in Sec.~\ref{sec:microstructure}), to pave the way in describing how the final $\sim 20$~seconds of inspiral may appear. Sec.~\ref{sec:inspiralbasics} covers the gravitational aspects of this: how does inspiral occur, at what order do post-Newtonian (PN) effects come into play, and most importantly how tides influence the dynamics. The resonant QNMs that are excited lead to a gravitational dephasing of the waveform that can be studied and used to infer properties about the inspiralling constituents. These tides however may also be important for (at least modulating) precursor emissions. The observational elements of these precursors are described and collated in Sec.~\ref{sec:precursors} with theoretical explanation(s) and modelling in Sec.~\ref{sec:prectheory}. These ideas in the context of multimessenger astrophysics more generally are reviewed in Sec.~\ref{sec:multimessenger} with a summary given in Sec.~\ref{sec:conclusions}.

The hasty reader who is familiar with neutron stars but cares about tidal-interaction theory and/or GWs can skip to Sec.~\ref{sec:inspiralbasics}, while a reader most interested in precursors observations or theory can head to Secs.~\ref{sec:precursors} and \ref{sec:prectheory}, respectively.

\subsection{Remarks on notation}
\label{sec:notation}
We use subscripts $A,B$ to denote primary and companion elements of a binary. Here ``primary'' will usually mean the heavier object, though for some theoretical discussions it may simply refer to the object for which finite size effects are resolved when ignoring (as is typical) the multipolar structure of the companion. In cases where no ambiguity can arise, the subscripts are dropped for presentation purposes. We adopt the indicator $\alpha$ for arbitrary mode quantum numbers in reference to a radial and angular decomposition; see Sec.~\ref{sec:tidesgeneral}. If only a right-hand subscript is written (e.g. $g_{i}$), this refers to the radial node number $n$ with $\ell = m =2$. Most of this Review is carried out in a Newtonian language, where Latin indices (aside from $n,m,$ or $\ell$) refer to spatial components of a tensorial object (typically with respect to  a spherical basis). Linear frequencies are written with $f_{\alpha} = \omega_{\alpha}/(2\pi)$ for angular frequencies $\omega$. When discussing object spin and the orbit, the symbol $\nu$ is used instead for the linear frequency (e.g. $\Oorb = 2 \pi \nu_{\rm orb}$). A superscript asterisk indicates complex conjugation. An overhead dot denotes a time derivative.

\section{Neutron star macrostructure}
\label{sec:macrostructure}

The macrostructure of a neutron star, such as its mass and radius (via the EOS), impacts greatly on premerger phenomena. In order to be self-contained, this section provides a brief review of that macrostructure with a view of the key aspects that will be relevant for us later in describing GWs and precursors.

\subsection{Equations of state}
\label{sec:eos}

Neutron stars exhibit an astonishing range of multiwavelength phenomena, from steady radio pulsing to rare storms of powerful soft-gamma flares \cite[e.g.][]{mcg14}. From a nuclear physics perspective however, it is thought that neutron-star matter comes into beta equilibrium shortly after birth \cite{hoyos08}. This implies that their cores are described by a unique EOS moderated by small, thermodynamic perturbations. Proto-stars, especially those born in merger events \cite{alf18}, may be exempt however as new particle production pathways can be opened at super-MeV temperatures. For mature stars a barotropic EOS is typically used though, where the hydrostatic pressure $p$ is just a function of the rest-mass density, $\rho$. This latter quantity is distinct from the \emph{energy density} $\epsilon$, though sometimes in the literature one can find ``EOS'' of the form $p = p(\epsilon)$; see Section 2 of \citealt{OBoyle20} for a discussion.

\begin{figure*}
	\centering
	\includegraphics[width=0.97\textwidth]{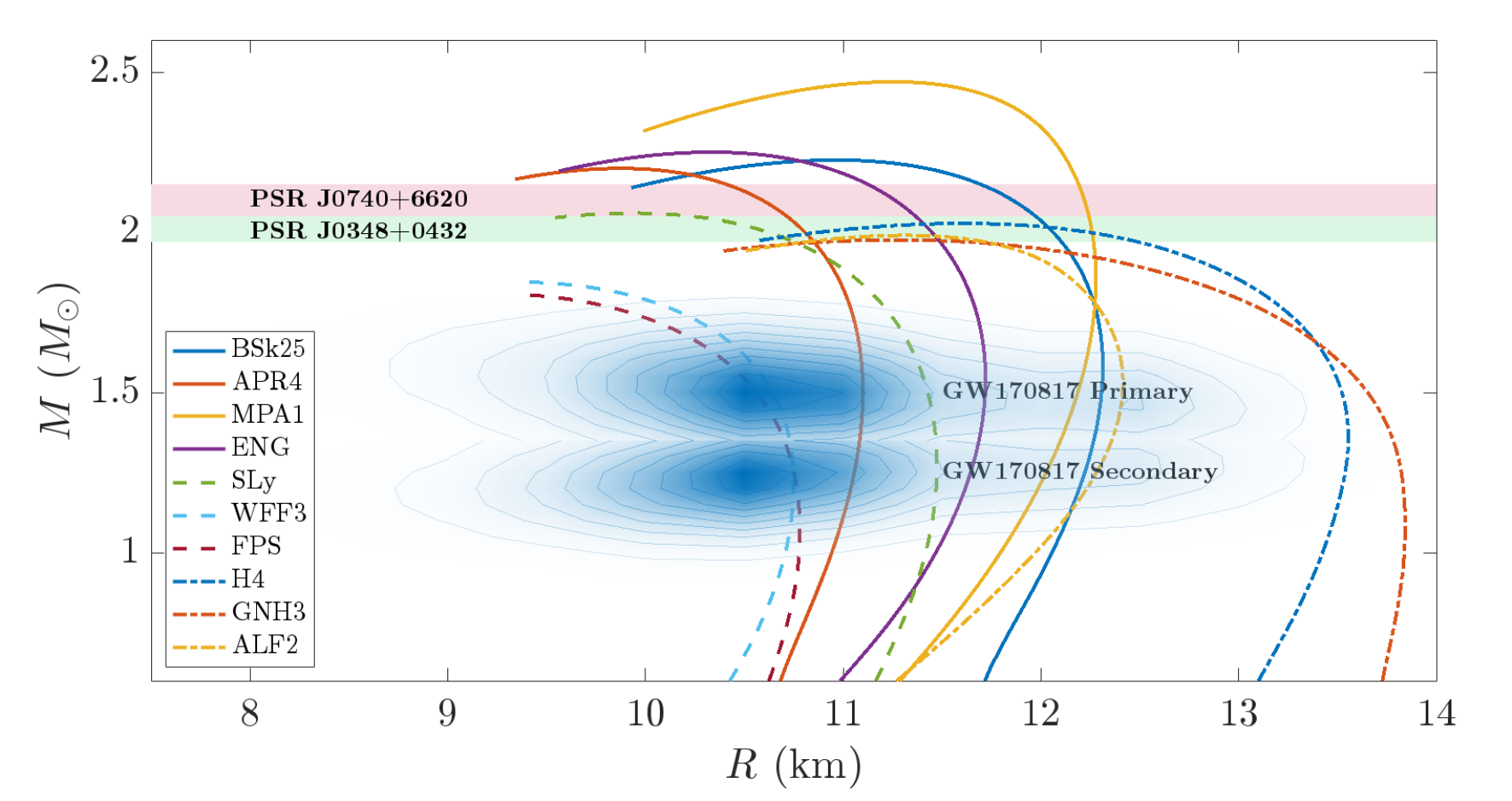}
	\caption{Mass-radius diagram for select EOS (see plot legends). These EOS belong to three different groups, distinguished by curve style (solid, dashed, dot-dashed; see main text). Inferences on the mass and radius of pulsars involved in binaries (PSR J0740+6620 and PSR J0348+0432) and in the binary merger event GW170817 are overplotted in the shaded regions.
 }
	\label{fig:EOS}
\end{figure*}

Many EOS families have been considered in the literature \cite{of16,2021ARNPS..71..433L}, primarily owing to our collective ignorance regarding the behaviour of matter at supra- and hypernuclear densities at low temperatures. The EOS families we consider in an effort to make this Review self-contained --- mass--radius curves of which are displayed in Figure~\ref{fig:EOS} --- cover stars with purely $nep\mu$ compositions to those exhibiting phase transitions to free-quark regimes \cite[i.e. the hybrid stars, like ALF2;][] {alfo05}. Although some of the considered EOS cannot support the heaviest (confirmed\footnote{For an up-to-date catalogue of neutron-star mass measurements, at least where there are no substantial systematics due to  mass transfer or mass loss, see \url{https://www3.mpifr-bonn.mpg.de/staff/pfreire/NS_masses.html}. There are supposedly heavier ``spiders'', though these are highly dynamical and also rotating rapidly.}) neutron star observed to-date, namely PSR J0740+6620 with a mass of $\sim 2.08^{+0.07}_{-0.07}M_{\odot}$, we consider them for completeness (especially those that pass right through the heart of GW 170817 contours). Another object worth mentioning in this landscape is HESS J1731-347, with a mass of $\gtrsim 0.8M_{\odot}$ and $R \approx 10.4$~km \cite{doro22}; if indeed this was representative of the lower limit to the possible neutron star mass, it would impose serious restrictions to the EOS and formation mechanisms. J1731 measurements would, funnily enough, be totally consistent with the dashed-curve EOS in Fig.~\ref{fig:EOS} that do not reach $M \gtrsim 2 M_{\odot}$; see Figure 5 in \citealt{ofen24}.

Thermal and inviscid effects in a mature neutron star, though crucial for modelling buoyancy-restored oscillations (i.e. $g$-modes; see Sec.~\ref{sec:dyntides}) especially in the crust (see Sec.~\ref{sec:crust}), are often not present in many of the ``standard'' EOS tables because the effects are thought to be safely negligible below $\sim 0.1$~MeV where the chemical potential is tiny. Such corrections can be handled with a $\propto T^2$ thermal pressure contribution. The ``general purpose EOS'' (following the naming of the catalogue) available on the CompOSE catalogue\footnote{These and many other EOS can be obtained in a tabulated form at \cite{compose17}: \url{https://compose.obspm.fr/home/}.} provide such information. Although a few general purpose EOS do extend to a lower cutoff at $10^7$~K (e.g. SLy4 and APR), many of them have their tables truncated at minima of $10^9$~K (still much higher than expected of a mature star seconds before merger; though cf. Sec.~\ref{sec:thermal}) since lower temperatures are well-modelled by cold EOS with some perturbative corrections, as described above. 

The impact of realistic thermal profiles on the $g$-spectra in this perturbative sense has been detailed by \citealt{Kruger14} and by \citealt{kuan23} with an independent code, finding consistent results.
As found by \citealt{kuan22gmode}, the $g$-spectra are linked to the ``slope'' of the EOS $M$ vs. $R$ curve, and are especially sensitive to central densities in the range of 1--3 $\rho_0$ for nuclear saturation value $\rho_0 \approx 2 \times 10^{14} \text{ g cm}^{-3}$ \cite{lp01}. The same slope impacts on the universality of some relations between bulk properties and dominant peaks in postmerger waveforms \cite{rait22}. \citealt{kuan22gmode} also found that $g$-mode scaling relations can be grouped according to EOS, with hadronic ones that can support heavy stars (solid lines in Fig.~\ref{fig:EOS}), those that cannot (dashed), and EOS involving phase transitions leading to either hyperon nucleation or quark deconfinement (dash-dotted); we refer the reader to \citealt{armen23} for a review on hyperonisation. These three families each abide by a different set of ``asteroseismological relations'' that can, in principle, be distinguished by observations. 

Detailed analyses of the EOS population, their historical development, together with astrophysical applications and constraints are found in \cite{lp01,2017RvMP...89a5007O,of16,2021ARNPS..71..433L}. 
Observationally, masses can be constrained from a variety of methods, ranging from orbital timing of binaries to the phased-resolved tracking of thermal hotspots atop an isolated star to GWs. Observations of the moments of inertia can also be made from periastron advance in binaries \cite{Greif20}. We now turn to describing a few EOS families. Note that some of the EOS appearing in Fig.~\ref{fig:EOS} are not covered below despite their consistency with data, though could adequately be described as ``relativistic models.'' For example, MPA and ENG are based on Dirac-Brueckner-Hartree-Fock models for purely nucleonic matter, while GNH(3) is a relativistic mean-field EOS that also contains hyperonic matter. In one of the considered EOS (ALF2) a hadron-quark hybrid core is present, described through the MIT bag model.

\subsubsection{WFF family}
\label{sec:WFF}

The WFF family (WFF1-3) was developed in the late 1980s \cite{WFF} based on realistic nucleon-nucleon interactions calculated using many-body methods. 
This family treats nucleons as non-relativistic particles, utilising the Argonne v14 (so-named because it uses 14 operator components to model nucleon-nucleon interactions) and Urbana vII potentials which account for three-nucleon forces. The family is generally considered to be a softer EOS, especially WFF1. According to the WFF family of EOS, the maximum mass of a neutron star typically ranges up to $~2.2 M_\odot$ (WFF1-2) in agreement with the aforementioned observations. Still, recent results from GW170817 and the Neutron Star Interior Composition Explorer (NICER) suggest a stiffer EOS, thus WFF1 is excluded as being too soft; it is for this reason we have excluded it from Figure~\ref{fig:EOS}. 
Despite these limitations, many works still use the WFF family of EOS as representatives of softer EOS. 
The WFF family has improved over time by including additional interactions and so on, with many modern EOS stemming from it \cite[see, e.g.][]{1995PhRvC..51...38W}.

\subsubsection{APR family}
\label{sec:APR}

The APR EOS is named for the authors Akmal, Pandharipande, and Ravenhall who developed it in the late 1990s \citep{APR}. It is widely adopted, despite the fact that at high densities relativistic corrections are large and using the non-relativistic Schr{\"o}dinger equation at $\rho \sim \text{several} \times \rho_{0}$ is questionable, as it meets the maximum mass observational limit in the static limit (and beyond to $\sim 2.3 M_{\odot}$), models the cooling evolution (implicitly via proton-fraction predictions and beta-decay rates in superfluids) as observed in young and old neutron stars \cite[see, e.g.][]{anz22,marino24}, and can meet the constraints set by GW observations of neutron-star mergers; it is not necessarily unique in these respects though. The APR EOS is derived from a microscopic nuclear many-body theory using the variational chain summation method. It predicts a phase transition from pure nucleonic matter to a state including nucleons and a neutral pion condensation at high densities \cite{2019PhRvC.100b5803S} and incorporates realistic nucleon-nucleon interactions based on the Argonne v18 potential (which, as per the discussion above, includes 18 interaction operators). In addition to two-body interactions, it includes three-body forces using the Urbana IX potential: these are essential for providing \emph{repulsive} interactions at high densities. It is through these repulsive interactions that one may construct heavier neutron stars. For up-to-date discussions, see \citealt{2019PhRvC.100b5803S, 2021PrPNP.12003879B}.

\subsubsection{SLy family}
\label{sec:SLy}

The SLy EOS is based on a semi-empirical approach that parameterises the interactions between nucleons (protons and neutrons) in terms of density-dependent coefficients of the nuclear interactions \cite{1997NuPhA.627..710C,1998NuPhA.635..231C,2001A&A...380..151D,2015PhRvC..92e5803G}. It is derived from the Lyon-Skyrme energy-density functional (SLy) and uses a parameter set of coefficients (e.g. terms weighting the kinetic energy of nucleons) that are chosen to describe the properties of both symmetric nuclear and neutron-rich {matter through fittings to six nuclear masses (spherical nuclei with an even number of neutrons and even number of protons).}  Neutron star models constructed using the SLy EOS can reach a maximum mass of  $\sim 2.1 M_\odot$ and have a radius of $11 \lesssim R \lesssim 13$~km for typical stars with 1.4 solar masses; it is therefore a relatively stiff EOS. The SLy EOS is designed to be valid for a wide range of densities, ranging from the crust up to a few times the nuclear saturation density. 

{Although the SLy EOS can accommodate existing mass and radius constraints and is widely used in simulations of supernovae and neutron star mergers, especially updated versions with finite-temperature corrections \cite{2015PhRvC..92e5803G,Schneider:2017tfi,2019NuPhA.983..252R}, it does not allow for the direct Urca or other ``enhanced cooling'' processes to operate. This places it in strong tension with observations of thermal emissions from accreting and isolated neutron stars \cite{marino24}.}

\subsubsection{BSk family}
\label{sec:BSk}

The BSk EOS is developed through the use of the Brussels-Skyrme (BSk) energy-density functionals\footnote{The BSk EOS table can be generated by the resources provided in \url{https://www.ioffe.ru/astro/NSG/BSk/} {or the CompOSE catalogue}.}.
Like SLy, BSk is based on Skyrme-type functionals that use density-dependent coefficients to parameterise nucleon-nucleon interactions. These functionals are fitted to match experimental data on finite nuclei and nuclear matter, resulting {in a sequence of incrementally named} models \cite{pear11,pear12,pote13,gori13,pear18}. {The BSk functionals are, however, fitted to a much larger set of experimental nuclear data relative to that of SLy, namely all nuclear masses with $Z,N \geq 8$ with a model root-mean square deviation of $\approx 0.5$~MeV. The deviation from mass data (considering the restricted set of even-even nuclei) for the SLy EOS is an order of magnitude larger \cite[$\approx 5.1$~MeV;][]{dob04}. In general, fitting terms using the largest available sets of nuclei, even and odd, spherical and deformed, is crucial to properly calibrate the functional(s)}. 

{In contrast to some other EOS, members of the family need not assume strict beta equilibrium; see \citet{fant22} for a description of accreted crusts with X-ray burst ashes made of iron for BSk19--21. Finite-temperature corrections are included but only for the surface layers at densities below $10^{6} \text{ g cm}^{-3}$}. The predicted mass-radius relationship varies depending on the model {number}, but typically one has a radius of 11--13 km for 1.4 $M_\odot$ neutron stars, as for SLy. The BSk models also predict a relatively high maximum mass of between 2 and 2.4 $M_\odot$ In Fig.~\ref{fig:EOS} we show the $M$-$R$ relation for one member of this family, BSk25, for which the maximum mass is 2.23 $M_\odot$ and the radius of the model with $1.4\,M_\odot$ is $\approx$ 12.3~km. {BSk21--25 --- unlike SLy4 as well as BSk19,20,26 --- \emph{do} allow for
the direct Urca process in dense matter and conform better to cooling observations \cite{marino24}.}

\subsection{Rotation and binary alignment}
\label{sec:rotation}

Angular momentum is ubiquitous in nature. In the case of neutron stars, approximate angular momentum conservation during the supernova process implies that proto-stars can be born spinning fairly quickly, even for relatively slow progenitors. Depending on the progenitor and the impact of fallback accretion \cite{stock20}, some neutron stars may be born with millisecond spin periods \cite{2022ApJ...924L..15F}; though see also \citealt{varma23,2023MNRAS.526.5249V}.

Even for ``slow'' objects (e.g. magnetars performing a full revolution at rates less than once per few seconds), it is essential to account for rotation when considering electromagnetic observables. Depending on the rotational phase the signal can be broadened or lost due to beaming effects, which impacts interpretation. Observationally, neutron stars seem to be anything between practically static up to spins of $\sim 800$~Hz. Why there is such an observational upper limit is a topic of active research, invoking various explanations from GW-enhanced spindown to centrifugal propellering either from a companion or fallback \cite[see, e.g.][for a discussion]{pat12}.

One would like to know what the likelihood of having a ``rapidly'' rotating star taking part in a merger is. A review of spin properties in binaries, and the formation of double neutron-star systems in general, can be found in \citealt{tauris17}. The situation is somewhat complicated however, and it has been argued that there may be three sub-populations of double neutron-star systems with distinct spin distributions, pertaining to (i) short-period and low-eccentricity binaries, (ii) wide binaries, and lastly (iii) short-period, high-eccentricity binaries \cite{jeff19}. The canonical formation channel involves a symbiotic binary (i.e. without dynamical captures) where there are two supernovae. These supernovae are separated by a timescale that is sensitive to a number of evolutionary specifics related to common-envelope evolution and how susceptible the companion is to Roche-lobe overflow, which in turn depends primarily on the mass ratio \cite[e.g.][]{van17}. 

Prior to double neutron-star formation, accretion by the first born from the non-degenerate star can be either of a disk or wind-fed nature, and will tend to spin-up the neutron star thereby leading to a ``recycled'' object with a spin that is (close to) aligned with the orbital angular momentum. A discussion on predictions for alignment angles can be found in Section 2.3 of \citealt{kuan23} and references therein: several binaries have non-negligible misalignment angles as measured through geodetic precession and optical polarimetric measurements \cite[e.g. PSR B1534+12 in a double neutron-star system with a misalignment of $\sim 27^{\circ}$;][]{fons14}. Such misalignment is relevant for the excitation of non-axisymmetric modes during inspiral (see Sec.~\ref{sec:spinacc}). Either way, for a spin-up rate of $\dot{\nu} \gtrsim 10^{-14} \text{ Hz s}^{-1}$, not unusual from observations or torque theory \cite{gs21}, the first-born neutron star could attain spins of $\gtrsim 100$~Hz within significantly sub-Hubble times $\lesssim 10^{2}$~Myr \cite[see table 1 in][]{tauris17}. Given that the dipolar magnetic field may be buried by large factors via epochs of mass infall \cite[][see also Sec.~\ref{sec:magneticfields}]{mp05,vig09,suvm16,suvm19}, the spindown that sets in, once the secondary collapses and stops gifting material, may not be sufficient to erase spinup before merger. 

\citealt{zhu18} estimate that between $15$ and $30\%$ binary neutron-stars will have spins measurable via GWs at $\gtrsim 90\%$ confidence. This implies the importance of accounting for spin in tidal modelling, as all mode eigenfrequencies are skewed by even a modest degree of rotation; see equation \eqref{eq:inertialframe} and also Appendix A of \citealt{suv24}.

\subsection{Magnetic fields}
\label{sec:magneticfields}

It has been recognised since the dawn of pulsar astronomy that rotating magnetic fields, as an induction-conduit for electric fields, are not only responsible for radio activity but also instigate the star's gradual slowdown and moderate crustal activity. In order to accurately interpret neutron star activity to unveil the stellar EOS and other aspects, models of their magnetic fields, together with their evolution and dynamics, have been constructed in the literature from a variety of techniques. Magnetic fields are crucial for all of the electromagnetic, premerger phenomena considered here; we therefore feel it is appropriate to provide a brief discourse on current understanding of magnetic structure. If strong enough, magnetic fields may even alter the GW signals associated with inspiral or EOS \cite[e.g.][]{dex17} in a variety of ways; see Sec.~\ref{sec:magacc}.

Although a complete survey of how the magnetic field impresses on observables associated with the neutron star population at large lies well beyond the scope of this Review (see \citet{teru19} for a thorough exposition), many observations have proven that magnetic multipoles and topologically complicated structures are a reality:

\begin{itemize}

\item[$\boldsymbol{\star}$]{Data from NICER and other experiments indicate that `hotspots' atop millisecond pulsars (e.g. PSR J0030+0451) are not antipodal \cite{bil19,riley21}. This implies equatorially-asymmetric heating, and therefore a magnetic field composed of a mix of odd and even-order multipoles \cite{suvm20,kal20}.}

\item[$\boldsymbol{\star}$]{Models of pulsar radio-activation cannot explain the bulk of the population if the surface and magnetospheric field geometries are dipolar, i.e. dipole ``death lines'' cut right through the middle of the population on the $B$-$P$ diagram \cite{cr93,ha01}.}

\item[$\boldsymbol{\star}$]{The morphology of pulsed emissions are highly varied, with some systems displaying long term epochs of nulling or interpulses. Interpulse phenomenology in radio pulsars can be qualitatively explained by an oblique rotator with a multipolar magnetic field, as the emissions are then composed of multiple components \cite{barn82}. The X-ray light curve from the magnetar SGR1900+14 also displays interpulse-like phenomena, which may be contributed by multiple hotspots on the surface \cite{bt07}. \citet{zhang07} suggest that starspots (i.e. localised, multipolar fields) may emerge through Hall evolution near the poles of neutron stars that are hovering around the death line, sporadically allowing the hosts to pulse and possibly explaining nulling \cite[see also][]{suv16}.}

\item[$\boldsymbol{\star}$]{{Multipoles are typically generated via cascade phenomena through Hall drift \citep{suv16} or plastic flows \citep{g22} in neutron star crusts if the ``magnetic Reynolds number'' is sufficiently large \cite[$B \gtrsim \text{few} \times 10^{12}$~G;][]{gl21}, and likely also through ambipolar diffusion in the core \cite{gus17}. For crustal fields with a high-degree of relative helicity \emph{inverse cascades} can instead operate \citep{deh24b}, where energy is transferred from small to large scales, though this presupposes an initially multipolar field. Either way, Hall-Ohm simulation outputs are supported by observations of neutron-star cooling and (more speculatively) field decay.}}

\item[$\boldsymbol{\star}$]{Many pulsar \emph{braking indices} differ from the canonical dipole value of 3, which points towards a complicated field geometry, anomalous braking torques, and/or mass-loaded winds initiating a different spindown behaviour \cite{mel99,bars10}. Glitch activity \cite{jg99}, Hall waves launched from superconducting phase transitions \cite{levin24}, or inclination angle evolution \citep{2017MNRAS.467.3493J} may also notably affect this index, $n = \nu \ddot{\nu} / \dot{\nu}^2$.}

\item[$\boldsymbol{\star}$]{Simulations of accretion show that even small accretion columns (or `magnetic mountains') warp field lines far from the column itself \citep{mp05,suvm19,suvm20}, with field line compression within the equatorial belt persisting over long, Ohmic timescales \citep{vig09}. Given that all neutron stars born from core collapse exhibit some degree of fallback accretion at birth from a temporary disc of bound material, one might expect all stars to have `buried' and multipolar components \citep{mp14}. Even ignoring this possibility, particle production and backflow in the magnetosphere will gradually advect field lines, instigating some (small) degree of multipolarity surviving over diffusion timescales. Such considerations were initially motivated by the observation that neutron stars in low-mass X-ray binaries (LMXBs; undergoing Roche-lobe overflow) tend to possess unusually low magnetic field strengths \citep{bhatt91}.}

\item[$\boldsymbol{\star}$]{Cyclotron resonant scattering line energies demand that a number of accreting and isolated neutron stars possess local fields (much) stronger than those implied by global, dipole-field observations \citep{stau19}.}

\item[$\boldsymbol{\star}$]{Current bundles injected into the magnetosphere by crustal motions twist the fields there, inducing multipolarity \cite{belo09}. {Models} of neutron star activity often invoke crustal failures as seeding events for outbursts \cite[such as glitches or flares;][]{baym71,suvk19,dal22}, and therefore such injections may be common. In fact, crust failures are one of the mechanisms proposed for premerger precursors; see Sec.~\ref{sec:prectheory}.}

\end{itemize}

Aside from \emph{poloidal} multipoles, the neutron star magnetic field also likely contains a multipolar and geometrically-complicated \emph{toroidal} component:

\begin{itemize}

\item[$\boldsymbol{\star}$]{Precession in magnetars, such as 4U 0142+61 \cite{mak14} and SGR 1900+14 \cite{mak21}, are most straightforwardly explained through a (sub-)crustal toroidal field of strength $\gtrsim 10^{16}$~G: such a field introduces a prolate distortion along the magnetic axis, which then becomes misaligned with the rotation axis, causing free precession \cite[see also][]{mast15,suvk20p}.}

\item[$\boldsymbol{\star}$]{Mixed fields are necessary for the stability of the star. Magnetohydrodynamic (MHD) models and stability analyses demonstrate that purely poloidal/toroidal fields are unstable over dynamical timescales \cite{1973MNRAS.161..365T,akg13,herb17}.}

\item[$\boldsymbol{\star}$]{{Dynamo models aiming to explain the intense fields of magnetars rely on poloidal-toroidal amplification cycles. More generally, differential rotation will wind up poloidal lines. Given that some differential rotation is inescapable as the spindown torque applied to the surface cannot instantaneously influence the core \cite[i.e. there is a crust-core lag;][]{melatos12}, toroidal components should be generated.}}

\item[$\boldsymbol{\star}$]{A universal feature of magnetothermal evolutions is that energy swaps between the toroidal and poloidal sectors, at least for sufficiently strong fields \cite[e.g.][]{pons11,deh23}. Multipolarity is thus inevitably generated in the toroidal sector from each of the considerations listed above.}

\end{itemize}

\begin{figure*}
	\centering
	\includegraphics[width=0.8\textwidth]{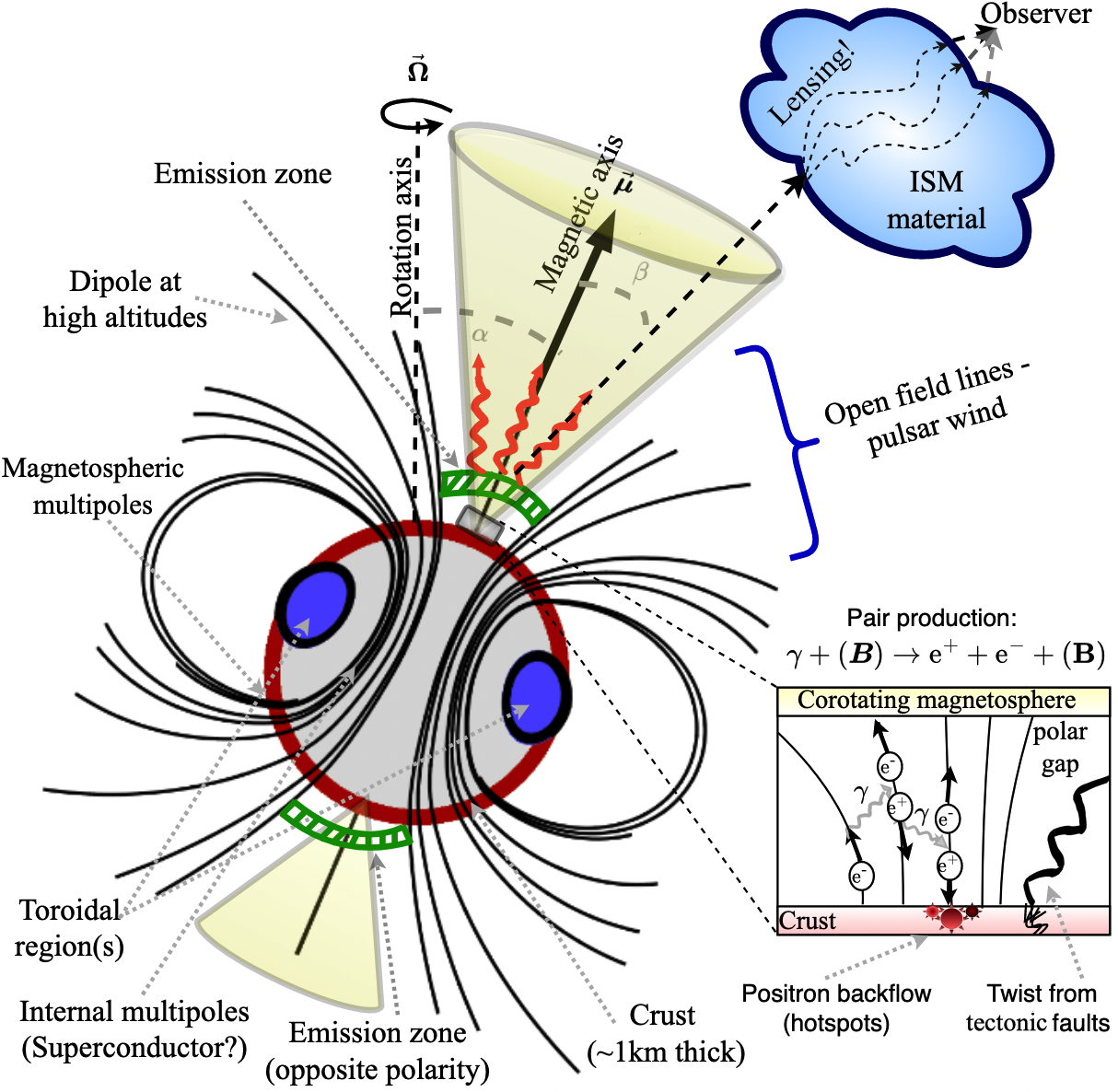}
	\caption{Conventional view of pulsar operation with a ``twisted torus'' magnetic field \cite{bra06} and polar gap(s) \cite{lor04}. The presence of complicated magnetic substructure is highlighted.}
	\label{fig:pulsarop}
\end{figure*}

Figure~\ref{fig:pulsarop} illustrates the conventional picture of pulsar operation and the presence of magnetic substructure generally (including the likelihood of internal multipoles and a superconductor; see Sec.~\ref{sec:superfc}). Interpretation of neutron-star phenomena is complicated by the fact that emissions must traverse the interstellar (or intergalactic) medium, which can lens photons or GWs \cite[see, e.g.][]{suv22}.

Perhaps more importantly for us here, magnetic fields are subject to decay. The exact way in which this occurs within a neutron star can be subtle. For example, evolution in the crust is thought to be governed by a coupled Hall-Ohm system \cite{gr92}, at least for sub-magnetar ($\lesssim 10^{15}$~G) fields where plastic flow does not enter the picture \cite{gl21}. Hall evolution is strictly conservative though and so even though the decay is only driven by Ohmic terms, Hall drift tends to accelerate such a decay by cascading: energy is transferred from large to small scales whereupon it is more susceptible to decay. 

If we ignore plastic flows \cite[though see, e.g.][]{gl21,g22,suvm23}, the overall crustal field strength evolution may be written, after solving a simplified volume-averaged version of the induction equation following \citet{ag08}, as 
\begin{equation} \label{eq:hallohm}
B(t) = B_{0} \frac{\tau_{\rm H}} {e^{t/\tau_{\Omega}} \left( \tau_{\rm H} + \tau_{\Omega} \right) -\tau_{\Omega}},
\end{equation}
with $\tau_{\rm H} = 4 \pi e n_{e} L^2 / c B_{0}$ and $\tau_{\Omega} = 4 \pi \sigma_{\rm e} L^2 /c^2$ for length-scale $L \lesssim R$ (smaller for higher-order multipoles), initial field strength $B(0) = B_{0}$, electron number-density $n_{e}$, and electrical conductivity {parallel to the magnetic field} $\sigma_{\rm e}$. For typical parameter choices in the crust, we may have therefore that $\tau_{H} \sim \text{few} \times 10^{4} (B_{0}/10^{14}\text{ G})^{-1}$~yr and $\tau_{\Omega} \sim \text{few}\times$~Myr. Given, however, that the GW inspiral time is orders of magnitude longer than either of these times unless the field strength is $\sim 10^{10}$~G, one anticipates weak fields at time of merger. This has important implications for the precursor flares we discuss in Sec.~\ref{sec:precursors}. 

It is important to note however that there are some ways around the decay issue, such as via Hall-stalling \cite[i.e. the system reaches a quasi-equilibrium where $\tau_{\rm H}$ tends to some much larger number;][]{gc14}, an absence of crustal impurities \cite[which increase the conductivity and hence the Ohmic time;][]{ip18}, plastic flows \cite[which can work to halt electron flows and thus suppress cascading and the formation of multipoles;][]{g22,suvm23}, or dynamical capture (thereby avoiding issues related to decay altogether). {Another interesting possibility relates to superconductivity (see Sec.~\ref{sec:superfc}). A strong core field may be retained if the decay timescale lengthens as the star cools \cite{ho17}, while a strong crustal field could exist if the expulsion time of core flux into the crust is long}. Whether it is at all possible for magnetar-level fields to persist over cosmological timescales remains an active area of research \cite[see, e.g.][for discussions]{ip18,suvg23}.

\section{Neutron star microstructure}
\label{sec:microstructure}

In some cases of relevance, the stellar microstructure can also influence premerger observables.  Again in the interests of being self contained, this section delves into some relevant aspects of microphysics, with an emphasis on the crust.

\subsection{Basic elements of crustal physics}
\label{sec:crust}

Although a neutron star may be born with temperatures in excess of $\gtrsim$~MeV, especially in a merger, the system begins to cool rapidly through a sequence of beta decays (Urca reactions) as neutrinos flood outward from what is currently an envelope \cite[e.g.][]{eich89,ross03,seki11}. The details of these very early stages in the star's life are a matter of active research and lie beyond the scope of this Review; see, for instance, \citet{sarinlasky21}. Still, this early behaviour \emph{is} important as concerns some early post-merger phenomena within our purview, so some details are touched on in subsequent sections.  

Nevertheless, for the mature stars taking place in a merger, it is expected that the outer layers long ago formed an elastic solid that should be distinguished from the liquid core. These low-density regions, with the exception of the very outer layers comprised of a thin ocean and atmosphere, are called a crust because nuclei are cold enough to freeze and form a crystalline solid which can sustain stress. The crust is important for all electromagnetic phenomena from neutron stars, as it is ultimately the region where external field lines are anchored. During late inspiral, a great deal of heat is generated via the tidal field and resonant modes; this heating (see Sec.~\ref{sec:thermal}), in addition to the stresses exerted by resonant pulsations (see Sec.~\ref{sec:breaking}), and despite the fact that the crust constitutes only $\sim 1\%$ of the total stellar mass, plays a central role in the precursor phenomena covered in later Sections. 

The nuclear phase of the matter in the envelope can be understood through the Coulomb coupling parameter for ions 
\begin{equation} \label{eq:Gamma}
    { \Gamma} = \frac {Z^2 e^2} {a_{i} k_{B} T},
\end{equation}
where $Z$ denotes the number of protons in the ion, $a_{i}$ is the ion sphere radius (i.e. the Wigner-Seitz cell radius) such that $4\pi a_{i}^3/3$ equals the volume per ion, $1/n_{i}$, $e$ is the elementary charge, $k_{B}$ is the Boltzmann constant, and finally $T$ represents the temperature. Aside from fundamental constants, each ingredient defining $\Gamma$ varies with depth in a complicated way \cite{cham08}. Once the temperature drops sufficiently ($\sim 10^{8}$~K) such that ${\Gamma} \gtrsim 175$ \cite[e.g.][]{far93}, the liquid envelope solidifies via a first-order phase transition into an elastic material --- the crust. 

In general, because of the density dependence in expression \eqref{eq:Gamma}, the crust may not entirely encompass the final $\gtrsim 1$~km of the star. It is expected at least that there will be an `ocean' separating the crust from the magnetosphere \cite[e.g.][see also Sec.~\ref{sec:oceans}]{gitt23}, the physics of which depends strongly on temperature, meaning first of all that a zero-temperature EOS cannot apply \cite{GPE83} but also that no stress can be supported there. The crust-core transition takes place at a (baryon) density, $n$, {that depends on the EOS \cite{1973NuPhA.207..298N,lor93,2001A&A...380..151D}; for instance, $0.07 \lesssim n \lesssim 0.09$~$\text{fm}^{-3}$ for the BSk EOS \cite{pear18}, i.e. $\rho \sim 10^{14}$~g/cm$^3$ within a factor of a few}. Crust-ocean and ocean-atmosphere transitions occur at much lower densities which are highly temperature and composition dependent \cite[see Figure 2 in][for instance]{harut06}.

\subsubsection{Supporting stress}

In elasticity theory, the extent to which a solid can withstand stress is mathematically encapsulated by the so-called Lam{\'e} coefficients relating stress and strain \cite{wells}. At a linear level, these are just the shear and bulk modulii, the latter of which is expected to be dynamically negligible in the crust \cite{cham08}. Given that the majority of work regarding oscillations or restorative forces in the crust are discussed at a linear level, and the difficulty of microphysical calculations, little is known about the higher-order coefficients in the crust. We do not discuss nonlinear elasticity further here \cite[though see][]{andcom21,sot24}.

Still, the shear modulus, $\mu$, is the leading-order quantity relevant for elastically-supported oscillations and stress support. Even though there really are multiple shear modulii which depend on the shape of nuclei \cite{PP1998}, we neglect such complications (e.g. related to the possibility of nuclear ``pasta'') and assume a single contributor. For the standard (i.e. spherical-nuclei) shear modulus, the often-quoted expression valid at low temperatures comes from~\citet{SHOII1991}, viz.
\begin{equation}
\label{eq:mu}
  \mu_{\text{S91}} \approx 0.1194\frac{n_i(Ze)^2}{a_{i}} ,
\end{equation}
which is proportional to ${ \Gamma}$. More sophisticated variants can also be found, such as that due to \citet{hh08} who fit results to molecular dynamics simulations,
\begin{equation}
  \mu_{\text{HH08}} \approx \left( 0.1106 - \frac{28.3}{\Gamma^{1.3}} \right) \frac{n_i(Ze)^2}{a_{i}} .
\end{equation}

Next-to-leading-order temperature corrections are discussed by \citet{baiko11} and others, as are various physical corrections to these formulae \cite[again see][]{cham08}. At the linear (i.e. Hookean) level, $\mu$ is the proportionality factor relating (shear) stress, $\boldsymbol{\sigma}$, to the elastic strain, $\boldsymbol{s}^{\rm el}$, viz. \cite[e.g.][]{bell14}
\begin{equation}
    s^{\rm el}_{ij} = - \frac{\sigma_{ij}}{\mu}.
\end{equation}
In GR, arriving at a similar relationship is rather more involved, though be can achieved through the \citet{1972RSPSA.331...57C} relations defining the elastic stress tensor through a Lie derivative. A modern discussion on relativistic elasticity can be found in \citet{andcom21}. The shear stress can be related to the Lagrangian eigenfunction associated with generic motions (see Sec.~\ref{sec:calc}) through
\begin{equation} \label{eq:sigmarel}
    \sigma_{i j} = \tfrac{1}{2}( \nabla_{i} \xi_{j} + \nabla_{j} \xi_{i}),
\end{equation}
where index symmetry is manifest. The perturbed metric and Christoffel symbols also appear in the GR variant of \eqref{eq:sigmarel}. Although often the Maxwell stress is used directly to define $\boldsymbol{\sigma}$ in works involving magnetar events, this is strictly-speaking invalid: one must instead model the perturbative viscoelastic response to magnetic pressues, which tends to induce significantly lower strains than the Maxwell terms alone \cite[see][for recent emphases]{koj24,levin24}.

\subsection{Breaking strain}
\label{sec:breaking}

\begin{table*}
	\centering
	\caption{Estimates for the breaking strain, $\sigma_{\rm max}$, obtained from the literature, in ascending order according to publication date. This list is not exhaustive, though roughly illustrates the plausible range of values depending on assumptions made on crustal microphysics. The parameter $s \approx 0.185 Z^{1/3} c/v_{\rm F}$ has been introduced by \citealt{ky20} for electron Fermi velocity $v_{\rm F}$.
}
	\begin{tabular}{ccc}
	    \hline
	    \hline
	    Physical model/simulation setup & $\sigma_{\rm max}$ (dimensionless) & Reference(s) \\
     \hline
     Imperfect (alkali) metals  & $\gtrsim 10^{-5}$ & \citealt{smol70} \\
          Perfect one-component crystal & $10^{-2} \lesssim \sigma_{\rm max} \lesssim 10^{-1}$ & \citealt{kit76} \\
     Li crystals  ($57\%$~Mg, $T \lesssim 100$~K) & $10^{-5} \lesssim \sigma_{\rm max} \lesssim 10^{-2}$ & \citealt{rud91} \\
     Energy and event rates of magnetar flares & $\lesssim 10^{-3}$ & \citealt{td95} \\
		Perfect, defective, and poly-crystal ($\Gamma\sim 834$) &  $\sim 0.1$  & \citealt{hk09} \\
  	Perfect body-centred cubic crystal & $\left( 0.0195 - \frac{1.27}{\Gamma - 71} \right) n_{i} \frac{Z^2 e^2}{a_{i}}$ & \citealt{ch10} \\
   Pure and imperfect crystals (various compositions) & $\gtrsim 0.1$ & \citealt{hh12} \\
      Maximum strain set by spin limit ($\lesssim 1$~kHz) & $0.008 \lesssim \sigma_{\rm max} \lesssim 0.089$ & \citealt{fhl18} \\ 
	  Polycrystalline crust (anisotropic, variable) & $\sim 0.04$ ($< 0.3$) & \citealt{bc18} \\
   Idealised nuclear pasta ($T=1$~MeV) & $\sim 0.3$ & \citealt{cap18} \\
   Deformed mono-crystals & $\approx 0.02 \frac{\sqrt{1+1.451s^2}}{1+0.755s^2}$ & \citealt{ky20} \\
   Multi-ion (strongly ordered) crystal ($T \ll $~MeV) & $0.02 \lesssim \sigma_{\rm max} \lesssim 0.08$ & \citealt{kozh23} \\
   Near-equilibrium, stretched lattice (various composition) & $\sim 0.05$ & \citealt{b24} \\
     \hline
	    \hline
	\end{tabular}
	\label{tab:sigmamax}
\end{table*}

One anticipates that for ${ \Gamma} \gtrsim 175$ the outer layers (except for the \emph{very} outer layers being oceans and atmosphere) of the neutron star will solidify an elastic solid that can support stress up until a point where a ``failure'' event occurs. Such stresses can develop through multiple channels, being a general mass quadrupole moment or ``mountain'' \cite[e.g.][]{suvm19,gitt23}, deformations due to gradual spindown \cite{baym71,km22}, differential rotation between the crust and interior neutron superfluid \cite[i.e. spherical Couette flow;][]{rud76}, magnetic field evolution \cite[e.g.][]{suv23}, or resonant pulsations (see Sec.~\ref{sec:prectheory}). What this threshold is --- the topic of this subsection --- has important implications for a variety of phenomena therefore \cite[e.g.][]{td93,td95,lan15}.

The reason for quotations around the word \emph{failure} above is that exactly how an overstraining event manifests in the crust is not well understood. As discussed in Sec.~\ref{sec:resfail}, this may have applications for precursors also. The simplest type of picture one may have of failure is that of a brittle material. The elastic maximum is breached, and suddenly the region ``cracks'' like glass. In this way one can envision an immediate and large release of magnetoelastic energy: field lines that were once held fixed (cf. Alfv{\'e}n's flux-freezing theorem) are now free to reconnect, as the stress holding everything together falters. \citealt{jones03} and other authors have argued against this picture, effectively because the hydrostatic pressure in the crust exceeds the shear modulus by $\gtrsim 2$ orders of magnitude, inhibiting the formation of true voids. It is likely that a more realistic picture is that of \emph{plastic} deformations: the crust becomes overstrained and undergoes a permanent but not immutable deformation \cite[e.g.][]{lev16}. In this case,  heat is released as the region fails and twist is injected into the magnetosphere via plastic motions, which prime it for reconnection and magnetic eruptions leading to energy release. 

The critical strain, $\sigma_{\rm max}$, has been estimated through a variety of different techniques and approximations over the last $\sim 50$ years, as collated in Table~\ref{tab:sigmamax}. It is a very difficult problem, in general, to estimate global features of the crust via molecular dynamics or other simulations, which are inherently local. For instance, the simulations of \citealt{hk09} apply for $\sim 10^{11}$ \emph{femtometers} of material. As emphasised by \citealt{km22}, while the former authors found a global failure mechanism once the critical strain of $\sigma_{\rm max} \sim 0.1$ is reached, it is probable that in a real crust the failure mechanism differs because of lattice dislocations, grain boundaries, permanent or temporary deformities due to previous failures, and other mesoscopic imperfections.

Aside from deducing $\sigma_{\rm max}$, there are multiple criteria that have been considered as to how strain leads to failure. Arguably the most popular is that of the von Mises criterion, where one stipulates that 
\begin{equation}
\label{eq:vM}
    \sigma = \tfrac{1}{2} \sqrt{\sigma_{i j} \sigma^{i j}} \geq \sigma_{\rm max} \implies \text{failure}.
\end{equation}
Another mechanism, perhaps more physical as discussed by \citealt{ch10} and others, is the Zhurkov model \cite{wells}. The main stipulation is that thermodynamic fluctuations should exceed some threshold energy in order for the breaking to occur. It is arguably more realistic since it accounts for the fact that stress is applied over a finite time interval, which leads to a more gradual deformation of the material in question, rather than in an abrupt sense as predicted by \eqref{eq:vM}. Nevertheless, because of its simplicity, the von Mises criterion is often used in the literature \cite[e.g.][]{lan15,suvk19,suv23} and is that which we adopt in this Review.

Based on the values of maximum stress that can be sustained by the crust, the most recent and sophisticated estimates of which are in the range $\sim 0.04$ to $\sim 0.1$ (see Tab.~\ref{tab:sigmamax}; at least when discounting the possibility of pasta structures in the inner mantle), one can attempt to probe the interior from multimessenger measurements (see Secs.~\ref{sec:inspiralbasics} and \ref{sec:prectheory}).

\subsubsection{Mountains}
\label{sec:mountains}

Another way in which the critical breaking strain impacts on neutron star predictions is through the maximum mountain size (i.e. mass quadrupole moment or ellipticity). While not especially relevant for premerger phenomena per se, whether or not a star had a history of GW emission could influence evolutionary channels and especially spin (Sec.~\ref{sec:spinacc}). For example, GW radiation-reaction contributes to spindown, and thus it may be that if the maximum mountain size is small \cite[as is predicted by modern approaches;][]{gitt23}, a larger mean spin frequency in late-stage inspirals may apply. This could impact on mode frequency distributions, and the plausibility of late-stage dynamo activity (as described in Sec.~\ref{sec:dynamo}). If a neutron star taking part in a merger already has a mountain through an old pile-up of accreted material \cite[e.g.][]{mp05} or some other means, the effective $\sigma_{\rm max}$ necessary to instigate failure would be reduced. Typically though, models assume an initially relaxed (elastic) state for the crust with $\sigma(t=0) = 0$.

\subsection{Stratification gradients}
\label{sec:stratification}

The imprint of composition or temperature can be quantified introducing  the convenient parameterisation
\begin{equation}
    	\tilde{\Gamma}=\gamma(1+\delta),
\end{equation}
where $\gamma$ is the adiabatic index associated with the beta equilibrium star, $\gamma = \frac{\epsilon +p}{p}\frac{dp}{d\epsilon}$. The function (parameter) $\tilde{\Gamma}$ --- not to be confused with the Coulomb coupling parameter $\Gamma$ ---  is that associated with the perturbation, generally computed in the ``slow-reaction limit'' where one assumes that the composition of a perturbed fluid element is frozen, and is related to the sound speed via $c_s^2=p\tilde{\Gamma}/(\epsilon+p)$. As such, by enforcing that the Lagrangian variation of some particle (proton) fraction is zero, the neighbourhood of perturbed fluid elements are no longer in beta equilibrium and the system supports buoyancy modes \cite{2010aste.book.....A,unno}. Generally, $\delta$ is both a function of time and space as the star heats and has position-dependent temperature and composition, with the matter such that $\delta \geq 0$ to satisfy the Schwarzschild criterion for convective stability. Compositional impacts are described above, while thermal ones can be estimated following \citealt{Kruger14} and others
\begin{equation}\label{eq:deltatemp}
	\delta_{T}(t,\boldsymbol{x}) \approx \left[ \frac{k^2\pi^2}{6}\sum_x \frac{n_x(\boldsymbol{x})}{E_F^x(\boldsymbol{x})} \right] \frac{T(t,\boldsymbol{x})^2}{p(t,\boldsymbol{x})},
\end{equation}
for particle species $x$, with number density $n_{x}$ and Fermi energy $E^{x}_{F}$, where the sum runs over the species list. Note these latter quantities could also be treated as functions of time if chemical reactions are accounted for. The subscript $T$ indicates the thermal contribution, which clearly vanishes as $T \to 0$.

\begin{figure}
	\centering
	\includegraphics[width=0.49\textwidth]{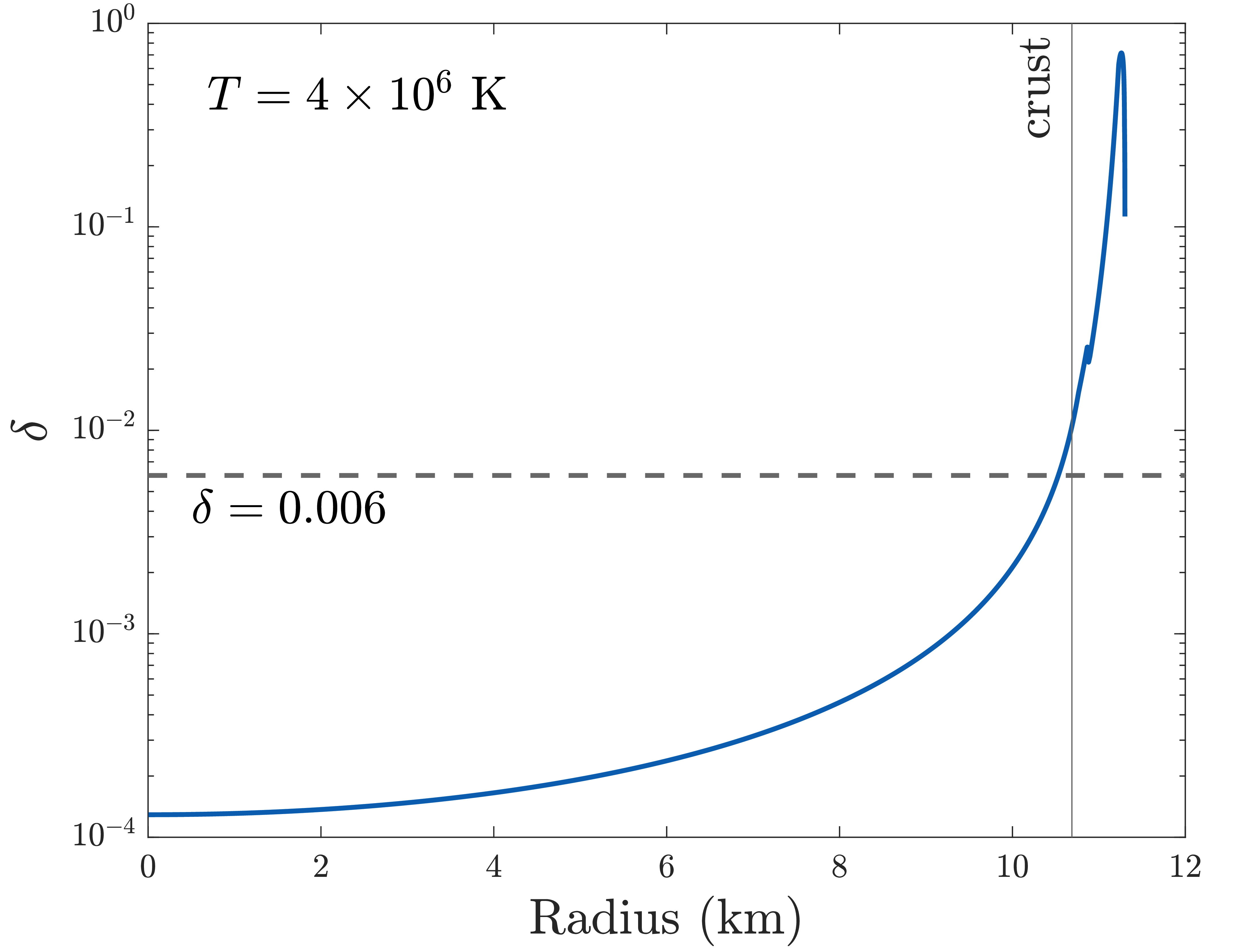}
	\caption{Stratification profile for an isothermal star for the (cold) SLy EOS with a mass of 1.41 $M_\odot$, where the crust-core boundary is indicated. A value of $\delta = 0.006$ is marked, considered roughly as a canonical average in a number of studies (see text). The curve terminates at the stellar surface, where we do not consider ocean or atmospheric layers. Adapted from \citealt{kuan23}.
	\label{fig:deltafn}}
\end{figure}

\begin{figure}
	\centering
	\includegraphics[width=0.49\textwidth]{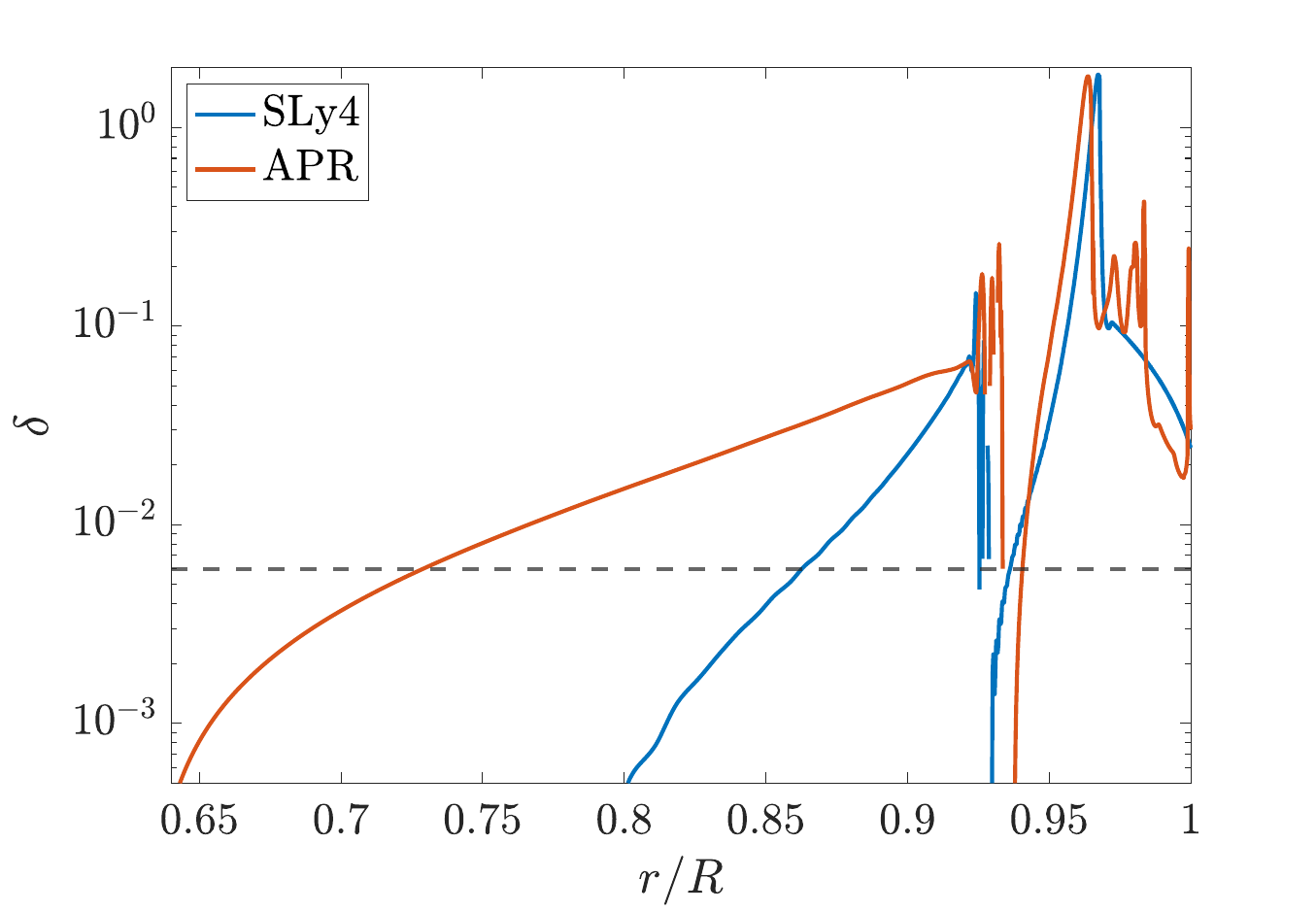}
	\caption{Stratification profile for an isothermal ($T= 10^{7}$~K) star with a mass of 1.41 $M_\odot$ for the finite temperature SLy4 (blue) and APR (red) EOS accessible within the CompOSE catalogue. The radial coordinate is normalised by the stellar radius for the respective stellar models. 
	\label{fig:deltafinite}}
\end{figure}

As such, a particular temperature profile and equation of state implies some value for $\delta$ that can be calculated self-consistently. Figure~\ref{fig:deltafn} shows one such case for an isothermal star with $T = 4 \times 10^{6}$~K. Taking the finite temperature SLy4 and APR EOS from the CompOSE catalog, we also show $\delta$ for a $1.41\,M_\odot$ star with constant temperature $T=10^7$~K in Figure ~\ref{fig:deltafinite}. A line corresponding to $\delta=0.006$ is shown for reference: such a value for the stratification is considered typical for pre-merger neutron stars in the literature \cite{xu17,kuan22gmode,ande20,ho23}.
We see that a constant $\delta=0.006$ approximately depicts the stratification in the outer regions near the crust-core interface, where the $g$-modes are mostly supported.

Within Fig.~\ref{fig:deltafinite} we normalise the (coordinate) radius by the stellar radius to compare stars with two different EOS. We truncate the domain (i.e. $r/R \ge 0.65 $) to show the $\delta$ profile in the outer core and crust only, as the stratification becomes minute at lower radii (cf. Fig.~\ref{fig:deltafn}).
The spiky behaviour near $ 0.9 \lesssim r/R \lesssim 0.95$ corresponds to densities where nuclei crush after reaching their respective saturation values \cite{2001A&A...380..151D}. Although certain types of smoothing are often performed \cite[see, e.g.][]{haen04,pote13}, we did not implement such filtering to the directly-accessible data from CompOSE. A comparison between these two figures effectively demonstrate the validity of the thermal-pressure approximation \eqref{eq:deltatemp} as compared to the use of a precise, finite-temperature EOS. Strong magnetic fields could also affect the effective stratification \cite{dex17}.

In reality, perturbations of a neutron star will lead to a departure from beta equilibrium, possible changes to the local viscosity, and heat propagation \cite{isra76,stew77,isra79,hisc83}.
The local effect of beta gradients can kill off $g$-modes with a frequency lower than the reaction rate as a result of the suppressed compositionally supported buoyancy \cite{Andersson19,coun24}, combined with the fact that heating leads to a shift in mode frequencies (see Sec.~\ref{sec:thermal}). The influences of diffusive physics in the oscillation spectrum and in the tidal interaction, in the context of coalescing binary neutron stars, are still not well understood. Some recent attempts have been made with simplified models \cite{hamm21,ripl23,sake2024}; in particular, \citealt{ripl24} suggested that these effects lead to a measurable phase shift in the GWs for high signal-to-noise-ratio events so that constraints on the bulk viscosity can be placed.

\subsection{Superfluidity and superconductivity}
\label{sec:superfc}

Mature neutron stars tend to be relatively cold with $k_{B} T \ll$~MeV. In this context, ``cold'' means that their thermal energies are well below the corresponding (core) Fermi energies, $10 \lesssim E_{F} \lesssim100$~MeV \cite[noting that the Fermi energy of neutrons is $E_F\approx 100\,(\rho/2\rho_0)^{2/3}$~MeV;][]{st83}. Because of the high degree of degeneracy, the neutrons and protons (and perhaps some exotica) occupying the core of the star are expected to become both superfluid and superconducting within $\lesssim 100$~yr \cite[see][for a reveiew]{hask18}. {This was speculated on first by \citet{mig59}, with the first proper calculations carried out by \citet{ginz64} providing support. Since then, more elaborate calculations have been
performed; see, for instance, \citealt{gand22,dris22,krot23} for recent contributions based on different many-body methods.}

The exact temperature at which such a transition occurs is both density- and pairing-mechanism dependent and a matter of active research, though is in the neighbourhood of $T_{c} \approx 10^{9}$~K (see Figure 1 in \cite{and21} for example). With the possible exception of some rare dynamical capture events \cite[][see also Sec.~\ref{sec:othereff}]{ye20}, stars taking part in a merger should be below this temperature. Various phases of the crust are also expected to be in such low-resistance states. As the strong interaction has an attractive component and neutrons and protons are fermions, they are expected to form a Cooper-pair condensate at low enough temperatures and thus neutrons in the crust are in fact likely to be superfluid \cite{wolf66}; by contrast, simple estimates all but confirm that that the electrons living in the neutron star crust are not superconducting (the critical temperature is practically zero).

One important aspect of superfluid+superconducting states in premerger objects concerns $g$-modes, and thus dynamical tides more generally. If the neutrons are superfluid they do not contribute to the buoyancy that other fluid constituents experience following some perturbation, as they are free to ``move out of the way'' \cite{pass16}. As described by \citet{kg14} and others, the neutron component is thus essentially decoupled from the oscillations and so the mass of some given oscillating fluid element is \emph{smaller} by a factor that depends on the relative particle abundances. Less inertia for the same force implies a greater oscillation frequency; typically, superfluid $g$-modes have factor $\approx 4$ larger eigenfrequencies than their normal counterparts \cite{yw17,and21}. This scaling of course depends sensitively on the exact EOS, the presence of temperature gradients, spin, and so on. 

Treating the system with a realistic, multi-fluid approach and noting that leptons (electrons and muons) are the main distributors of entropy in a superfluid core \cite{pass16}, modes supported by leptonic buoyancy exist and may be significant also \cite{kg14}. The leptonic Brunt-V{\"a}is{\"a}l{\"a} frequency does not exist in the crust however and thus these $g$-modes are unlikely to be relevant for precursors (Sec.~\ref{sec:prectheory}), though could be for GWs as their linear frequencies are several hundred Hz \cite{rau18}.

Larger frequencies imply \emph{later} resonances as far as dynamical tides are concerned, which generally means the overlap integrals will be larger but conversely that the window in which the resonances are active will be smaller (see Sec.~\ref{sec:dyntides}). Nevertheless, \citealt{yw17} found the net energy siphoned from the orbit into the oscillations is $\sim 10$ times larger than the normal fluid case. This clearly may be important as concerns inferences on the nuclear EOS from GW measurements of dephasing, as even the normal-fluid $g$-modes can be significant (see Figures ~\ref{fig:mode_evol} and \ref{fig:mode_evol20} in later sections, and also \citet{ho23}). Such an investigation was carried out by \citealt{yu23}, with superfluid $g$-mode dephasing reaching $\mathcal{O}(1\text{ rad})$ \cite[see also][]{yu24}. Later resonances may, however, have a more difficult time in explaining \emph{early} precursor flare onset times (see Sec.~\ref{sec:onset}).

Superconductivity has been studied less in the premerger context. This could in principle distort the star significantly away from spherical symmetry \cite[][which influences the tidal coupling]{land13} and shift the mode spectra through magnetic corrections as the Lorentz force scales like $\sim H_{c1} B$ which can be large even if $B$ is relatively small (see Sec.~\ref{sec:spectral} and \citealt{suvucxbs}). The lower critical field can be estimated through [see equation (7) in \citet{glamp11}]
\begin{equation} \label{eq:critfield}
    H_{c1} \approx 4 \times 10^{14} \left( \frac{m_{p}} {m^{\ast}_{p}} \right) \left( \frac{x_{p}} {0.1} \right) \left( \frac {\rho} {10^{14} \text{ g cm}^{-3}} \right) \text{G},
\end{equation}
for proton fraction, $x_{p}$, and effective (entrained) mass, $m^{\ast}_{p}$. In the cores of some heavy stars under some EOS, equation~\eqref{eq:critfield} could reach $\sim 10^{16}$~G. Nevertheless, in instances where Figures are shown in this work, the impacts of superfluidity and superconductivity have been ignored in calculating mode properties.

\section{The mechanics of late inspirals: gravitational waves}
\label{sec:inspiralbasics}

Consider a binary, involving at least one neutron star, with component masses $\MA$ and $\MB$. Ignoring complications about how the objects may reach short orbital separations $a$ \cite[this issue is reminiscent of the famous ``final parsec problem'', though for neutron-star mergers the solution is likely rooted in common-envelope theory;][]{taam00}, the (quadrupole formula) GW inspiral time reads
\begin{equation}
\label{eq:gwinspiral}
    \tGW \approx 5 \left[ \frac{a \left(1-e^2\right)^{7/8}} {3 \times 10^{10} \text{ cm}} \right]^{4} \left( \frac{ M_{\odot}^3} {\MA^3 q (1+q)} \right) \text{Myr}.
\end{equation}
Since the pioneering works of \citealt{Peters63,pet64}, it is anticipated that compact binary-mergers circularise well before merger, so that the eccentricity $e \ll 1$. Note the normalization in expression \eqref{eq:gwinspiral}: although it would only take light $\sim 1$~second to travel between two such stars, GW radiation-reaction takes $\sim$~Myr. Narrowing our attention immediately to late stages where the separation is $\lesssim 20 R_{A}$ for ``canonical values'' of the stellar radius $\RA = 12$~km and mass $\MA = 1.4 M_{\odot}$ (see Sec.~\ref{sec:eos}), we find $\tGW$ from \eqref{eq:gwinspiral} corresponds to tens of seconds, or more precisely that
\begin{equation} \label{eq:inspiral2}
    \tGW \approx 24.5 (a/20 \RA)^4 q^{-1} (1+q)^{-1} \text{ s}.
\end{equation}
The convergence to coalescence, occurring roughly when $a \lesssim 3 (\RA+\RB)/2$ \cite[e.g.][]{ls94a,ls94b}, accelerates rapidly in the final stages. The orbital separation can be inversely related to the orbital frequency $\Oorb$ through a Keplerian or quasi-Keplerian relationship, and thus $\Oorb$ grows in time. This sweep-up behaviour culminates in what is known as the ``chirp'' in the GW community.

As we explore throughout the remainder of this Section, the simple result \eqref{eq:inspiral2} is an overestimate \cite[e.g.][]{koch92}. When considering PN or finite-size effects pertaining to \emph{tides}, the inspiral time is reduced as additional means of energy depletion become available: the impact of tides increases as the orbital frequency sweeps up because the \emph{tidal forcing terms} similarly increase in frequency. Getting a precise estimate of the tidal \emph{dephasing} is of critical importance when trying to connect to observational data, since without accurate inference of the time relative to merger (in some appropriate frame of reference) that some high-energy events occurs, one cannot hope to extract the maximum amount of information. More generally though, the tides encapsulate details about stellar structure and thus matching dephasing templates to data can be used to learn about many areas of fundamental physics. 

\subsection{Tides: general theory}
\label{sec:tidesgeneral}

In the final stages of a binary inspiral, tidal effects become significant. These are usually separated into two distinct effects (i) those associated with the ``equilibrium'' or ``adiabatic'' tides (Sec.~\ref{sec:eqtides}), and (ii) those associated with ``dynamical'' tides (Sec.~\ref{sec:dyntides}).

\begin{figure*}
	\centering
	\includegraphics[scale=.6]{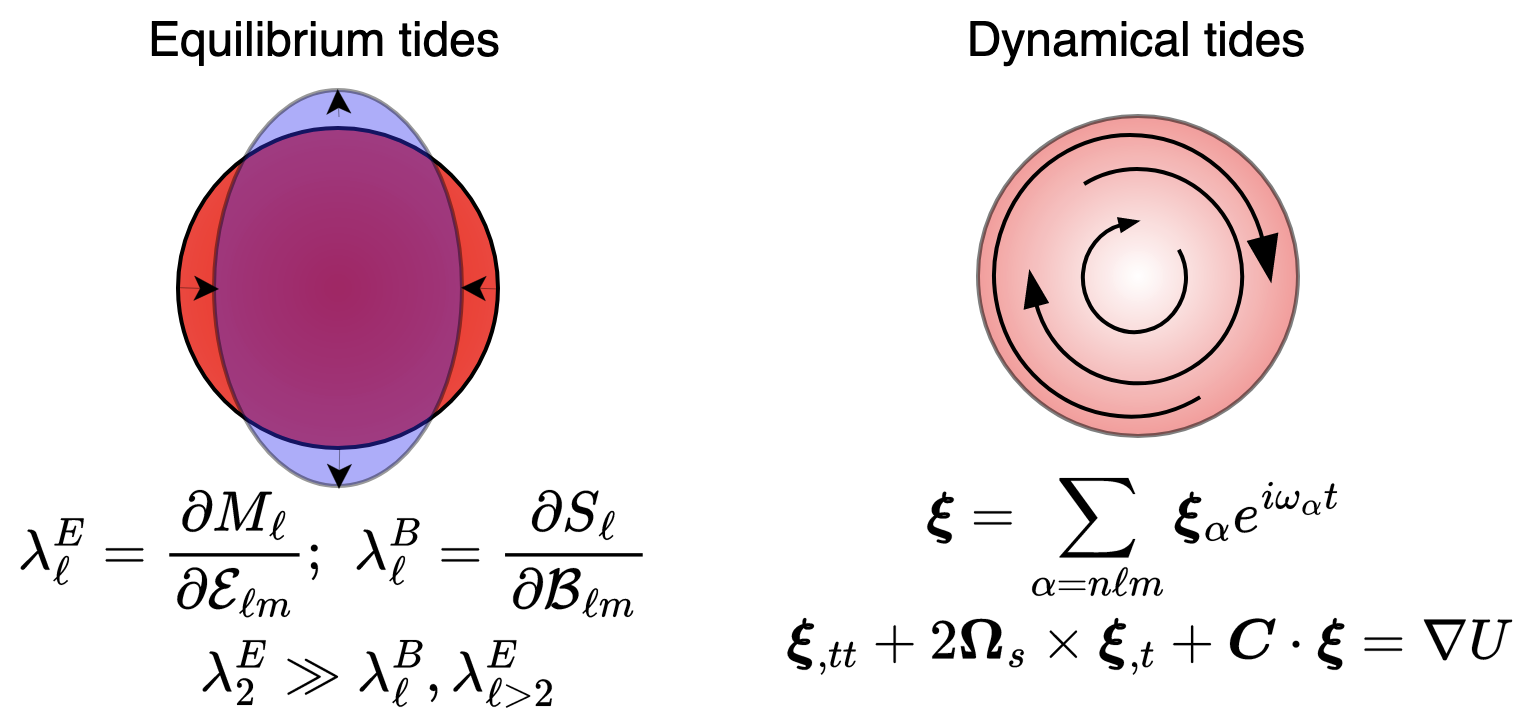}
	\caption{A cartoon demonstration of the difference between equilibrium and dynamical tides. The system on the left exhibits the typical (though greatly exaggerated) quadrupolar deformation, meaning that the quadrupolar, electric Love number $\lambda^{E}_{2}$ is very large. In general, this implies that the mass quadrupole moment is strongly susceptible to tidal perturbations. The system on the right, by contrast, remains (approximately) spherical though has internal fluid motions excited to a large degree (i.e. strong dynamical tides). These are described by a coupled set of forced oscillation equations, as detailed in Sec.~\ref{sec:calc}.}
	\label{fig:tidesfig}
\end{figure*}

Within the literature, one can find a few different definitions for the qualifiers ``adiabatic'', equilibrium'', and ``dynamical'' in the context of tides. For example, following \citealt{yu24}, the word ``adiabatic'' emphasises that one considers deformations in the zero-frequency limit $\Oorb \rightarrow  0$. This is more typically called the equilibrium tide, though \citealt{yu24} make the distinction by allowing the latter to include $\Oorb >0$ influences. Finite-frequency terms involve both non-resonant and resonant excitations, the latter of which these authors reserve for ``dynamical'' tide status. In other works, the inclusion of non-resonant, but time-dependent terms leads to corrections of the equilibrium quality factors which are then called ``effective'' \cite[e.g.][]{pnig24}. In this Review, the word ``dynamical'' conveys $\Oorb > 0$, though we are mostly interested in resonances and their connections to high-energy phenomena. We emphasise at this stage that many of the formulae that we present are done so in a Newtonian language. When one reaches the level of wishing to actually compare results with data in a serious way, it will generally be necessary to consider fully GR expressions insofar as possible\footnote{Subtle issues arise in GR, making this non-trivial. For example, even the notion of eccentricity is ambiguous \cite{knee22}. An all-encompassing notion of ``tide'' is also hard to define via the Weyl tensor or other means; see Section I.C in \citealt{pit24}.}. Given that our main purpose here is pedagogical, we avoid writing out formalisms in this way, though point the interested reader toward relevant literature where appropriate. A depiction of these two class of tides is given in Figure~\ref{fig:tidesfig}.

In the Newtonian language, the essential basics of tides can be formulated as follows. Our neutron star primary, of mass $\MA$ and radius $\RA$ and spin angular momentum vector $\bOs$, is in an orbit with a companion of mass $\MB$. Although it is generally expected that spin-orbit misalignment is small in late-stage binaries (cf. Sec.~\ref{sec:rotation}), we introduce a spin-orbit inclination angle $\Theta$ --- the angle made between $\bOs$ and the orbital angular momentum, denoted ${L}$. In principle the problem can be further complicated with several additional angles if the magnetic field becomes dynamically important (magnetic inclination angles; cf. Sec.~\ref{sec:magneticfields}), the finite size of the secondary is also considered (subtended angles), or the secondary is spinning ($\Theta_{B}$). These effects are almost always ignored in the literature.

Following \citealt{Lai06} and others, we orient a spherical coordinate system around (the centre of mass of) $\MA$ with the ``$z$-axis'' directed along ${L}$. The gravitational potential produced 
by $\MB$ can then be expanded in terms of spherical harmonics $Y_{\ell m}$ \cite[e.g.][]{Press77,reis94} 
\begin{equation}
\label{eq:ueqn}
U(\bx,t)
 =-G \MB \sum_{\ell m'}\frac{W_{lm'}r^\ell} {a^{\ell+1}} e^{-im'\Phi(t)} Y_{\ell m'}(\theta_L,\phi_L),
\end{equation}
for orbital phase $\dot{\Phi}(t)= \Oorb$, and we define
\begin{equation}
\begin{aligned}
W_{\ell m'} &=  (-)^{(\ell+m')/2}\left[\frac{4\pi} {2\ell+1}(\ell+m')!(\ell-m')!
\right]^{1/2} \\
& \times \left[2^\ell\left(\frac{\ell+m'} {2}\right)!\left(\frac{\ell-m'} {2}
\right)!\right]^{-1}.
\label{eq:wlm}
\end{aligned}
\end{equation}
Here, the symbol $(-)$ is zero if raised to a non-integer power and one otherwise. Although one can work with the full harmonic expansion here, it is generally only the $\ell = 2$ and $|m| \leq 2$ terms that are observationally relevant \cite{zahn66,zahn77}; for $\Theta = \pm \pi/2$, it is often only the $m'=0$ (equilibrium tide) and $m' =2$ (dynamical tide) terms that are relevant. 

The above-described coordinate system can be related to the more natural one for describing fluid motion in the primary star. Consider now angular coordinates with respect to the corotating frame of the neutron star, with the ``$z$-axis'' now oriented along $\bOs$. Fluid variables, decomposed into different sets of spherical harmonics, can thus be related through
\begin{equation} \label{eq:harmonicWigner}
Y_{\ell m'}(\theta_L,\phi_L)=\sum_{m}{\cal D}^{(\ell)}_{m m'}
(\Theta)Y_{\ell m}(\theta,\phi_s),
\end{equation}
where $\mathcal{D}^{(\ell)}_{mm'}$ is the so-called Wigner $\mathcal{D}$-function \cite[e.g.][]{Ho98,xu17,Kuan21a}, and $\phi_s=\phi+|\bOs| t$.

This small bit of theory is actually sufficient to specify the problem (again, at the Newtonian level): one wishes to model the response of the internal neutron star fluid to an external acceleration given by $\nabla U$, with $U$ given by \eqref{eq:ueqn}. In practice within the literature, this is achieved in a few steps using some mathematical trickery.

\subsubsection{Mathematical description and calculation methods}
\label{sec:calc}

First, one wishes to solve for the ``background'' (magneto-)hydrostatics. For cold, mature, and not ultra-magnetised stars this involves employing a barotropic EOS, $p = p(\rho)$, popular candidates of which are described in in Sec.~\ref{sec:eos}; in principle, thermal contributions can play a role at late times but these are often ignored (see Sec.~\ref{sec:thermal}). Once the relevant background has been constructed, a set of free perturbations are introduced for each variable $X$ (e.g. density, pressure, gravitational potential, velocity field, $\ldots$) through an Eulerian scheme
\begin{equation}
    X \rightarrow X + \delta X + \mathcal{O}(\delta^{2} X) .
\end{equation}
The linear equations of motion are conveniently expressed using displacement vector
\begin{equation} \label{eq:lagdisplacement}
    \delta \boldsymbol{u} = \boldsymbol{\xi}_{,t},
\end{equation}
such that one obtains a ``master equation''
\begin{equation}
\label{eq:master}
    \boldsymbol{\xi}_{,tt} +2 \boldsymbol{\Omega}_{s} \times \boldsymbol{\xi}_{,t} + \boldsymbol{C} \cdot \boldsymbol{\xi} +\mathcal{O}(\xi^2) = 0,
\end{equation}
from which all\footnote{This is no longer strictly true in GR (or when including complicated microphysics capable of independent, secular responses) as the metric variables, even in the Regge-Wheeler gauge for example, cannot be uniquely expressed in terms of this displacement (cf. $w$-modes, which exist even in the no-fluid limit where $\boldsymbol{\xi} \to 0$).} other variables follow [e.g. $\delta \rho = - \rho \nabla \cdot \boldsymbol{\xi} - (\boldsymbol{\xi} \cdot \nabla) \rho$]. In the above,  $C_{ij}$ is some spatially-dependent tensor (defining some self-adjoint operator) that depends on the particulars of the problem under consideration (Newtonian, GR, Cowling, etc). Once some boundary conditions are imposed (e.g. regularity at the centre as $r \to 0$), the free-mode problem is fully specified.

Again in practice however, equation \eqref{eq:master} is solved through decomposition. In the irrotational case, the spatial dimensions of the problem are separated out using the spherical harmonics \eqref{eq:harmonicWigner} where $\phi_{s} \to \phi$, depending on the \emph{polarity} of the eigenfunction, via
\begin{align}\label{eq:eigenfn}
	&\xi_{\alpha}^{r}(\boldsymbol{x},t) = r^{\ell-1} W_{n \ell}(r,t) Y_{\ell m}(\theta,\phi)\\
	&\xi_{\alpha}^{\theta}(\boldsymbol{x},t) = -r^{\ell-2}V_{n \ell}(r,t)\partial_{\theta}Y_{\ell m}(\theta,\phi)\\
	&\xi_{\alpha}^{\phi}(\boldsymbol{x},t)= -r^{\ell}(r\sin\theta)^{-2}V_{n \ell}(r,t)\partial_{\phi}Y_{\ell m}(\theta,\phi),
\end{align}
for radial ($W$) and tangential ($V$) functions, where we have introduced  the shorthand ``$\alpha$'' to mean a generic set of quantum numbers $n, \ell,$ and $m$ and we have 
\begin{equation}
    \bxi = \sum_{\alpha = n\ell m} \boldsymbol{\xi}_{\alpha}(\boldsymbol{x},t).
\end{equation}
Modes with $n>1$ are referred to as ``overtones'', as $n$ is defined by counting the number of radial nodes in the eigenfunction \cite[see, e.g.][]{unno}. Rotation, however, generally makes a separation of variables as above impossible, and a further sum over a dummy azimuthal index becomes necessary unless one further sets up a hierarchy in powers of $|\boldsymbol{\Omega}_{s}|$ \cite[e.g.][]{stroh91}. 

Finally, it is usually numerically more straightforward to evolve \eqref{eq:master} in the Fourier domain, where one further introduces a temporal decomposition through
\begin{equation}
\label{eq:fourier}
    \boldsymbol{\xi}_{\alpha}(\boldsymbol{x},t) = \boldsymbol{\xi}_{\alpha}(\boldsymbol{x}) e^{i \omega_{\alpha} t},
\end{equation}
where the abuse of notation is apparent (and the carry-over to $V$ and $W$ in the static case is straightforward). Here we remark that $\omega_{\alpha}$ is the (angular!) mode frequency in the co-rotating frame (that is, ``according to the star''); the inertial-frame (``laboratory'') frequency instead reads
\begin{equation} \label{eq:inertialframe}
    \omega_{\alpha,i} = \omega_{\alpha} - m |\boldsymbol{\Omega}_{s}|.
\end{equation}
In solving mode problems one must also take care to note that complex conjugates generally also solve the master equation, though we ignore this complication here for pedagogical purposes. \cite[see][for a recent discussion]{pnig24}.

At this stage, the amplitude of the modes have not yet entered, as these fall out of the homogeneous equation \eqref{eq:master}. Enter the tides. Formally this amounts to instead solving the inhomogenous version of the master equation,
\begin{equation}
\label{eq:inhommas}
    \boldsymbol{\xi}_{,tt} +2 \boldsymbol{\Omega}_{s} \times \boldsymbol{\xi}_{,t} + \boldsymbol{C} \cdot \boldsymbol{\xi} = \nabla U.
\end{equation}
Solving this equation is conceptually straightforward: project $U$ into a set of spherical harmonics, as we have already done in expression \eqref{eq:ueqn}, and repeat the above procedure making use of orthogonality relations (taking care to ensure that one does not confuse the mode quantum numbers with the tidal field quantum numbers). This does not quite work as easily as one might hope in practice however because the symmetries of $U$ are not shared by $\boldsymbol{\xi}$, meaning that the tidal field distorts the spectrum ($\omega_{\alpha}$ and $\boldsymbol{\xi}$) as well as driving the system. In fact, the expansion \eqref{eq:fourier} may not produce anything useful because a non-harmonic forcing term equation typically forbids harmonic time-dependencies and orthogonality, and thus the ansatz involving harmonics (in both time and space) is not even necessarily well-defined.

Fortunately, except possibly at very late stages in the inspiral, the tidal distortion of the spectrum is weak \cite{denis72}, as can be formally estimated with the \citealt{unno} formula; see Sec.~\ref{sec:spectralmods}. One thus considers a ``volume-averaged'' problem where the amplitude evolution, $q_{\alpha}(t)$, of each mode is considered in isolation. The problem is thus reduced from 1+3 to 1+0 dimensions, and we will end up with some ODEs for the amplitude evolution. The key step involves introducing an inner-product,
\begin{equation} \label{eq:orthog}
\langle \boldsymbol{A},\boldsymbol{B}\rangle = \int d^3x\,\rho\, \left( \boldsymbol{A}^\ast\cdot \boldsymbol{B} \right),
\end{equation}
between the modes ($\boldsymbol{A}$) and some angular harmonic of the tidal potential ($\boldsymbol{B})$. Crucially, this inner-product defines some-kind of weighted integral over \emph{space} only. Applying this bracket to both the left- and right-hand sides of \eqref{eq:inhommas}, one finds \cite{Lai06}
\begin{equation}
\label{eq:ampevol}
\begin{aligned}
{\dot q}_\alpha+i\omega_{\alpha} q_\alpha &= 
{\frac{i} {2\varepsilon_\alpha}}\langle\bxi_\alpha,-\nabla U\rangle \\
&= \sum_{m'}f_{\alpha,m'}\,e^{im|\boldsymbol{\Omega}_s| t-im'\Phi},
\end{aligned}
\end{equation}
with
\begin{equation}
f_{\alpha,m'}=\frac{iG\MB}
{2\varepsilon_\alpha}\sum_{\ell}{\frac{W_{\ell m'}} {a^{\ell+1}}}{\cal D}^{(\ell)}_{mm'}
Q_{\alpha,\ell m},
\end{equation}
where the \emph{overlap integrals}\footnote{Two ways of generalising overlaps to GR have been proposed in \citealt{Kuan21a} and \citealt{miao24}. The orthogonality between modes and the sum rule for tidal overlaps \cite{Reisenegger94} are equally respected by both definitions, while only the former predicts a vanishing $g$-mode overlap in the zero stratification limit (at least for a simple, constant $\delta$ law; see Sec.~\ref{sec:stratification}). This issue is related to overlap ``leakage'' discussed in Sec.~\ref{sec:imodes}. Throughout we adopt the definition of \citealt{Kuan21a}.} read
\begin{equation}
Q_{\alpha,\ell m}=\left\langle\bxi_\alpha,\nabla (r^\ell Y_{\ell m})
\right\rangle,
\label{eq:overlap}
\end{equation}
and a spin-offset is introduced through
\begin{equation}
\varepsilon_\alpha=
\omega_{\alpha}+\langle\bxi_\alpha,i\bOs\times\bxi_{\alpha}\rangle,
\end{equation}
and we have used the normalization $\langle\bxi_\alpha,\bxi_\alpha\rangle
=1$.

\begin{table*}
	\centering
	\caption{A non-exhaustive list of Post-Newtonian effects that influence the inspiral for compact objects. {The order counting presented in the first column is quoted with reference to the power of $c$ of the associated coefficient(s) in the equations of motion for binaries (i.e., $N$ PN corresponds to a $c^{-2N}$ scaling).}
 Constraints on each of these orders can be found in the violin plots produced by the LVK consortium; see \citealt{abb19}. In principle, stellar structure and multipole moments corrections together with additional couplings enter at ever-higher orders \cite{thorne80}. {Some additional modified gravity terms may also enter at lower PN orders; see Table 3 in \citet{yu16}.}
}
	\begin{tabular}{cll}
	    \hline
	    \hline
	    Post-Newtonian order & Effect(s) & Reference(s) \\
     \hline
		0 & Energy deposited into modes & \citealt{alex87,alex88} \\
      		1 & Stellar structure; pericentre advance & \citealt{wago74,ande23} \\
  		{1.5} & Scalar-field contributions to dynamics (non-GR) & \citealt{Mirshekari13,yagi16} \\
          		1.5 & Spin-orbit coupling; tail backscatter & \citealt{blan95} \\
              		2 & Self-spin, spin-spin, quadrupole-monopole couplings, mag. dipoles & \citealt{pois98,ioka00} \\
  		2.5 & Quadrupole formula for GWs & \citealt{Peters63} \\
  3 & Gravitational tails of tails & \citealt{blan99} \\
  4 & Dissipative tidal number & \citealt{ripl23,sake2024} \\
  5 & Gravitoelectric quadrupole Love number & \citealt{Damour12,yagi14} \\
    		$6$ & Gravitomagnetic quadrupole Love number & \citealt{abde18,henry20} \\
    		$6.5$ & Spin-tidal coupling & \citealt{abde18,cast22} \\
	    \hline
	    \hline
	\end{tabular}
	\label{tab:pneffects}
\end{table*}

Thus, provided the free mode spectrum can be constructed, one need ``only'' solve equation \eqref{eq:ampevol}. This, however, is still not quite the end of the story as the orbital phase $\Phi$ must be evolved simultaneously. By modelling the fluid motion in the neutron star as a set of harmonic oscillators \cite{alex87,alex88,scha24} and incorporating mode kinetics into the Hamiltonian of the binary, \citet{schafer85,1995MNRAS.275..301K,Kuan21a} have shown how this can be achieved with a high-order PN method, though with the modes themselves calculated in full GR without Cowling (i.e. including the metric perturbations). While not all are taken into account \cite[such as those occurring in gravities with non-metric fields where dominant mono- or dipolar radiation exists;][]{yagi16}, we list PN orders at which various effects occur in Table~\ref{tab:pneffects}. We remark in this respect that although, strictly speaking, radiation-reaction is a 2.5 PN effect, often this is said to appear at \emph{Newtonian} order when using a Newtonian definition for the system's quadrupole moment(s) (see below). 
{To avoid confusion between different conventions for PN order counting, we focus on effects on binary motion and do not refer to the hierarchical imprints on GWs.}

PN effects in shaping stellar structure have been examined in several references \cite[e.g.][]{wago74,suv18,ande23} and starts already at first order, though no systematic analysis of how these directly influence the inspiral has been carried out. Tides and spins contribute hierarchically at several PN orders accounting, for instance, for the breakdown of the point-particle approximation. Tides absorb orbital energy in the Newtonian theory already (i.e. at $0^{\rm th}$ order), while spin couples to the orbital angular momentum at the 1.5$^{\rm th}$ order and to self-spin and companion spin at 2$^{\rm nd}$ order. At 2.5$^{\rm th}$ order enters the leading-order dissipation due to GW emission [cf. equation \eqref{eq:gwinspiral}]. The electric- (magnetic-) type stationary deformations factor into the dynamics at 5$^{\rm th}$ (6$^{\rm th}$) order, as introduced in Sec.~\ref{sec:eqtides}, and the coupling between this deformation to the spin follows at the next half order. GW170817 placed constraints on each of these PN orders, as detailed in \cite{abb19}.

With the above in mind, the notion of \emph{dephasing} can be made precise. In the stationary phase approximation, the frequency-space GW phase, $\Psi$, can be written as \cite[e.g.][]{Finn93,Cutler94}
\begin{equation}\label{eq:psi_lagendre}
	\Psi(f_{\rm gw}) = 2\pi f_{\rm gw} t_{\rm ref} -\phi(t_{\rm ref}) -\frac{\pi}{4},
\end{equation}
where $t_{\rm ref}$ is a given reference time and $\phi(t)$ is the \emph{time-domain} phase associated with $f_{\rm gw} = \Omega_{\rm orb}/\pi$ and the shift $\pi/4$ is conventional. The quantity $\Psi$ can be shown to satisfy \cite{Damour12}
\begin{equation}
	\frac{d^2\Psi}{d\Omega_{\rm orb}^2}= \frac{2Q_{\omega}}{\Omega_{\rm orb}^2},
\end{equation}
for some dimensionless $Q_{\omega}$ (a quality factor akin to the overlap integrals), measuring the phase acceleration, viz.
\begin{equation}\label{eq:phiacc}
	Q_{\omega}=\Omega_{\rm orb}^2\left(\frac{d\Omega_{\rm orb}}{dt}\right)^{-1}.
\end{equation}
The dephasing, called $\Delta \Phi$ here, is thus just the difference between the calculated $\Psi$ from \eqref{eq:psi_lagendre} when relevant terms are kept (i.e. the PN ones described above) as compared to when they are de facto ``switched off''.

\subsubsection{Tides: equilibrium}
\label{sec:eqtides}

The equilibrium tide simply corresponds to the ``$\Phi(t) = 0$'' portion of the dynamics. In this case, the quasiharmonic time dependence of \eqref{eq:ueqn} drops out, and our interest shifts to bulk, geometric deformations of the stellar surface (``zero-frequency oscillations''). The extent to which the stellar interior is suspectible to a time-independent tidal potential can be encapsulated by the (shape) \emph{Love numbers}; effectively, much like the $Q_{\alpha}$ defined previously, these measure the extent of orthogonality between the stellar fluid (or solid) and some angular portion of the tidal field (i.e. $\ell, m'$). In general, therefore, there is an infinite set of Love numbers, though often in the literature one will find the term ``Love number'' just to mean the quadrupolar, $m'=0$ Love number. In the context of compact binaries in GR, \citealt{dam83} first quantified how tidal Love numbers influence inspiral. Note also that if the star is static and axisymmetric at background order, the index $m'$ falls out of the equations and thus the static, or even effective, Love numbers are characterised by a single harmonic number $\ell$. These reduced Love numbers are typically denoted $k_{\ell}$. 

Again working in the static limit for simplicity, the Love numbers define proportionality factors weighting the multipole moments, $I_{\ell m}$, of the previously-spherical star when affected by the tidal field. One has \cite{dam09,bp09}
\begin{equation} \label{eq:massquaddef}
    G I_{\ell m} = \frac{2 \ell + 1}{2 \pi \ell!} k_{\ell} R^{2 \ell +1} \mathcal{E}_{\ell m},
\end{equation}
where the tidal moments are defined implicitly by the relations
\begin{equation}
    \mathcal{E}_{a_{1} a_{2} \ldots a_{n}} x^{a_{1}} x^{a_{2}} \cdots x^{a_{n}} = \sum_{m} \mathcal{E}_{\ell m} r^{\ell} Y_{\ell m}(\theta ,\phi)
\end{equation}
and
\begin{equation}
    U = - \sum^{\infty}_{\ell =2} \frac{1}{\ell!} \mathcal{E}_{a_{1} a_{2} \ldots a_{\ell}} x^{a_{1}} x^{a_{2}} \cdots x^{a_{\ell}},
\end{equation}
which is the same $U$ from \eqref{eq:ueqn}. Note that the multipole moments $I_{\ell m}$ are also implicitly defined at some PN order; see Section V in \citealt{thorne80}. At Newtonian order, $I_{\ell m}$ is defined through an integral over the mass-density weighted by a spherical harmonic and radius to power $\ell$:
\begin{equation}
    I_{\ell m} = \frac{16 \pi}{(2 \ell + 1)!!} \left[ \frac{\ell^2 + 3\ell +2}{2 (\ell^2 -\ell)} \right]^{1/2} \int d^{3} x \rho Y^{\ell m \ast} r^{\ell}.
\end{equation}

In a GR setting, the theory becomes somewhat more involved. The Love numbers now acquire a magnetic counterpart. This is essentially because all forms of energy gravitate, and thus angular momentum deposits made by the tidal field influence the spacetime through \emph{current multipoles}: the ``gravitomagnetic'' portion of the field can excite some current-like components provided the star is not static and axisymmetric \cite{pani18}. Essentially one can find the correspondence \cite{yagi14,pani15}
\begin{equation}
\label{eq:tidmoms}
    \lambda^{E}_{\ell} = \frac {\partial M_{\ell}} {\partial \mathcal{E}_{\ell m}} \qquad ; \qquad \lambda^{B}_{\ell} = \frac {\partial S_{\ell}} {\partial \mathcal{B}_{\ell m}} ,
\end{equation}
where the (axisymmetric) mass ($M_{\ell}$) and current ($S_{\ell})$ multipole moments can be defined either via the \citealt{thorne80} or Geroch-Hansen \cite{hansen74} formulae. These formulations are totally equivalent by a theorem due to \citealt{gur83}. In the above we have the odd and even parity sectors of contractions of the trace-free Weyl tensor \cite[e.g.][]{damo91,pani15}. Using the schemes introduced by \citealt{pap15,suvm16} to generalise the Geroch-Hansen definitions to theories beyond GR, one could try to  extend the correspondence \eqref{eq:tidmoms} to some other theory of gravity. This has not yet been attempted.

\begin{figure*}
	\centering
	\includegraphics[width=\textwidth]{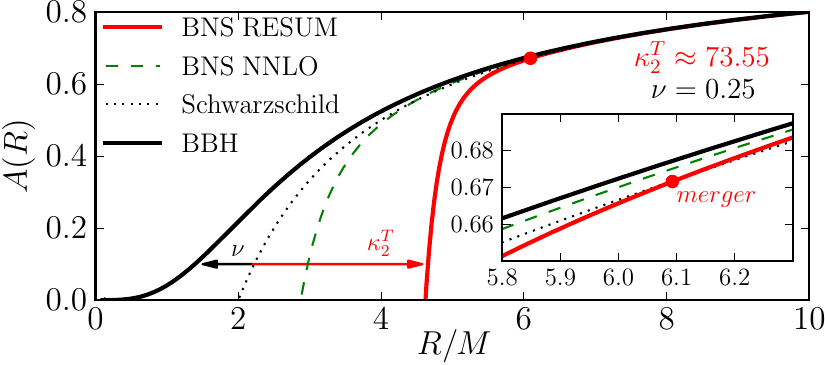}
	\caption{The main radial gravitational potential $A(R)$ --- the lapse function of the effective-one-body spacetime over which the dynamics of a test particle describes the dynamics of the real binary system \cite{buon99,buon00} ---
as a function of the radial coordinate $R=a$ with $M = M_{A} + M_{B}$ (for this Figure only) in a tidal-EOB model.
The binary black hole sector of this model (solid black) shows a deviation from the lapse of the Schwarzschild spacetime (dotted), indicating the dynamics of a test particle on the effective spacetime is different from that of a (plunging) orbit around a Schwarzschild black hole due to the finite symmetric mass ratio $\nu = M_AM_B/M^2$ of the binary (which should not be confused with the our use of $\nu$ as stellar spin used elsewhere).
For the neutron star case, the gravitational potential including up to next-to-next-to-leading-order tidal effects (i.e. up to 7 PN; green dashed) is shown together with the result resumming the gravitational-self-force information up to 7.5 PN \cite[][red]{bini14}.
The specific equal-mass neutron-star binary considered here has a tidal deformability (equation \ref{eq:kappaT}) of $\kappa_2^T\approx73.55$. 
From \citealt{bern15} with permission. }
	\label{fig:teob_A}
\end{figure*}

Tides effectively enhance the ``attraction'' between the components of the binary, leading to earlier merger relative to binary black-holes or point-particles. The tidal dynamics are elegantly described by a radial interaction potential in the effective-one-body (EOB) formalism \cite{bini12,bern15,hind16}. PN models of tides in the EOB formalism were initially verified against numerical simulations by 
\citealt{bai10,bern12}, who demonstrated that the models can be improved with such calibrations but that it is not possible to accurately describe tides close to merger with a PN model. Such a restriction was lifted by \citealt{bern15} through \emph{resumming} techniques. A schematic depiction of this attractive feature is shown in Figure~\ref{fig:teob_A}, where the red curve describes the radial potential for a neutron-star binary with tidal resumming. A significant drop in the potential occurs at larger radius, indicating the stronger attraction felt by the binary than the equivalent binary black hole.
The dominant term of the tidal imprints on waveform is delivered by the quadrupole quantity \cite{Damour12}
\begin{align}\label{eq:kappaT}
\kappa_{2}^T = \frac{3}{13}\left[ \frac{(M_A+12M_B)M_A^4\Lambda_A+(M_B+12M_A)M_B^4\Lambda_B} {(M_A+M_B)^5} \right],
\end{align}
for $\Lambda_{A}=\frac{2}{3}k_2^{A}(R_A/M_A)^5$ with the same definition for object $B$ \cite[see also][]{hind08,flan08,Hinderer09,damo10}. Note $k_2$ is defined in expression \eqref{eq:massquaddef}, effectively being the constant of proportionality between the quadrupole moment and the quadrupolar tidal potential (appropriately generalised to GR via the Weyl tensor; though cf. Footnote 4).

\subsubsection{Tides: dynamical}
\label{sec:dyntides}

Pulsation modes inside compact stars, tidally-forced or otherwise, are generally characterised according to the nature of the restoring force that ultimately damps the oscillations. The theory of mode oscillations is an active and rich area of research, which we cannot do justice in this Review. We therefore refer the reader to, e.g. \citealt{1998MNRAS.299.1059A,kk01,Pratten19,ande21}, though a few QNM groups are particularly important in the context of GWs and tides, so we introduce them briefly. 

Modes that are restored by pressure are known as $p$-modes, with the lowest radial quantum-number ($n=0$) mode referred to as the ``fundamental'', or $f$-mode. These modes remain non-degenerate in the spectrum in the limit that all physical ingredients (rotation, magnetic fields, stratification, \dots) are discarded except for the hydrostatic pressure. Including more physics in the model will augment the $p$-spectrum, but the classification remains the same. Some other modes obtain a hybrid-like character when additional physics is included, in the sense that the spectrum is strongly codependent on more than one variable, such as the torsional (magneto-elastic) modes \cite{col12,gab12,gab16} and the inertial-gravity modes \cite{papa81,lock99,Lai06}. The excitation of these modes during inspiral comes at the expense of the orbital energy, as described mathematically in Sec.~\ref{sec:calc}, which can be computed using numerical techniques \cite{Lai94,1995MNRAS.275..301K,yw17,Passamonti20,will22,prat22,kk22,kk23,yu24}.

Modes that are instead restored by buoyancy are referred to as $g$-modes (i.e. ``gravity'' modes, not to be confused with GWs). The $g$-modes, in contrast to $f$-modes, do depend sensitively on the internal composition of the star (see Sec.~\ref{sec:stratification}) and thus may be able to reveal a different kind of information. Depending on the chemical composition of the star, the $g$-modes tend to follow different relations as described by \citealt{kuan22gmode}. That is to say, whether the EOS is purely hadronic or hybrid (for instance) has a strong impact on the scaling of the modes with the mean density, temperature, and other, microphysical parameters. Generally speaking, the $g$-mode spectrum will be influenced by both composition and temperature (entropy) gradients. The former can be calculated from a EOS that provides the speed of sound and $d\epsilon/dp$ in a tabulated form so that the Brunt-V{\"a}is{\"a}l{\"a} frequency can be determined \cite{2010aste.book.....A,unno}. {A related class of modes are that associated with interfaces: realistic stars with crust-core \cite{Tsang11}, crust-envelope \cite{sul24}, and possibly pasta transitions, where the state of matter changes, also host a class of interface ($i$-) modes with eigenfunctions that peak strongly around the boundary layer (see Sec.~\ref{sec:imodes}).}

The $p$-modes, and $f$-mode in particular, are especially important for astrophysical processes involving neutron stars owing to their compactness, including the tides: it can be shown that the $f$-mode couples most strongly to the tidal potential $\nabla U$ out of all other modes (at least for astrophysical stars, e.g. without near break-up rotation rates or virial-strength magnetic fields). The $f$-modes tend to have a (linear) frequency in the range of $2 \lesssim \omega_{f}/2 \pi \lesssim 3$kHz depending on the EOS \cite[see][for a review]{kk99}, and thus are primarily relevant in the late stages of an inspiral. The dephasing induced by the $f$-mode oscillation is a problem that has attracted considerable attention, where the effects are usually incorporated via the Love number as an effective dressing \cite{hind16,stein16,ma20,ande21b,Steinhoff21,gamba23}. Considering rapidly rotating stars leads to a more complicated picture, since the retrograde mode may actually come into resonance much earlier as the inertial-frame frequency of the mode is $\omega_{i} = \omega_{0} - m \Omega_{s}$ [equation~\eqref{eq:inertialframe}; see \citealt{kk22,kk23}]. 

In \citealt{Kuan21a}, the method presented in \citealt{1995MNRAS.275..301K} to simultaneously evolve the inspiral and mode amplitudes was extended to 3.5 PN order. In particular, the conservative dynamics, leading-order gravitational radiation, and mode excitations (see Tab.~\ref{tab:pneffects}) are incorporated into a total system Hamiltonian. Solving the associated equations of motion allows one to quantify the above effects, including that of spin-modulations in mode excitation (deferred to Sec.~\ref{sec:spinacc}). The aforementioned physics can also be incorporated into the effective Hamiltonian of the EOB formalism \cite{hind16,stein16}, where spin can also be included \cite{Steinhoff21}.
We note that only modes whose pattern speed is along the orbital motion can be considerably excited while there is another set of modes rotating in the opposite direction. Hereafter, we always discuss modes belonging to the former class unless stated otherwise. {An important caveat with respect to these results notes the absence of an elasticity terms within the equations of motion, which could skew the results for $g$- and other modes to some degree, as discussed in Sec.~\ref{sec:gmodes}; \cite[see also][]{Passamonti20,Gittins24}.}

\begin{figure*}
	\centering
	\includegraphics[width=\columnwidth]{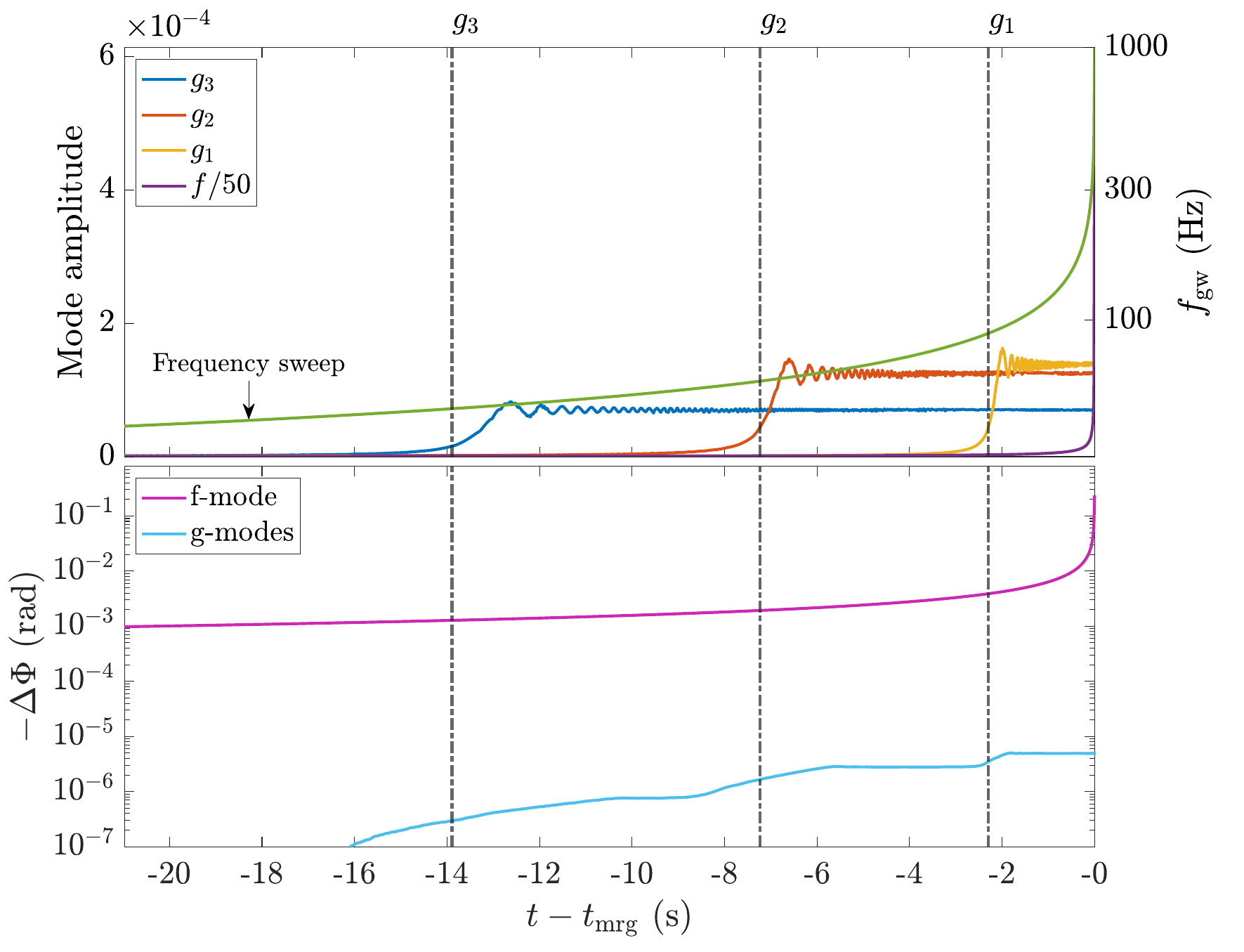}
    \includegraphics[width=\columnwidth]{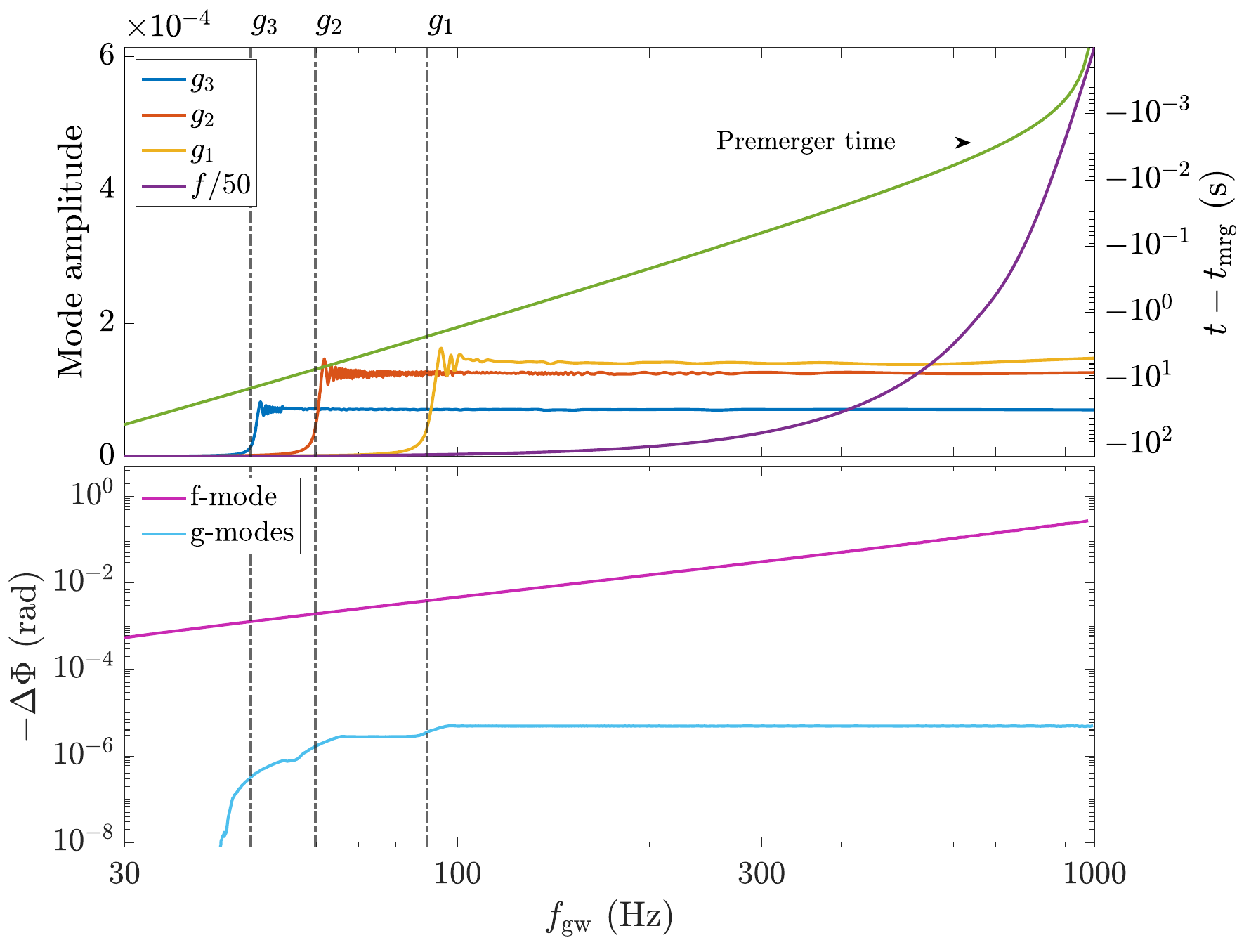}
	\caption{\emph{Left panel}: self-consistent mode amplitude evolutions of $f$- and $g_{1,2,3}$-modes due to tidal forcing, together with the dephasing (lower; see equation~\ref{eq:psi_lagendre}) associated with the $f$- (magenta) and $g$-modes (other colors; see plot legends) as functions of time relative to the merger, $t-t_{\rm mrg}$. The GW frequency ($f_{\rm gw}$) is overplotted as the green line, corresponding to the vertical axis on the right-hand axis. \emph{Right}: Same information as per the left panel though instead as functions of $f_{\rm gw}$, with $t-t_{\rm mrg}$ overlaid as the green line, corresponding to the vertical, right-hand axis. The evolution is carried for a $1.35+1.35\,M_\odot$ binary neutron-star (i.e. $q=1$) with the generalised piecewise-polytropic approximation of the APR EOS from \citealt{OBoyle20} (the original version of the piecewise-polytropic approximation in \citealt{read09} has discontinuous $\tilde{\Gamma}$, preventing an accurate determination of $g$-modes' tidal properties when $\delta$ is small) with a constant stratification $\delta=0.005$. The vertical lines mark the frequency of $g_{1,2,3}$-modes, which are 90, 60 and 48~Hz, respectively.
 }
	\label{fig:mode_evol}
\end{figure*}

Figure \ref{fig:mode_evol} shows how the mode amplitudes evolve due to the tidal coupling that were solved for simultaneously with orbital motion (top), and the associated dephasings compared to point-particle approximation of the inspiral. The $g$-mode amplitudes increase rapidly when the GW frequency ($f_{\rm gw}$, twice as the orbital frequency; green) sweeps through their characteristic frequencies (vertical dash-dotted), and remains roughly unchanged afterward. In this specific case, the $f$-mode's frequency is higher than $f_{\rm gw}$ at the end of evolution, and thus the mode (purple) is never resonantly excited. Although there are only 4 modes involved in the evolution shown in Fig.~\ref{fig:mode_evol}, an arbitrary number can be considered, in principle. 
Still, the dephasing yielded by the $f$-mode excitation [$\mathcal{O}(10^{-1})$ radian] is more than one order of magnitude greater than that collectively caused by $g_1$- to $g_3$-resonances ($<10^{-5}$ radian with $60\%$ attributing to $g_1$, $\lesssim40\%$ from $g_2$, and only a feeble contribution from to $g_3$).
We note that we terminate the numerical computation when $f_{\rm gw}$ reaches $10^3$~Hz since the PN scheme is obviously invalid in that strong-gravity regime. However, the excitation of $f$-modes can be even more important than shown above: given that
the merger frequency can be approximated as \cite{damo10}
\begin{align} \label{eq:mergtimefreq}
    f_{\rm gw, mrg} \simeq \frac{4000\,{\rm Hz}}{(M_A+M_B)/M_\odot},
\end{align}
$f$-resonances could be reached shortly prior to merger.

On the other hand, if the neutron star has a spin, the frequencies of retrograde modes will be reduced \cite{1949ApJ...109..149C,1951ApJ...114..373L,yoshi02,gaer08}, pushing the onsets of resonance earlier in the inspiral. Of the most relevant effects in terms of waveform is the $f$-mode excitation in a neutron star possessing an anti-aligned spin with the orbit \cite{Ho98,ma20,Steinhoff21}, as also demonstrated in numerical simulations \cite[e.g.][]{dudi22,gamba23}; effects of spin are described in more detail in Sec.~\ref{sec:spinacc}.

Current analytic waveform models include either implicitly or explicitly dynamical tides to introduce some effective dressing. In particular, \citealt{ande20} show that the Love number can be expressed as a sum of the contributions of various mode with the $f$-mode being the primary source. As the orbital frequency increases, the tidal interaction will enter a dynamical regime, where the Love number varies with tidal frequency and is often referred to as an effective Love number \cite{hind16,stein16}. Another stream to model the dynamical tidal effects is to introduce amplification factors to the leading-order tidal effects to blend in higher-order PN contributions \cite{damo10,bini12,naga18,akca19}. Phenomenological fittings to numerical simulations have also been developed \cite{diet17,diet19a,diet19b,abac24} \cite[see also][]{will24}.

\subsection{Spectral modulations: general considerations}
\label{sec:spectralmods}

In reality, the free-spectrum consisting of all the modes described above and more, will also be perturbed due to the tidal hammering as the interior structure of the star is now also perturbed; the relative shifts for the \emph{angular} frequencies, $\delta \omega_{\alpha}$, can be deduced from the leading-order \citealt{unno} formulae. Given some perturbing force $\boldsymbol{F}$, the eigenfrequency correction reads, in the Newtonian context \cite[see][for GR generalisations]{Kuan21a,miao24},
\begin{equation} \label{eq:unno}
    \delta (\omega_{\alpha}^2) = 2 \omega_{\alpha} \delta \omega = \frac {\int dV \boldsymbol{F} \cdot \bxi^{\ast}_{\alpha}} {\int dV \rho |\bxi_{\alpha}|^2}.
\end{equation}

The above can be evaluated for any given perturbing force which is subdominant with respect to that of the hydrostatics. Given that spin, magnetic fields, and thermodynamics play a significant role in premerger phenomena beyond just mode modulations, these are covered in their own subsections, Secs.~\ref{sec:spinacc}, \ref{sec:magacc}, and \ref{sec:thermal}, respectively.

\subsubsection{Tidal corrections}
\label{sec:tidalcorrections}

The perturbing tidal force is given by
\begin{equation} \label{eq:tidfff}
\boldsymbol{F}^{T} = \rho \nabla U.
\end{equation}
It is straightforward to evaluate expression \eqref{eq:unno} for \eqref{eq:tidfff} for any particular QNMs, as was done for example by \citealt{suvkprec19} for $f$- and $r$-modes and \citealt{Kuan21a} for $g$-modes. The former found for Maclaurin spheroids the simple result
\begin{equation} \label{eq:tidalf}
\begin{aligned}
\frac{ \delta \omega^{\rm Tidal}_{f}} {42.3 \text{ Hz}} =& - q \left( \frac {\alpha_{f}} {0.1} \right)^{-1} \left(\frac{\MA}{1.4M_{\odot}}\right)^{3/2} \left(\frac{\RA}{13\text{ km}}\right)^{3/2} \\
&\times\left( \frac{a}{100 \text{ km}}\right)^{-3} \Bigg[ 1 - 0.21 \left(\frac{\nu}{300\text{ Hz}}\right)^{-1} \\
&+ 0.055  \left(\frac{\nu}{300\text{ Hz}}\right)^{2} \Bigg],
\end{aligned}
\end{equation}
for mode amplitude $\alpha_{f}$ directly proportional to the overlap integral \eqref{eq:overlap} \cite[see equation 10 in][]{suvkprec19}. The result matches that of more realistic EOS to within a factor $\sim 2$, and is typically small; see also \citealt{denis72}. For $g$-modes, the shift is expected to be of order $\sim 0.01\%$ and can be safely ignored \cite{Kuan21a}.

\subsubsection{Curvature (frame-dragging)}
\label{sec:framedragging}

Accounting for the fact that frequencies differ between the neutron-star and laboratory frames because the star is embedded in a region of strong curvature can be important \cite{detw08,stein16,blan24}. The gravitational redshift factor can be deduced from the metric lapse function, if available from a numerical simulation. To PN order though, and ignoring spin corrections (i.e. Lense-Thirring and quadrupolar corrections to the timelike component of the metric), one finds \cite[see equation 3.6 in][]{stein16}
\begin{equation} \label{eq:redshift}
    z_{B} \approx 1 - \frac{5 G \MB}{4ac^2} = 1 - 0.03 \left(\frac{\MB}{1.6 M_{\odot}}\right) \left(\frac{100\,\text{km}}{a}\right),
\end{equation}
where one anticipates $\omega_{\alpha,i} \mapsto (1+z_{B}) \omega_{\alpha,i}$ for an inertial-frame frequency $\omega_{\alpha,i}$. For resonances applying some seconds prior to merger,  we expect separations within the resonance window to be of order $a\gtrsim100$ km. As expression \eqref{eq:redshift} shows, a frequency shift of at most a few per-cent would apply therefore. By contrast, the redshift at moments closer to merger may seriously impact the high-frequency (e.g. $f$- or superfluid $g$-) mode resonances \cite{stein16}. On the other hand, frame-dragging effects counterbalance this static effect: \citealt{Steinhoff21} found that at late inspiral there are near cancellations and the effective $z_{B}$ may be small.

\subsection{Spin effects}
\label{sec:spinacc}

\begin{figure*}
	\centering
	\includegraphics[width=\columnwidth]{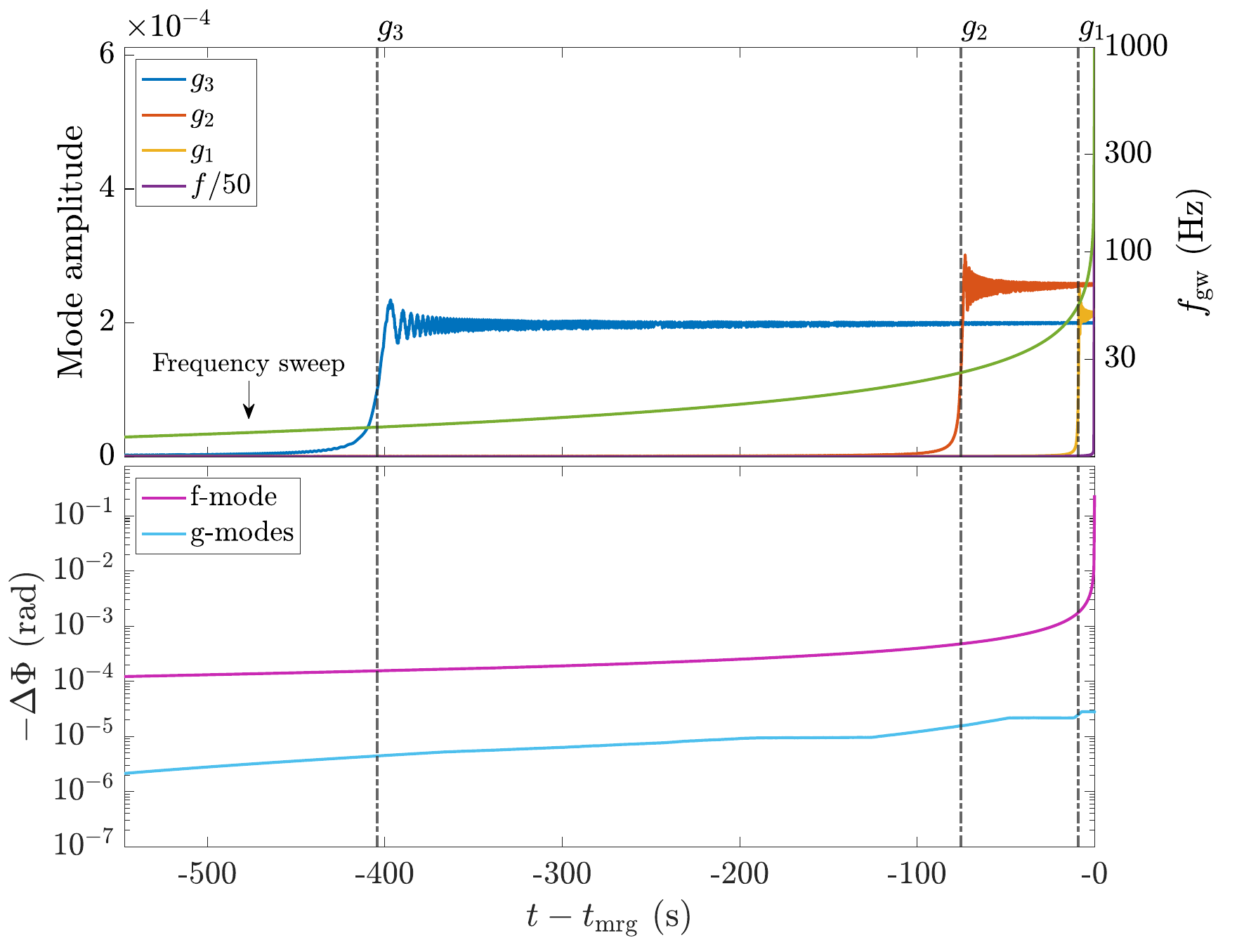}
    \includegraphics[width=\columnwidth]{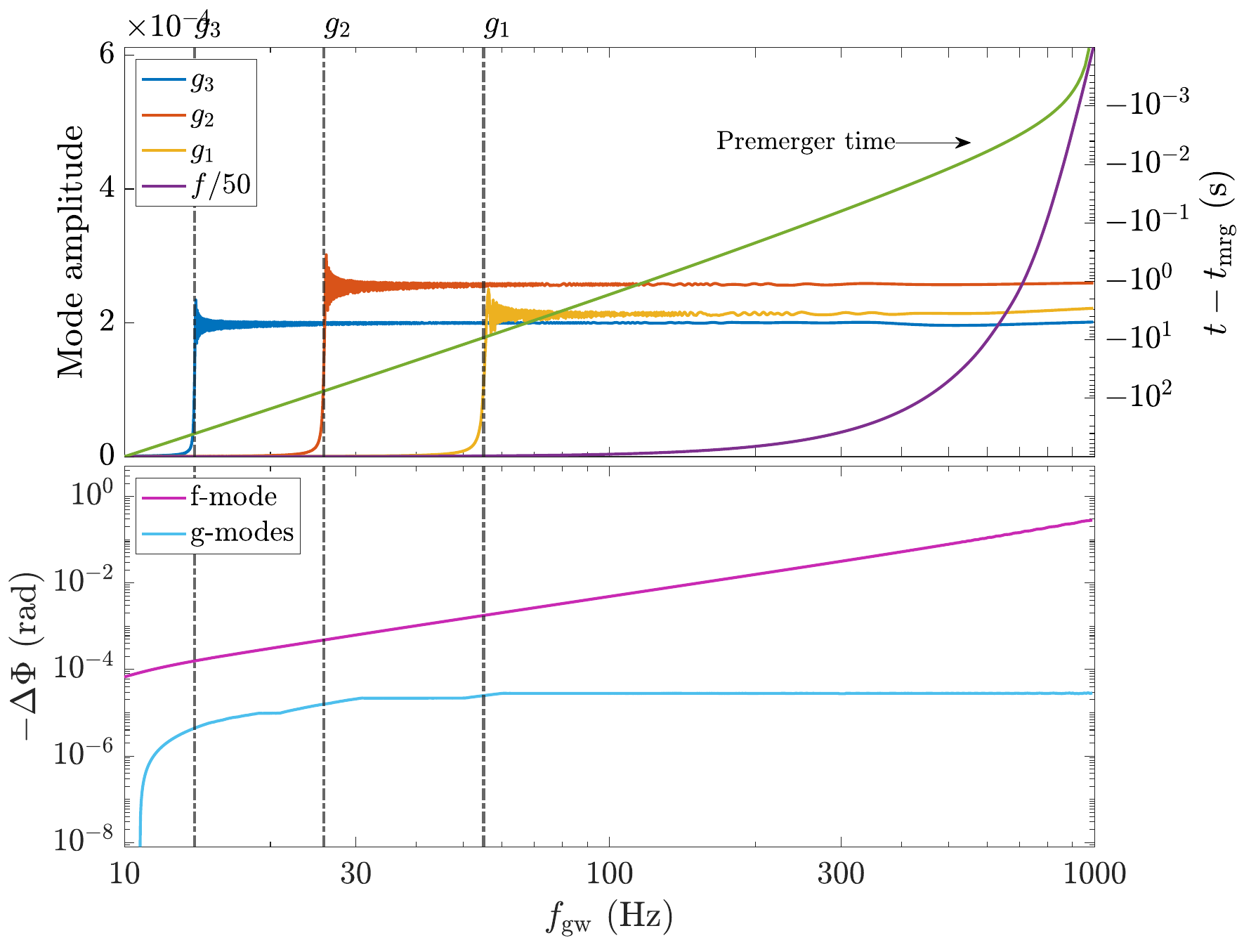}
	\caption{Similiar to Fig.~\ref{fig:mode_evol} though assuming the primary neutron star spins at 20~Hz in the opposite direction to the orbital angular momentum (i.e. anti-aligned with $\Theta = - \pi$). 
 }
	\label{fig:mode_evol20}
\end{figure*}

As far as premerger phenomena are concerned, rotation is no less important than in other instances of neutron-star astrophysics. Primary effects are due to the facts that intrinsic angular momentum (i) alters the equilibrium shape of the neutron star; (ii) introduces a Zeeman-like splitting of the modes, which introduces prograde and retrograde bifurcations, the latter of which can potentially be unstable to the \citealt{1970PhRvL..24..611C,1978ApJ...221..937F,1978ApJ...222..281F} (CFS) instability; (iii) impacts on the efficacy of tidal couplings depending on spin-orbit (mis)alignment; and also (iv) influences the inspiral directly through spin-orbit, spin-spin, and self-spin couplings. This list is not exhaustive, as rotation also influences the evolutionary track of stars in complicated ways, though the above effects are what we concentrate on in this work as they are directly applicable to the premerger phase. We give a brief description of the above points (i)--(iv) here, commenting more throughout as appropriate.

\begin{itemize}

\item[{(i)}]{Centrifugal forces deform the star starting at order $\mathcal{O}(\Omega^2)$. For a uniform density object, the rotational oblateness is estimated by $\epsilon_{\rm rot} = 5 \Omega^2 / 4 \Omega_{k}^2$ where $\Omega_{k}$ is the Keplerian break up value \cite[e.g.][]{2021FrASS...8..166K}. In much the same way that equilibrium tides impact on the evolution through a quadrupolar deformation, so too does the rotation \cite[e.g.][]{done13}.}

\item[{(ii)}]{Modes are generally split into prograde and retrograde families through the azimuthal number, $m$. That is to say, when $\Omega_{s} \neq 0$, the mode frequencies and eigenfunctions depend on $m$ which can be either positive or negative. This splitting leads to a more complicated set of couplings and dynamical tides (as described in Sec.~\ref{sec:dyntides}). Moreover, if the star is spinning sufficiently fast such that $\omega_{\alpha,i} < 0$ while $\omega_{\alpha} >0$, an allegedly retrograde mode instead appears as prograde in the inertial frame \eqref{eq:inertialframe} --- from which radiation reaching infinity is measured --- and thus is subject to the Chandrasekhar-Friedman-Schutz (CFS) instability \cite{1970PhRvL..24..611C,1978ApJ...221..937F,1978ApJ...222..281F,2001IJMPD..10..381A}. 
}

\item[{(iii)}]{Spin-orbit misalignment changes the Wigner coefficient and hence the tidal coupling, which couples in with point {ii} above.}

\item[{(iv)}]{Spin-orbit coupling, independent of tides, is important and modifies the dephasing of GW from 1.5 PN order, in a way that a spin antialigned with the orbital motion tends to accelerate the merger while an aligned spin works to delay the inspiral \cite{kidd93,cutl93,kidd95,blan95}.} 

\end{itemize}

As soon as one accounts for spinning stars, the geometrical description becomes considerably more complicated. A few additional angles are introduced relating to the directionality of the orbital angular momentum versus that which is intrinsic to the stars. If we consider just a single star to have spin, we need only introduce a single angle, introduced as $\Theta$ in expression \eqref{eq:harmonicWigner} \cite[e.g.][]{Lai06}, representing the single surviving misalignment angle. If we have alignment but a significant spin, the effect is that of earlier resonances. In the anti-aligned case, it is the lower inertial-frame frequency branch of modes that are most strongly excited by the tides \cite{Ho98}. This generally implies a stronger tidal influence, as earlier resonances imply a longer period of time to drain the orbit. On the other hand, a later resonance for some modes could imply a greater dephasing since, again typically, larger overlaps are achieved for higher eigenfrequencies. Larger amplitudes are especially relevant for precursor observations, since the crust is likely to be more susceptible to a greater strain exerted over a short period of time rather a moderate strain over a marginally longer window. This is the case for the von Mises criterion at least, which only considers the maximum strain (see Sec.~\ref{sec:breaking}).

Simultaneous mode and orbital evolutions for arbitrary misalignment angles $(\Theta \neq 0, \pi)$ were calculated by \citealt{kuan23} for $g$-modes and \citealt{kk23} for $f$-modes. Physically, non-zero misalignments allow for the excitation of odd azimuthal mode-numbers $(m=1,3,\ldots)$ depending on the Wigner coefficient, and can reach relatively large amplitudes. For a system spinning at even a modest rate, $\nu \gtrsim 2$~Hz, but with significant misalignment angle, $\Theta \sim 80^{\circ}$, it was shown that the resonant pulsations from both $m=1$ and $m=2$ modes could theoretically break the crust, leading to the emission of two time-separated precursors (as observed in GRB 090510, for instance; discussed in more detail in Sec.~\ref{sec:precursors}). Whether multi-resonances allowed by significant misalignment are relevant for GW dephasing has not yet been calculated in detail in the literature.

Figure~\ref{fig:mode_evol20} demonstrates the impact that moderate rotation ($\nu = 20$~Hz) can have with respect to resonance timings and associated dephasing. Noting the difference in scale between both time and the vertical axes on the plots relative to Fig.~\ref{fig:mode_evol}, several effects are visible. The $g$-mode are particularly affected by earlier excitations of modes: for the static case it is clear that the $g_{1}$-mode amplitude is largest, while that title instead goes to $g_{2}$ in terms of raw amplitude when $\nu = 20$~Hz. At the same time, retrograde $g$-modes may be smeared out if their frequencies are reduced by spin to be slower than the chemical reaction rate of neutron star matter (Sec.~\ref{sec:stratification}), which is about $\lesssim 10^2$~Hz ($\sim$~mHz) assuming that the direct (modified) Urca reaction is the dominant beta-decay channel \cite{Andersson19,coun24}. That is, locally the Brunt-V{\"a}is{\"a}l{\"a} frequency should be higher than the chemical balancing rate. In principle, one could also use the above ideas to test whether the direct Urca channel can operate in neutron stars, a topic that has received much attention lately \cite[e.g.][]{anz22,marino24}. That is, if $g$-mode dephasings at low frequencies were conclusively ruled out that would provide strong evidence for direct beta-decay channels. Such ideas have yet to be thoroughly investigated in the literature.

\begin{figure}
	\centering
	\includegraphics[width=0.49\textwidth]{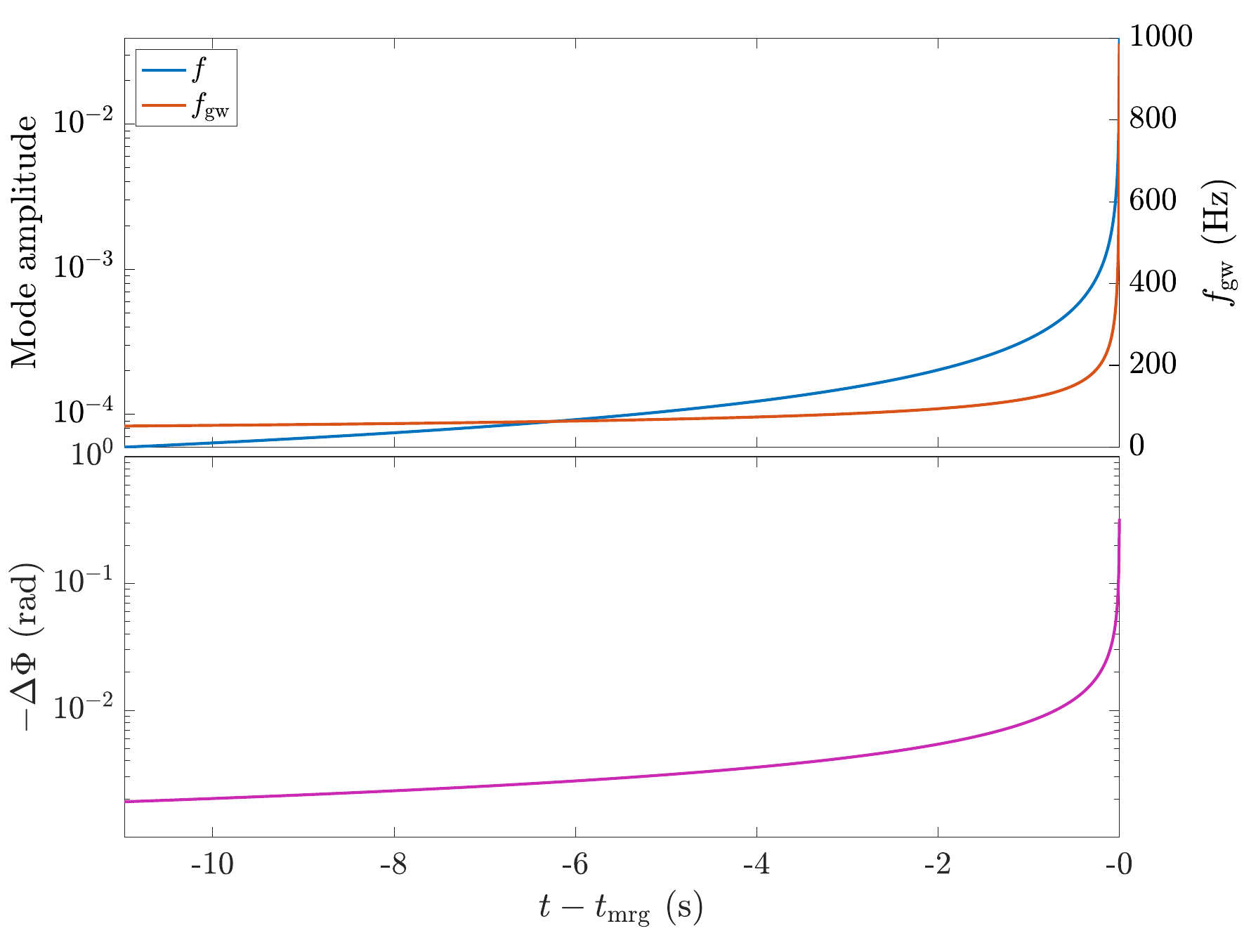}
	\caption{Similar to Fig.~\ref{fig:mode_evol} though for a star spinning with $\nu = 200$~Hz in the opposite direction of the orbit. In this case, the (retrograde) $g$-modes are totally washed out of the spectrum due to reaction-rate considerations (see text).}
	\label{fig:spin_200}
\end{figure}

For the spinning case depicted in Fig.~\ref{fig:mode_evol}, the $g_3$-mode is probably not realistic in light of the above \cite[though see also][]{pass09}. That said, if being conservative to assume that the chemical reaction rate is at the mHz level, then the $g_1$- to $g_3$-modes will be eliminated when the neutron star spins at 51, 35, and 28~Hz, respectively. Therefore, the only relevant mode in a rapidly spinning neutron star ($>10^2$~Hz misaligned with the orbit) is the $f$-mode, and possibly the interface and/or $r$-modes (see Sec.~\ref{sec:resfamilies}), though these latter modes are not incorporated here. This is demonstrated in Fig.~\ref{fig:spin_200} for a star with spin $\nu = 200$~Hz: the $g$-modes are washed out of the spectrum, based on the above predictions (even if not mathematically at the level we have setup the problem), and only the $f$-mode remains, whose frequency is reduced from the static value $1955$~Hz to $1727$~Hz.  Even in this case, however, the mode does not become resonant [see equation~\eqref{eq:mergtimefreq}]. The dephasing is therefore, in this case, the weakest of all considered thus far, though could increase with higher spins \cite{kk23,yu24}. This demonstrates the complicated, non-linear dependence that stellar spin has on GW observables. Including spin within merger simulations is especially difficult, though was first successfully implemented by \citealt{bern14}.

\subsubsection{Tidal spinup}
\label{sec:tidalspin}

In addition to the direct energy deposits from the orbit, tidal interactions torque each of the neutron stars. From a resonant interaction alone, \citealt{lai97} shows that a net angular momentum transfer of
\begin{equation}
    \Delta J_{\alpha, \rm res} = \int dt \int d^{3} x \delta \rho_{\alpha} \left( - \frac{\partial U^{\ast}}{\partial \phi} \right)
\end{equation}
can be expected. An upper bound to the bulk spinup of a given star can then be obtained by assuming that $\Delta J_{\alpha, \rm res}$ contributes only to the uniform rotation of the star, so that $\Delta \Omega_{\alpha,s} \sim \Delta J_{\alpha, \rm res}/ I_{0}$. For $f$-modes where the tidal overlap reads $Q \gtrsim 0.1$, the above estimate indicates a non-negligible spinup of order unity may occur, albeit only in the very final moments of merger [see also equation (6.2.6) in \cite{Lai94}]. For a $1.4 M_{\odot}$ pair of stars, the general result is \cite[e.g.][]{suvkprec19}
\begin{equation}
    \frac{\Delta \Omega_{\alpha,s}}{\Omega_{s}} \lesssim 0.2 \left(\frac{30 \text{ Hz}}{\nu}\right) \left( \frac{\omega_{\alpha,i}}{1 \text{ kHz}}\right)^{-5/3} \left( \frac{Q_{\alpha}} {0.01} \right)^{2}.
\end{equation}
 For $g$- and other non $f$-modes, the effect is likely sub-leading though could achieve $\Delta \Omega_{\alpha,s} / \Omega_{s} \gtrsim 0.01$ for favourable overlaps. The long-lived tidal-grinding provided by the non-resonant $f$-mode could also be significant. Either way, such effects complicate the dynamics of the system considerably, as one must, in general, account for a time-dependent spin factor in modelling the tidal dynamics.  Such modelling has not been attempted aside from the work of \citealt{lai97} for neutron stars, and in the different contexts of binaries involving a white dwarf by \citealt{rac07,fullerlai11}.

It is actually expected that the tidal torques will primarily induce differential rotations instead of bulk motions, essentially due to relation \eqref{eq:lagdisplacement}. This differential rotation can excite convective motions and invites the possibility of premerger dynamos; see Sec.~\ref{sec:dynamo}.

\subsection{Magnetic effects}
\label{sec:magacc}

Magnetic fields control all of the electromagnetic phenomena associated with neutron stars, including those occurring premerger (see Secs.~\ref{sec:magneticfields} and \ref{sec:breaking}). Dynamical magnetic fields can also generate electric fields via induction which work towards accelerating the binary inspiral. 

As described by \citealt{lai12} and others prior \cite[beginning with][]{gl69}, suppose that the magnetic field of the primary greatly exceeds of the companion, so that the latter may be treated effectively as a perfect conductor with vanishing dipole moment. The orbital motion of this companion through the rotating magnetosphere of the primary generates an electromotive force, driving a ``direct current'' between the constituents, with the closed magnetic field lines playing the role of wires in the circuit. This is the so-called ``unipolar inductor'' model, which will be covered in some more detail in Sec.~\ref{sec:magnetosphericinteraction}. The exact way in which EM energy is siphoned out of the orbit depends on the total resistance of the circuit, $\mathcal{R}$, which includes contributions from the primary, companion, and the (intertwined) magnetosphere. \citealt{lai12} argues that the maximum energy dissipation rate reads
\begin{equation} \label{eq:magdiss}
    \dot{E}_{\rm EM} = - T_{\rm mag} (\Omega_{\rm orb} - \Omega_{s}) = \zeta_{\phi} (\Omega_{\rm orb} - \Omega_{s}) \frac{\mu_{d}^2 R_{B}^2} {2 a^5},
\end{equation}
for electromotive torque $T_{\rm mag}$, primary (equatorial) dipole moment $\mu_{d}$, and twist parameter $\zeta_{\phi}$. This latter parameter relates to the induced toroidal field due to kinetics, as is familiar from studies of low mass X-ray binaries \cite[e.g.][]{gs21}, though its value is highly uncertain: it depends on poorly-understood boundary-layer physics.

In fact, \citealt{piro12} suggested implicitly (i.e. in a different notation) that the twist could reach values much larger than unity, $\zeta_{\phi} \gg 1$, which could significantly accelerate the inspiral. Furthermore, the torque $T_{\rm mag}$ will generally spin up the primary when $\Omega_{\rm orb} > \Omega_{s}$; \citealt{piro12} argued this could lead to near synchronization, and thus rapidly rotating stars prior to merger if the dipole moment is sufficiently large (cf. Sec.~\ref{sec:rotation}). On the other hand, if the magnetic pressure exerted by the toroidal field is too large, the flux tubes defining the circuit will be broken by unstable kink modes. This is the essence of the argument put forward by \citealt{lai12} regarding the limiting value of the twist ($\zeta_{\phi} \sim 1$), and hence a relatively weak electromagnetic contribution \eqref{eq:magdiss}. Still, quasi-periodic circuits may be established if after such a flux tube breaks, reconnection between the (inflated) field lines re-links the two stars so that the cycle may repeat. Understanding the dynamics of such a system requires 3D simulations which have not yet been conducted; though see \citealt{Crinquand19}, who argue via numerical simulations that $\zeta_{\phi}$ could greatly exceed unity in cases where the stellar spins are anti-aligned with respect to each other. 

\begin{figure}
	\centering
	\includegraphics[width=0.49\textwidth]{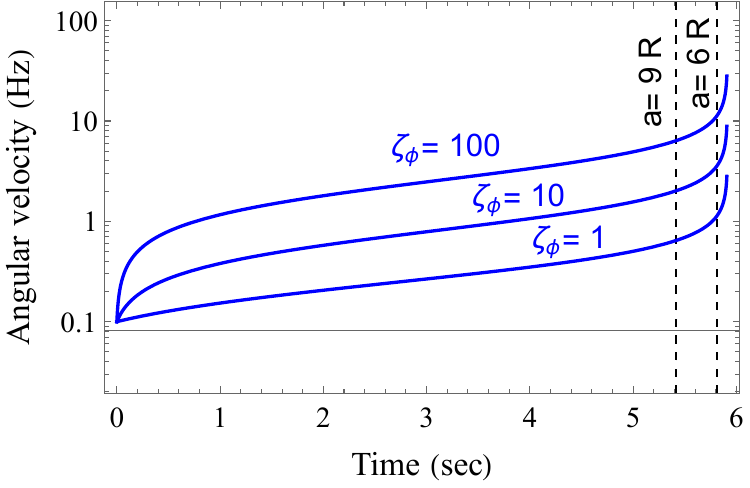}
	\caption{Estimates for electromotive spinup in the final seconds of inspiral, following expression \eqref{eq:spinupeqn}, for a magnetar with equatorial field strength $B \approx 10^{14}$~G and an unmagnetised companion. For large twists ($\zeta_{\phi} \gg 1$), spin-up could be significant and contribute to dephasing through the effects described in Sec.~\ref{sec:spinacc}.}
	\label{fig:emspinup}
\end{figure}

To get a rough sense of the impact though, we can integrate the torque-balance equation
\begin{equation}
  \frac{d}{dt}  \left(\frac{1}{2} I_{0} \Omega_{s}^2\right) = \dot{E}_{\rm EM},
\end{equation}
which, provided we take $\Omega_{\rm orb} \gg \Omega_{s}$, has solution
\begin{equation} \label{eq:spinupeqn}
    \Omega_{s}(t) \approx \left( \Omega_{s,0}^2 + \mu_{d}^2 \frac{R_{B}^2}{2 I_{0}} \int dt \frac{\zeta_{\phi} \Omega_{\rm orb} } {a^5}  \right)^{1/2},
\end{equation}
where we ignore the dipole spindown and field decay taking place in the final seconds or minutes. If we take a constant value for $\zeta_{\phi}$ and further use the leading-order GW orbital decay rate (cf. Sec.~\ref{sec:inspiralbasics}),
\begin{equation} \label{eq:adotleadingorder}
    \dot{a} = - \frac{64 G^3}{5 c^5} \frac{M^3 q(1+q)}{a^3},
\end{equation}
together with $\Omega_{\rm orb}^2 \approx G M(1+q)/a^3$ we can evaluate expression \eqref{eq:spinupeqn} easily. One set of results for various values of $\zeta_{\phi}$ are shown in Figure~\ref{fig:emspinup}, where we postulate a magnetar-level field at the equator of $\sim 10^{14}$~G, radius $12$~km, moment of inertia $I_{0} = 10^{45} \text{ g cm}^{2}$, and two $1.4 M_{\odot}$ stars ($q=1$). The evolution is tracked starting from a separation of $200$~km ($\Omega_{\rm orb} \approx 216$~Hz). The magnetic star is practically static initially with $\Omega_{s,0} = 0.1$~Hz, though half a second before merger ($a \approx 9 R$) can reach non-negligible spins if the twist parameter is large ($\zeta_{\phi} \gtrsim 1$). For the case $\zeta_{\phi} = 10^{2}$ --- still significantly less than that estimated by \citealt{piro12} for some systems --- the spin frequency could reach $\nu \gtrsim 1$~Hz when $a(t) \sim 9 R$. Larger field strengths amplify the effect further like $\sim B^2$. If some magnetars with larger fields take part in mergers, as hinted at by some precursor observations (see Sec.~\ref{sec:prectheory}), spin-up could be dynamically impactful. These spinups will combine with that of the tidal torques discussed in Sec.~\ref{sec:tidalspin}.

\subsubsection{Internal fields?}
\label{sec:internalfields}

Magnetic effects can also be associated with that of the interior, rather than the exterior. A strong tension can either reduce or increase the tidal deformability of a star depending on the nature of the magnetic geometry: poloidal (toroidal) fields tend to induce an oblate (prolate) distortion. This was studied by \citealt{giac09} and more recently by \citealt{zhu20} using complementary techniques, though the results are rather pessimistic: relative shifts in the deformabilities $\Lambda$ [see below equation \eqref{eq:kappaT}] amount to $\lesssim 0.1\%$ for $B \lesssim 10^{15}$~G for neutron-star EOS. For quark-star EOS, the effects are greater. In a superconducting core, the result may also be somewhat larger if $H_{1} \sim 10^{16}$~G [e.g. \citealt{land13} and equation \eqref{eq:critfield}], which leads to a factor $\gtrsim 10$ increase in the effective magnetic tension and hence a percent-level shift for magnetars. Tidal deformabilities for stars with superconducting components have not been calculated in the literature. Some additional effects could arise if the internal magnetic fields are large enough to distort the stellar EOS, though these must be typically larger than even $H_{c1}$ \cite[see, e.g.][]{dex17,suvg22}. A recent discussion on mathematical elements of GRMHD equilibria can be found in \citet{suvg23b}.

\subsection{Thermal and viscous effects}
\label{sec:thermal}

As quantified by equation \eqref{eq:inhommas}, tidal energy is siphoned into QNMs during inspiral. This energy is essentially in the form of the potential and kinetic energy associated with the master eigenfunction $\boldsymbol{\xi}$. As detailed by \citealt{lai12} and others, three ways in which these modes can be damped is through GW radiation-reaction, chemical relaxation through neutrino emission, and viscous dissipation. The first of these is largely irrelevant (though note the damping time of $f$-modes can be of the order $\sim 100$~ms, which could theoretically be fast enough if a very early resonance occurs because of rapid rotation; see Sec.~\ref{sec:spinacc}). 

Viscous heating is likely to play a significant role. As described in \citealt{kuan23} and elsewhere, the stellar stratification is generally a function of time because of this heating, implying that the $g$-mode \emph{spectrum} itself is time-dependent \cite[this gets even more complicated if accounting for nuclear reactions;][]{hamm21}. In particular, \citealt{Lai94} has shown that the \emph{non-resonant} $\ell=2=m$ $f$-mode in an aligned (or non-rotating) binary will increase the star's temperature by (see equation 8.30 therein)
\begin{align}\label{eq:heat}
    T_{\rm vis} \approx 3.6\times10^{7}\left( \frac{3R_{A}}{a} \right)^{5/4} \text{K},
\end{align}
where it is assumed that the heat content is determined solely by the thermal energy of a nonrelativistic, degenerate, free-neutron gas. For a superfluid star, the relativistic electrons dominate this latter quantity of interest and expression \eqref{eq:heat} is increased further by a factor $\gtrsim 2$. 

It has been suggested that $T$ could reach values considerably larger than \eqref{eq:heat}. \citealt{Arras18} --- who also provide an overview of the tidal heating literature in their Introduction --- find that chemical heating, driven by the induced fluid motions during inspiral, could raise the temperature to $\sim 2 \times 10^{8}$~K if the system has access to the direct Urca mechanism. This is because density perturbations excited via tides force the system out of beta equilibrium, instigating chemical heating. If the core contains ``strange'' matter such that nonleptonic weak interactions raise the bulk viscosity by a couple orders of magnitude (e.g. hyperonic matter), tidal heating could even raise the temperature to $\lesssim 10^{10}$~K \cite{ghosh24}. Such enormous temperatures would have a significant impact on the $g$-spectrum of the system leading to dephasing, aside from that which is associated with the heating directly. The latter effect would likely exceed $\sim 0.1$~rad and thus be detectable with the LVK network \cite{ghosh24}; see also Sec.~\ref{sec:melt}. In the limit of totally efficient tidal dissipation the core temperature would easily reach $\sim 10^{10}$~K before merger, as calculated by \citet{mezr92}.

{If the system were to reach even a fraction of such values, the effective Love number should decrease because extra thermal pressures cause the outer boundary ($p=0$) to extend. Using various EOS, \citet{kan22} calculated the quadrupolar Love numbers at fixed temperatures finding that for $T \sim 1$~MeV ($\sim 10^{10}$~K) there is a decrease in $k_{2}$ of order $\approx 20\%$ with a $\approx 70\%$ drop for $T \sim 5$~MeV; the tidal deformability, however, is largely insensitive to the temperature since the radius increases to compensate the reduced $k_2$ for the EOS considered therein. High temperatures would also impact on the elastic maximum of the crust (Sec.~\ref{sec:breaking}).}

A super-to-normal fluid transition due to intense heating ($T > T_{c}$) could also lead to a rapid shift in the already-excited $g$-mode spectrum. Such a rapid transition could lead to ``glitch-like'' phenomena in the waveform as the system undergoes a premerger phase transition. Practically speaking, crude estimates for how bulk heating will influence the $g$-modes can be made noting that variations in the stratification $\delta$ obey the relation $\Delta\delta/\delta=2\Delta T/T$ since $\delta\propto T^2$ from expression \eqref{eq:deltatemp}. One can then recompute spectra with $\delta \to \delta + \Delta \delta$ at each timestep. This could combine with a discontinuous jump in the frequencies of order $\sim 4$ from the above super-to-normal transition (see Sec.~\ref{sec:superfc}). Efforts to self-consistently account for such changes are in progress.

Overall, thermal elements in mergers are rather subtle and not fully understood beyond the above. For example, \citealt{hamm21} pointed out that out-of-equilibrium elements in a merger lead to a softening of the EOS in some density regions, and to composition changes that affect processes that rely on deviation from equilibrium, such as bulk viscosity, both in terms of the magnitude and the equilibriation timescales inherent to the relevant set of reaction rates. Whether the estimates described above may apply in a realistic merger environment, including the possibility of constraining them, is not obvious. Significant tidal heating of the stars can also induce mass loss via winds before merger \cite{mezr92}, which could lead to electromagnetic, premerger emissions if strong shocks develop in the winds. Such debris will also contribute to the dynamical and postmerger ejecta polluting the eventual crash site, which could inhibit jet formation and affect the breakout timescale (see Sec.~\ref{sec:jetform}). If the stars reach high-enough premerger temperatures, the bulk viscous timescales calculated in \citealt{alford20} could be utilised.

\subsubsection{Crust melting?}
\label{sec:melt}

Heat imparted due to tides reduces the effective Coulomb parameter $\Gamma$ in a space- and time-dependent way (see Sec.~\ref{sec:crust}). It has been suggested that actually the entire crust may melt prior to coalescence because $\Gamma < 175$ is achieved \cite{pan20} \cite[see also][]{hamm21}. This is plausible in some of the scenarios described above. If indeed the crust undergoes a kind of global elastic-to-plastic transition at late times, \citealt{pan20} argue the heating rate due to tides and (resonant) modes would increase significantly as a liquid is more susceptible, which can manifest at the level of the waveform through a $\mathcal{O}(0.1 \text{ rad})$ dephasing. However, according to estimate \eqref{eq:heat}, the crust may not reach its melting point even by coalescence, unless the viscosity assumptions made by \citealt{Lai94} underestimate the degree of heating. A melted crust obviously cannot yield either, meaning that the melting time, $t_{\rm melt}$, relative to coalescence, sets an upper limit to a resonant failure time to explain a precursor (see Sec.~\ref{sec:resfail}). This would rule out $f$-mode-induced failures altogether for large bulk viscosity. Increasing the width of the liquid layers near the surface of the star would also strongly impact the spectrum of ocean modes; see Sec.~\ref{sec:oceans} and \citealt{watts12}. A discussion on theoretical elements of mode-induced crust melting can be found in \citealt{lind00}.

\subsection{Residual eccentricity}
\label{sec:othereff}

Aside from canonical formation channels involving a symbiotic binary (see Sec.~\ref{sec:rotation}), mergers might occur in globular clusters where captures are not irregular. Although the recent Monte Carlo simulations of \citealt{ye20} estimate a globular-cluster merger rate of only $\sim 0.02 \text{ Gpc}^{-3} \text{ yr}^{-1}$, such systems may retain a significant eccentricity as the usual circularisation arguments made in Sec.~\ref{sec:inspiralbasics} following the \citealt{pet64} formulae do not apply. Tidal phenomena in eccentric inspirals acquire additional complexity because resonances can be triggered at different orbital phases (likely near periastron), and thus generally will happen earlier; multiple resonances can also be triggered each passage.

It has been shown that for eccentric binaries there can be chaotic growth of the modes, with the kinetic energy stored in a mode rivalling that of the orbital binding energy, $U \sim G M m /a$, after many cycles \cite{mard95,vick18}. Given that this energy greatly exceeds that which is typically expected to be stored in resonant modes, $E_{\rm kin,max} \sim 10^{45}$erg, it would be worthwhile to revisit such studies but for neutron stars specifically \cite{chir17,yang19,wl20}. One such study was carried out by \citealt{vick18b}, finding appreciable effects depending on a complicated relationship between the eccentricity and how much energy is siphoned into the $f$-mode; see also \citealt{tak24}.

\subsection{Remarks on merger simulations and future challenges}
\label{sec:mergerfuture}

Here we provide a brief discussion on merger simulation results \emph{with respect to mode excitations}; we direct the reader interested in GRMHD details more generally to \citealt{br17,kiuc24}. Although the mode evolution is not as transparent as in Figs.~\ref{fig:mode_evol}--\ref{fig:spin_200}, some fully relativistic simulations may resolve the excitation of the $f$-mode in the last $<100$~ms of inspiral given that the dephasing between the numerical waveform and the analytic model including realistic $f$-mode effects is small \cite{Steinhoff21,dudi22,gamba23}. Aside from the $f$-mode, numerically capturing the excitations of other (polar) modes that could experience a resonance (viz.~$g$-, $i$-, shear-, and ocean-modes) are currently out of reach due to a number of technical and theoretical challenges. This is important since $g$-mode dephasings in particular may not be totally negligible \cite[discussed recently by][for instance]{ho23} and represents a future challenge for the numerical-relativity community. Failing to account for such dephasings, if large, could lead to spurious inferences about neutron-star structure.

Spatial resolution is not the only technical requirement to see the excitation of lower frequency modes. In such cases, one would need to start the simulation at a time when the orbital frequency is less than $\sim50$~Hz to include their resonance windows. This corresponds to many seconds prior to merger (cf. Fig.~\ref{fig:mode_evol}). This is borderline impossible owing to both computational demand and numerical stability considerations when coupled in with the above resolution demand, at least in a full numerical-relativity sense where the spacetime is evolved self-consistently with (magneto-)hydrodynamics using the full Einstein equations. For higher quantum numbers ($g_{n\geq2}$, ...) even greater resolution is required, and the problem becomes more severe still. On top of this, $g$- and other modes' tidal response tend to be rather shallow compared 
to the $f$-mode \cite[see][and others]{Lai94,1995MNRAS.275..301K,Shibata94,Passamonti20}, and thus 
numerical dissipation can easily bias the results, i.e. convergence 
is expected to be much harder to obtain than the $f$-mode case. This is in addition to the usual separation problem of the real and imaginary components of $g$-modes \cite[e.g.][]{Kuan21a}.

From a theoretical perspective, most (though not all) of the long-term inspiral simulations adopt a cold EOS through a piecewise-polytropic approximation which does not include the crust structure and cannot support the thermal and compositional stratification gradients that exist in a real star. Therefore, all of the microphysically-dependent modes may not even exist in the first place within the computational setups employed. It is thus even difficult to estimate how many CPU hours and further code innovations would be needed to accurately simulate the evolution of these modes and claim it was either resolved or not resolved \cite[cf.][]{hamm21}. The highly-dynamical nature of the spacetime and fluid makes it also not obvious whether ``modes'' even exist in the usual sense of the word, as these are inherently linear while the Einstein equations are inherently nonlinear \cite[see, e.g.][for discussions]{gab09,sot24}.

\section{Precursor flares: observations}
\label{sec:precursors}

It is well-accepted now since the dual discovery of GW 170817 \cite{LIGOScientific17} and GRB 170817A \cite{gold17} that neutron-star binary mergers are at least in part responsible for \emph{short} GRBs. Some mergers may however also produce \emph{long} bursts, as described in Sec.~\ref{sec:longvsshort}. We thus use the phrase \emph{merger-driven} to encompass this wider class of GRBs.

\begin{figure}
	\centering
	\includegraphics[width=0.497\textwidth]{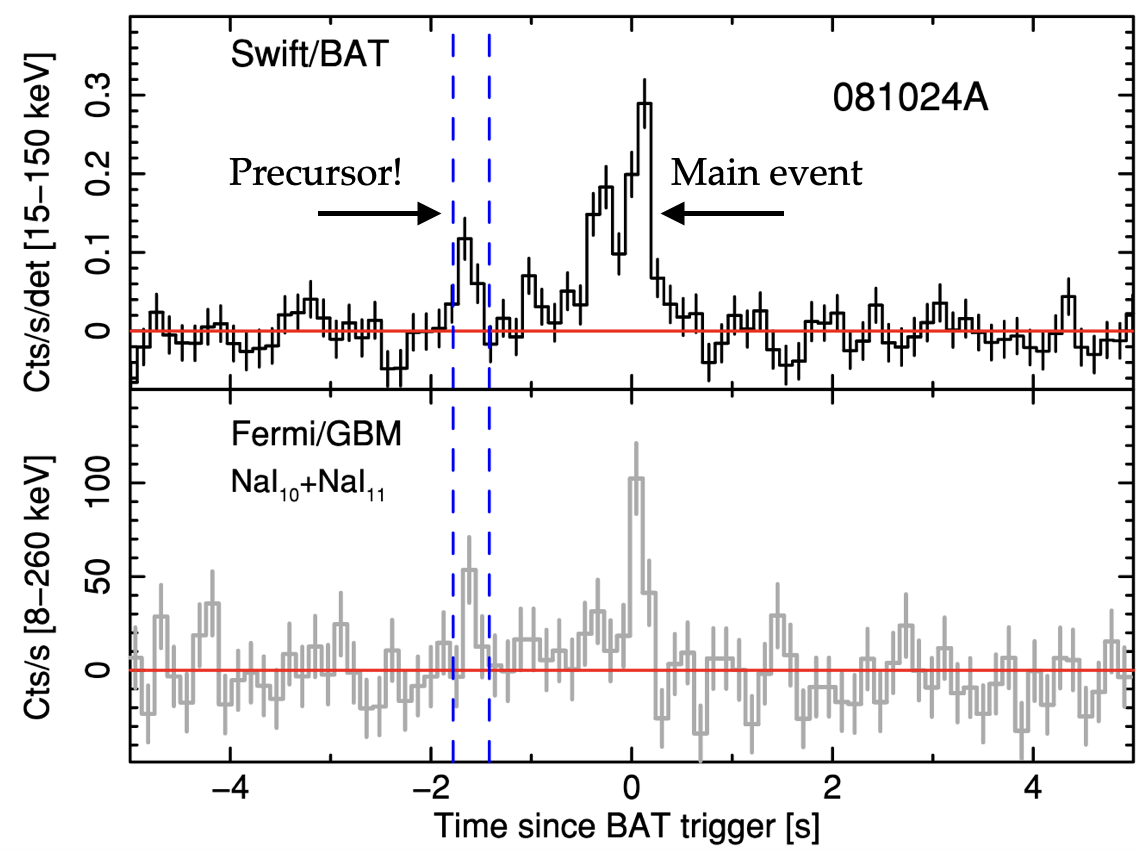}
	\caption{Light curves for GRB 081024A from Swift/BAT (top panel) and Fermi/GBM (bottom). The significant precursor event is delineated by the vertical blue lines, as marked by the arrow. Adapted from \citealt{troja10} with permission.}
	\label{fig:troja_burst}
\end{figure}

A small percentage (Sec.~\ref{sec:rarity}) of merger-driven GRBs display precursor phenomena, where energetic flashes are observed even many seconds prior to the main event in some cases \cite[e.g.][]{troja10,koshut95,zhong19,Coppin20}. Based on jet delay physics (Sec.~\ref{sec:jetform}), there is good reason to suspect that least some of these events, such as the GRB191221A precursor measured $\sim$~twenty seconds prior to the respective GRB \cite{Wang20}, ignite during the premerger stage. Some precursors may be attributed to post-merger phenomena however: the precursor from GRB150922A had only a $\sim 30$~ms waiting time, which is considerably less than the $\sim 1.7$~s delay observed in GW/GRB170817 \cite{gold17}. Such postmerger phenomena could be related to the nature of the jet, possibly interacting with the dynamical ejecta, launched by the post-merger remnant (e.g. in cocoon or choking jet models; see Sec.~\ref{sec:post-merger}). It is thus unclear what fraction of precursors occur before \emph{merger}, even though they occur before\footnote{This raises a subtle issue: it could be that the GRB is missed (e.g. beamed away) but a precursor is launched in the direction of Earth. The precursor may therefore be mistaken for the actual GRB, since such events can be spectrally similar in some instances (see also Sec.~\ref{sec:spectral}).} the \emph{GRB} by definition. 

Before proceeding to precursor observations, we remark that following LVK's O3b run, searches for GWs were carried out for a variety of GRBs that took place during the observational window, regrettably without any success \cite{grbgws22}. Some of these GRBs were associated with precursors, placing some (very mild owing to source distance) constraints on event nature.

\subsection{Statistical preliminaries}
\label{sec:precstats}

\begin{figure}
	\centering
	\includegraphics[width=0.497\textwidth]{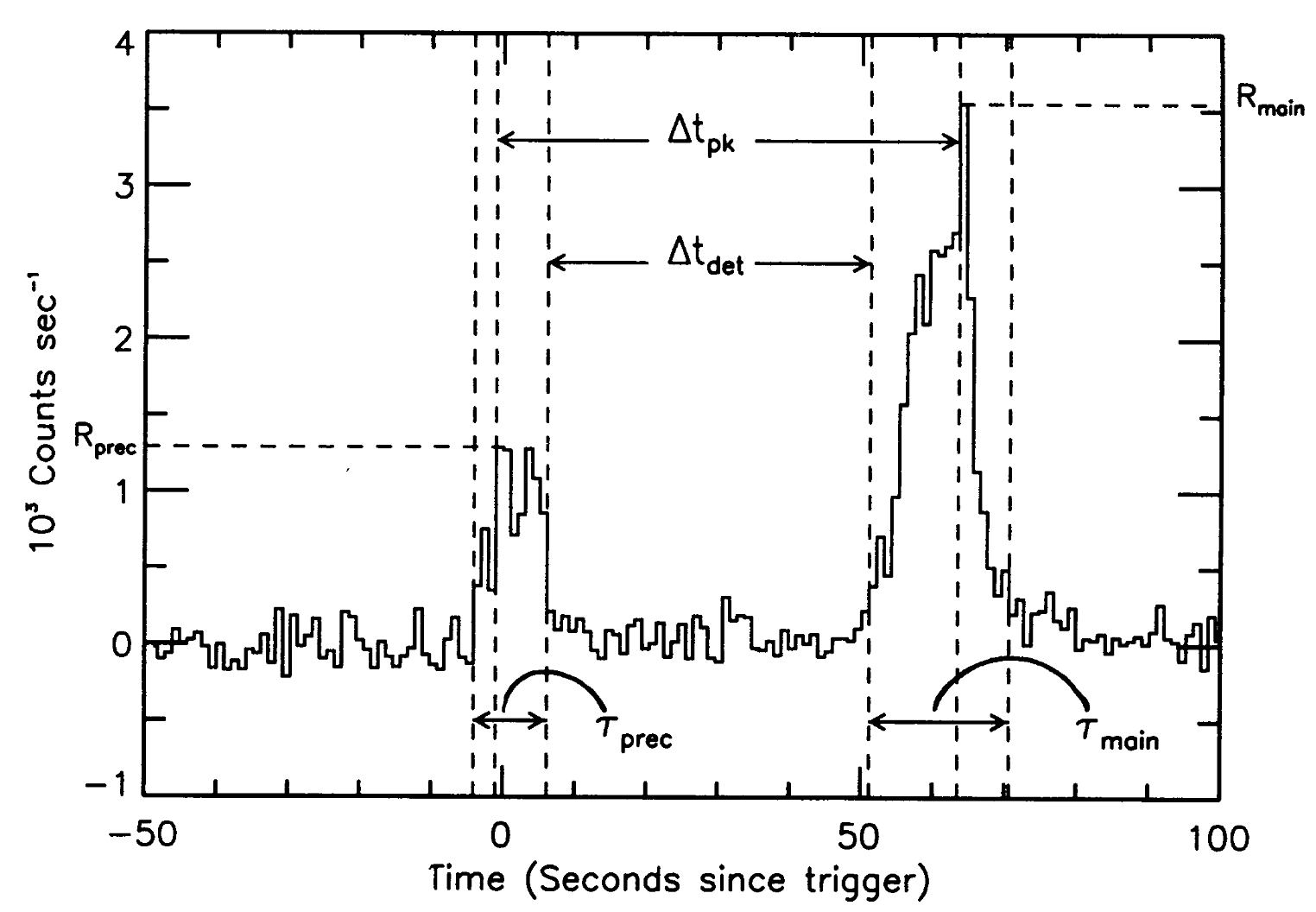}
	\caption{Diagrammatic representation of different types of timescales for precursors and GRBs. For short GRBs in particular, these distinctions can lead to significant factors separating reported waiting times. Reproduced from \citealt{koshut95} with permission.}
	\label{fig:koshut_burst}
\end{figure}

Light curves for a prototypical precursor --- that associated with GRB 081024A --- are shown in Figure~\ref{fig:troja_burst}, adapted from \citealt{troja10}. The event and precursor were observed by Swift and Fermi with a clear coincidence and separation time of $\sim 1.5$~s. What one ought to call the \emph{waiting time} (as is the terminology we use) is however not obvious and occasionally inconsistent in the literature. For instance, \citealt{zhong19} report a \emph{quiescent} time for 081024A as $\approx 0.91$~s, meaning the time from precursor end to main burst beginning, which is clearly shorter than that between peak precursor flux and main event beginning. These definitional differences are important to account for if one wishes to compare in earnest with real, astrophysical data. Although applying for long bursts, a visual depiction of the different timescales is shown in Figure~\ref{fig:koshut_burst} from the pioneering precursor observations described by \citealt{koshut95}. Though their nomenclature of peak ($\Delta t_{\rm pk})$ and detectable ($\Delta t_{\rm det})$ waiting times did not persist in the literature, it is clear that even factor $\gtrsim 2$ differences in waiting time can occur if different definitions are applied.

Aside from this issue, statistical algorithms used to hunt for precursors can offer rather different values for event lengths or waiting times. For the case shown in Fig.~\ref{fig:troja_burst} for 081024A, \citealt{zhong19} find the precursor took place between $-1.65$ to $-1.59$ seconds prior while \citealt{troja10} report $-1.70$ prior to $-1.45$. The event duration is different by a factor $\sim 4$ between these. While this Review cannot cover all the statistical nuance, readers should be aware that results shown by us and others in the literature are sensitive to a variety of systematics, including that associated with different instrument observations. For example, in Fig.~\ref{fig:troja_burst} it is clear that the waiting time may be different if one were to use only Fermi or Swift data: this could, in principle, be attributed to either instrument systematics or the bandwidth of emitted photons.

\subsection{GRBs: short, long, or ultra-long?}
\label{sec:longvsshort}

Historically, GRBs have been classified according to their $T_{90}$ durations: long have $T_{90}>2$~s while short bursts have $T_{90}\leq2$~s. It was long postulated that these two classes of burst were associated with different progenitors; the long ones with ``collapsars'' or some kind of active core-collapse \cite[e.g.][]{br17}, and the short ones with merger events involving at least one neutron star \cite[e.g.][]{burns20}.

As it often does, Nature proves more complex: in 2010 an ``ultra-long'' burst was detected, GRB 101225A (so-called ``Christmas Burst''), observed by Swift and boasting a staggering 28 minute duration. This led some authors to introduce a third classification for such ultra-long bursts. Additionally, it has been suggested that the Christmas burst originated from a merger, though not necessarily involving two neutron stars \cite[e.g.][]{xmas11}. Moreover, although predicted already by \citealt{pac98}, \emph{kilonovae} are now routinely observed \cite[including after GW170817;][]{val17}. These transients involve bright, broadband electromagnetic emissions due to the radioactive decay of heavy $r$-process nuclei that are produced and subsequently ejected quasi-isotropically during the merger process. They are thought to be a signature of a binary neutron-star coalescence and can be searched for with a variety of instruments owing to their broadband nature; a detailed discussion on kilonova detectability, with current and future detectors, is provided by \citealt{chase22}.

Some notably long bursts have been accompanied by kilonovae, thereby making it even less clear whether the usual $T_{90}$ classification can be used to distinguish between collapse or merger scenarios. For instance, \citet{Yang15} examined the late-time afterglow data from GRB 060614 --- a burst with $T_{90} \gtrsim 100$~s ---  finding a (candidate) $r$-process-powered merger-nova. Such complications are the reason we use the words \emph{merger-driven GRBs} when talking about precursors. Indeed, all GRB classes show precursors with varying statistics, with about $\sim 9\%$ of the global GRB population showing precursor activity \cite[see, e.g.][]{Coppin20}. Long bursts seem to display precursors more often than short ones though (cf. Sec.~\ref{sec:rarity}).

Some particularly relevant examples are GRBs 211211A \cite{troja22} and 230307A \cite{lev24}, with durations of 50 and 35 s, respectively. Both of these events are likely associated with binary mergers due to kilonovae and heavy-element production. They also both displayed precursor phenomena (see Sec.~\ref{sec:onset} and Table.~\ref{tab:precdata} below). It is probable therefore that some merger-driven precursors have historically evaded analysis because there was an inherent assumption that the GRBs were fuelled by an isolated object.  The above discoveries highlight the need for a new classification scheme for GRBs, which influence the precursor inferences covered in this Section; see \citealt{deng22} for ideas in this direction. With the recent advent of the James Webb Space Telescope (JWST), classification schemes are likely to improve in the coming years.

\subsubsection{Excluded events}
\label{sec:excluded}

In sections that follow regarding merger-driven precursor properties, a few candidate events, discussed in the literature, are excluded here. These are the (possible) precursors associated with GRBs 050724, 080702A, 091117, and BATSE trigger 2614 (3B 931101). 

The first two of these are described by \citealt{troja10}, and are remarkable in that the waiting times exceed $\sim 100$~s. The authors remark several times however that they ``are unable to confidently determine whether they are real features'' as they were not seen by instruments other than Swift and are of relatively low significance. Although the possibility of such a long waiting time could offer exciting insight into neutron-star structure (cf. Fig.~\ref{fig:mode_evol20}), we do not comment on these events further here. The final BATSE event is described by \citealt{koshut95}, also with a very long waiting time of $\sim 75$~s. The significance is, however, similarly unclear (as is perhaps visually evident from the fourth-to-last panel seen in their figure 3). Note that \citealt{Wang20} found a precursor from 080702A but much closer to the main event; it could be, therefore, that this is the second example of a known double precursor event after 090510 (see Sec.~\ref{sec:090510}). In the case of GRB 0911117, \citealt{troja10} find a significance of only $\sim 1.8 \sigma$ and \citealt{zhong19} remark that the data can no longer be downloaded from the Swift archives (although this event may have also been recorded by Suzaku). We exclude this otherwise unremarkable event also (waiting time $\sim 2.7$~s).

\subsection{Rarity}
\label{sec:rarity}

Uncertainties notwithstanding with respect to the long-short classification issues detailed above, classifications for \emph{merger-driven} precursors also depend on a variety of systematics. For instance, some authors require that a genuine precursor must precede the main burst by more than the burst $T_{90}$ \cite[e.g.][]{troja10,min17}. This would mean there is a clear separation of timescales (as in Fig.~\ref{fig:koshut_burst}), and not a possible blur period where it is unclear whether the precursor was related to burst ramp-up (cf. GRB 150922A). Some other studies allow for arbitrarily short waiting times relative to the main burst for early flashes to qualify for precursor status \cite[see ][for discussions]{zhong19,Coppin20,Wang20}. 

Another more obvious issue relates to sample size and instruments, with some events only being seen by Fermi, and/or Swift, and/or INTEGRAL \cite[some events even by Agile, KONUS, Suzaku, Messenger, and HEND-Odyssey;][cf. Table~\ref{tab:precdata}]{min17}. As such one has to make a statistical decision about the significance required to define a precursor \emph{with respect to one or more instruments}. Taking a high cut-off for some information criterion (e.g. Akaike or Bayesian) will naturally reduce the event rate and hence predict that precursors are rarer. As noted by \citealt{suvkprec19} and others, even these deceptively small systematics can change  event rate predictions from $\sim 0.4\%$ \cite{min17} to $\sim 10\%$ \cite{troja10}, with the most recent studies falling somewhere around a few percent. If we included some long bursts as part of the sample, as argued reasonable in Sec.~\ref{sec:longvsshort}, the event rate would adjust further. Without a clear criterion for separating merger-driven and formed-in-isolation GRBs, it is very difficult to formally estimate precursor rarity and thus this remains an open problem in our view. The problem also goes the other way: is it strictly necessary to have a merger to form a short GRB \cite[see][]{virgil11}? This issue has also been discussed by \citealt{Wang20} in light of the fact that even deducing $T_{90}$ is prone to systematics.

The next standout problem is deducing the rarity of the subset of precursors that are launched \emph{premerger}. It is likely that not all precursors are in fact launched pre-merger; if we took the GW/GRB 170817A waiting time of $\sim 1.7$~s as canonical, about $\sim 25\%$ of precursors would be associated with the inspiralling stars themselves (see Figure~\ref{fig:histogram} below). The delay time is however rather sensitive to the post-merger environment. High-resolution, numerical studies show that black holes typically launch jets within $\lesssim 10^{2}$~ms while metastable neutron stars (if able at all; see Sec.~\ref{sec:postmerger1}) tend to take longer \cite[see][and references therein]{sarinlasky21}. Black hole environments are typically cleaner as baryon pollutants from dynamical or post-merger ejecta, that could drag the jet and reduce its Lorentz factor, are swept under the event horizon, leading to shorter breakout timescales \cite[see][for a discussion]{pav21}. These issues are discussed in more detail in Sec.~\ref{sec:jetform}. Even so, combining the above, we may estimate that of order $\gtrsim 1\%$ binary mergers involving at least one neutron star display a \emph{premerger precursor}. 

\subsection{Spectral inferences}
\label{sec:spectral}

Spectral information about a precursor may be useful in reducing the uncertainties detailed above, as well as landing on a likely candidate for the ignition mechanism. As described in detail by \citealt{Tsang11,Tsang13}, one may anticipate less thermality in cases where magnetic fields are dynamically dominant, at least with respect to premerger models. Essentially, the matter boils down to how important backreactions are. If the local magnetic fields in the region where the flare originates are strong, energy may be more freely transported along open field lines (via Alfv{\'e}n waves; see also Sec.~\ref{sec:dynamo}) without scattering that would otherwise tend to thermalise the spectrum. If instead the field is relatively weak there, cross-field transport is hardly prohibited and scattering can take place en masse. A typical decider in this matter will be how strong the magnetic energy density is compared to the precursor luminosity, with \cite{Tsang11}
\begin{equation}
    B \gg 10^{12} \times \left( \frac{L_{\rm prec}}{10^{46} \text{ erg s}^{-1}}\right)^{1/2} \text{ G}
\end{equation}
leading to a strongly non-thermal spectrum.

In the above we have introduced the precursor luminosity, $L_{\rm prec}$. Deducing this quantity from observations is non-trivial, and few events in the literature have had luminosities reported. The difficulty stems from the fact that one needs to translate (i) photon counts into fluxes, and then (ii) fluxes into luminoisites. For the former this depends on the energy band of the instrument(s) and often requires delicate extrapolation between instruments \cite[see][for a discussion on common pitfalls]{meredith23}. The latter requires knowledge of source redshift, often not recorded at the times of publishing because it requires hunting through astronomical databases for coincident events (there may not be any). One must also be careful in deducing a kind of averaged luminosity as opposed to a peak one, since the gamma-ray flux will, of course, be time-dependent. Thus far, no clear understanding of the precursor luminosity distribution is available in the literature \cite[though see][for a recent study]{deng24}. It should also be understood that $L_{\rm prec}$ refers to an \emph{isotropic} luminosity, and thus could actually be lower if there is a non-negligible amount of beaming (see Sec.~\ref{sec:energetics}).

Many bright precursor flares tend to exhibit a non-thermal spectrum  \cite[see table 2 in][for instance]{zhong19}. This would be expected if Alfv{\'e}n waves propagating along open field lines are the primary means of the associated energy transport as detailed above, though this requires high field strengths and brings its own set of issues (cf. Sec.~\ref{sec:magneticfields}). Regardless, brightness combined with non-thermality has led directly to the suggestion that some mergers contain magnetars \cite[e.g.][]{troja10,suvkprec19,Kuan21b,tsang23,xiao24}. Precursor rarity may therefore be directly related to the prevalence of magnetars in mergers. Explaining how a strong field can persist into merger, even in $\lesssim 1\%$ of the population, remains an open issue under this interpretation. One possibility is that the field is not preserved at all but rather \emph{generated} just before merger from convective instabilities triggered by mode-driven differential rotation; see Sec.~\ref{sec:dynamo}.

As some specific examples, \citealt{Wang20} noted that the (bright) precursors in GRBs 111117A and 160804180 strongly favoured non-thermal fits. These both had short waiting times relative to the main event, and could thus favour a magnetospheric interaction or high-frequency mode resonance (see Sec.~\ref{sec:prectheory} for a review of theoretical elements). \citealt{zhong19} noted a strong favouring for a non-thermal, cutoff power-law in the (dim) precursor from GRB 130310A, which occurred $\sim 4$~s prior to the main event. The precursors of GRBs 081216 and
141102A, both with delay times of $\sim 1$~s, can be convincingly fit with a blackbody \cite{wl21}. They may therefore be associated with weak fields, or some kind of shock-breakout model. The absence of thermal components in the extremely bright precursor from GRB 211211A \cite[$L_{\rm prec} \approx 7 \times 10^{49} \text{ erg s}^{-1}$;][]{xiao24}, which occurred $\sim 1$~s prior to the main event could theoretically be expected from either a Poynting-dominated outflow from a remnant or possibly premerger magnetar interaction. Systematising such an analysis would constitute a useful addition to the literature, though spectral fittings can be notoriously complicated and this extends yet another observational problem therefore. Some recent efforts towards systematic, spectral classification were carried out by \citealt{deng24}.

\subsection{Waiting times}
\label{sec:onset}

\begin{figure*}
	\centering
	\includegraphics[width=0.8\textwidth]{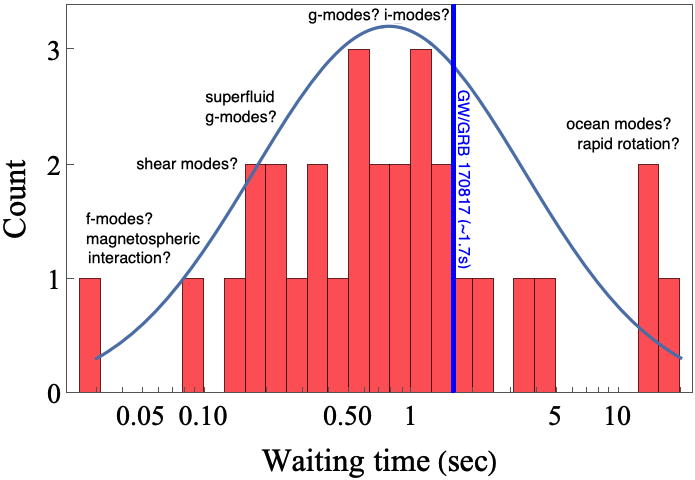}
	\caption{Histogram of precursor waiting times (cf. Sec.~\ref{sec:excluded}), overlaid with the best-fit lognormal distribution and several indicators for potential (premerger) precursor ignition mechanisms. Note these waiting times are \emph{relative to GRB not merger}: the solid blue line marks the delay observed in GW/GRB 170817. Data from Table~\ref{tab:precdata} and references therein.}
	\label{fig:histogram}
\end{figure*}

Merger-driven precursors exhibit a range of GRB-relative waiting times, from $\sim 30$~ms (150922A) to $\sim 20$~s (191221A). Table~\ref{tab:precdata} presents relevant data from the literature, including main GRB name, the duration of the precursor, the time delay between precursor and GRB, GRB duration, and some comments on remarkable features (if any). The waiting time distribution is illustrated graphically in Figure~\ref{fig:histogram}, showing that a reasonable fit is obtained with a lognormal distribution. Overlaid are a number of possibilities relating to premerger explanations, as described in Sec.~\ref{sec:prectheory}.

It is evident that there is a significant spread of precursor phenomena, with a couple orders of magnitude separating the shortest and longest waiting times and also durations. It is probable therefore that there are multiple ignition mechanisms, involving either pre- or post-merger phenomena; the physics pertaining to delays are discussed in Secs.~\ref{sec:jetform} and \ref{sec:launching}. We remark on the nature of some entries in Tab.~\ref{tab:precdata}, as some have error bars while others do not. In \citealt{Wang20}, from which many of our tabulated precursor results are quoted from, the authors did not given a confidence interval attached to the $\pm$ values they provided. As per standard convention, these error bars may correspond to $84.13 \%$ confidence upper and lower limits containing the $68.27\%$ confidence interval \cite{geh86}. In some other studies, such as \citealt{min17,zhong19}, waiting times are given as a kind of mean value from $T_{90}$ data (see Sec.~\ref{sec:precstats}), which we quote here using the $\sim$ symbol. Owing to these differences in notation and convention, on top of other systematics described thus far, caution should be applied when weighting precursor significance across different studies.

\begin{table*}
	\centering
	\caption{List of precursors (likely) associated with merger events, organised in descending order of waiting timing relative to main event, with some remarks on special features (if any). Key: XAM = X-ray afterglow with internal plateau, possibly suggestive of a magnetar remnant as identified by \citealt{rowl13} and others \cite[e.g.][]{suvk20p,suvkok21a,skk22}. Most data were compiled by \citealt{wl21}, though others not listed there come from \citealt{min17} (071030), \citealt{zhong19} (100717, 130310A, 170726794), \citealt{tsang23} (230307A), and \citealt{deng24} (180703B). The long-form Fermi GBM catalogue name is used for some GRBs. Some precursors may be postmerger. 
}
	\begin{tabular}{ccccc}
	    \hline
	    \hline
	    Source & Precursor duration (s) &  Relative delay (s) & GRB duration (s) & Remarkable features \\
     \hline
150922A & $0.05^{+0.01}_{-0.01}$ & $0.03^{+0.01}_{-0.01}$ & $0.08^{+0.01}_{-0.01}$ & Peak flux larger than that of the main pulse \\
100223110 & $0.02^{+0.03}_{-0.01}$ & $0.08^{+0.02}_{-0.03}$ & $0.12^{+0.01}_{-0.01}$ & - \\
080702A & $\approx 0.31$ & $\approx 0.13$ & $\approx 0.64$ & XAM; Stable magnetar? \\
160804180 & $0.16^{+0.02}_{-0.02}$ & $0.17^{+0.02}_{-0.02}$ & $0.26^{+0.02}_{-0.02}$ & - \\
170709334 & $0.46^{+0.01}_{-0.27}$ & $0.17^{+0.30}_{-0.07}$ & $0.15^{+0.07}_{-0.04}$ & Thermal precursor and main GRB \\
111117A & $0.18^{+0.05}_{-0.03}$ & $0.22^{+0.03}_{-0.06}$ & $\approx 0.46$ & XAM; Stable magnetar? Debated $T_{90}$; $z = 2.211$ \\
100702A & $\approx 0.04$ & $\approx 0.23$ & $0.16^{+0.03}_{-0.03}$ & XAM; $t_{c} \approx 178$~s \\
180703B & $\sim 1.5$ & $\sim 0.3$ & $\sim 1.54$ & Thermal spectra; long-duration precursor\\
060502B & $\approx 0.09$ & $\approx 0.32$ & $\approx 0.24$ & Debated $T_{90}$ \\
100827455 & $0.11^{+0.05}_{-0.04}$ & $0.34^{+0.06}_{-0.06}$ & $0.09^{+0.02}_{-0.01}$ & Debated waiting time \cite{zhong19} \\
230307A & $\approx 0.4$ & $\approx 0.4$ & $\approx 33$~s & LGRB but Kilonova? $L_{\rm prec} \approx 3.6 \times 10^{50}$~erg/s(!) \\
090510 (I) & $0.05^{+0.07}_{-0.03}$ & $0.52^{+0.04}_{-0.08}$ & $0.30^{+0.01}_{-0.01}$ & Double! XAM; $z=0.9$; peaks in 15–50 keV band \\
081216 & $0.15^{+0.05}_{-0.03}$ & $0.53^{+0.04}_{-0.05}$ & $0.24^{+0.02}_{-0.02}$ & Debated spectra \cite{deng24} \\
071112B & $\approx 0.01$ & $\approx 0.59$ & $\approx 0.27$ & - \\
150604434 & $0.17^{+0.25}_{-0.01}$ & $0.64^{+0.02}_{-0.29}$ & $0.21^{+0.03}_{-0.02}$ & - \\
100213A & $\approx 0.44$ & $\approx 0.68$ & $\approx 0.94$ & - \\
181126A & $\approx 0.72^{+0.18}_{-0.27}$ & $0.85^{+0.40}_{-0.29}$ & $0.46^{+0.11}_{-0.13}$ & - \\
081024A & $\approx 0.06$ & $\approx 0.91$ & $\approx 0.94$ & Debated $T_{90}$; XAM; Collapse time $\approx 125$~s? \\
211211A & $\approx 0.2$ & $1.08^{+0.20}_{-0.20}$ & $\approx 35$ & LGRB; Kilonova; QPOs main and prec(!); XAM  \\
140209A & $0.61^{+0.08}_{-0.08}$ & $1.10^{+0.08}_{-0.08}$ & $1.03^{+0.04}_{-0.06}$ & Debated $T_{90}$ ($\approx 2.4$~s?); LGRB?; Strongly thermal \\
101208498 & $0.17^{+0.12}_{-0.08}$ & $1.17^{+0.1}_{-0.14}$ & $1.03^{+0.03}_{-0.04}$ & - \\
141102A & $0.06^{+0.10}_{-0.06}$ & $1.26^{+0.11}_{-0.15}$ & $0.48^{+0.04}_{-0.04}$ & Thermal spectra \\
170726794 & $\sim 0.08$ & $\sim 1.53$ & $\sim 0.25$ & - \\
170802638 & $0.015^{+0.17}_{-0.11}$ & $1.85^{+0.14}_{-0.21}$ & $0.33^{+0.04}_{-0.04}$ & - \\
071030 & $0.9 \pm 0.2$ & $\sim 2.5$ & $2.7 \pm 0.5$  & Data appear lost \cite[see][]{zhong19}? Debated $T_{90}$ \\
100717 & $\sim 0.15$ & $\sim 3.32$ & $\sim 1.23$ & Strongly non-thermal. Debated $T_{90}$ \\
130310A & $0.9 \pm 0.32$ & $4.45 \pm 0.8$ & $\sim 2.15$ & Debated spectra \cite{qin21}; Magnetar flare? QPOs\\
180511437 & $2.80^{+1.38}_{-1.69}$ & $12.72^{+1.80}_{-1.57}$ & $3.33^{+0.18}_{-0.24}$ & LGRB? Debated $T_{90}$; Longest precursor \\
090510 (II) & $\approx 0.4$ & $\approx 12.9$ & $0.3^{+0.01}_{-0.01}$ & Double! Peaks around $\sim$300 keV \\
191221A & $0.03^{+0.59}_{-0.03}$ & $19.36^{+1.24}_{-3.19}$ & $0.37^{+0.26}_{-0.13}$ & - \\
     \hline
	    \hline
	\end{tabular}
	\label{tab:precdata}
\end{table*}

\subsection{Some exceptional precursors}
\label{sec:exceptional}

Here we go into some more detail about some precursors with remarkable properties. These events may be particularly useful in deducing information about the neutron-star EOS and other fundamental physics.

\subsubsection{GRB 211211A}
\label{sec:211211A}

GRB 211211A was accompanied by a kilonova \cite{raj22}. This by itself is  exciting: the GRB was long $(T_{90} \gtrsim 30$s), and yet the association with a kilonova identifies the origin as a merger event (see Sec.~\ref{sec:longvsshort}). Aside from this, the event showed a precursor flare $\sim 1$s prior to the main event \cite{xiao24}. In \citet{skk22}, the precursor timing was matched with a $g$-mode resonance with sufficiently large overlap that the crust may break, within the resonant failure model (see Sec.~\ref{sec:resfail}). Since magnetic fields skew the mode eigenfrequency, the magnetic field strength that was obtained $(\gtrsim 10^{14}$~G) was required to be consistent with that set by the extraordinarily high isotropic luminosity $(\sim 7 \times 10^{49} \text{erg s}^{-1}$) and the absence of thermal emissions in the precursor (see Sec.~\ref{sec:spectral}). 

{In addition to that described thus far, the duration of the \emph{main event} may also point towards magnetar participation in the merger. \citet{kiu14} found in GRMHD simulations with merging magnetars that the accretion disc that comes to surround the remnant is itself highly magnetised. Although not seen in their simulation, it has been proposed that this, combined with the magnetic field sourced by the remnant if a neutron star, can set up a barrier that episodically chokes the accretion flow thereby prolonging the event duration  \cite[see][and references therein]{xiao24}.}

The 211211A precursor was also modulated by quasi-periodic oscillations (QPOs) at $\sim 22$~Hz \cite{xiao24} \cite[see also][]{chirenti24}. Seismic excitations in the crust is a natural prediction of a failure: a deep quake relieves energy from the crust which then causes the surroundings to vibrate through aftershocks. These aftershocks are most conducive to torsional modes, as initially pointed out by \citealt{dunc98}, which happen to have frequencies in the $\sim 20$~Hz range. The combination of having mass estimates for the binary members from the kilonova \cite{raj22}, estimates for the internal properties and magnetic field strength from a $g$-mode {(or other)} resonance, \emph{and} estimates for the shear modulus in the crust from the torsional mode frequency proves intriguing. Although a thorough parameter space scan was not performed by \citet{skk22}, it was shown that conventional assumptions on the EOS (Fig.~\ref{fig:EOS}) and shear modulus from recent literature can accommodate all of the observed properties. More precisely, the frequency of the precursor QPO (interpreted as a torsional mode) and relative timing of the precursor itself (interpreted as a $g$-mode resonance) can be well-fit with a star that has kilonova-inferred mass $M= 1.25 M_{\odot}$ \cite{raj22} with the APR EOS (Sec.~\ref{sec:APR}) and shear modulus computed from nuclear physics (Sec.~\ref{sec:crust}). Other interpretations are valid, though. For example, the QPO could be associated with a magnetoelastic mode \cite[e.g.][]{gab12,gab16} and the precursor could be attributed to an interface-mode \cite[e.g.][see also Sec.~\ref{sec:imodes}]{Tsang11,Tsang13}.

\subsubsection{GRB 180703B}
\label{sec:180703B}

The 180703B precursor is unusual because of the event duration. As found by \citealt{deng24}, the precursor emissions began $\sim 1.8$~s prior to the main burst, lasting all the way until a lull at $\sim 0.3$~s, followed by the GRB of a similar duration. The delay could therefore be longer depending on what one calls the waiting time (again see Sec.~\ref{sec:precstats}). The authors however denoted this precursor as a ``gold sample'' burst of very high significance with a thermally-dominated spectrum. The main GRB was also well fit with strongly thermal components and overall high-degree of spectral similarity. These facts together may point towards a common origin, i.e. a postmerger model with a choking jet or similar (see Sec.~\ref{sec:post-merger}).

\subsubsection{GRB 180511437}
\label{sec:180511437}

The precursor in 180703B described above is not however the longest duration precursor; that title goes to that seen prior to GRB 180511437 with a duration of $\sim 2.8$~s \cite{Wang20}. Given that the $T_{90}$ duration of this event has been debated in the literature, with some stipulating a long burst (see Sec.~\ref{sec:longvsshort}), it is difficult to place constraints on the progenitor. However, given the long delay ($\sim 12$~s) and the lack of any obvious spectral similarity,  a long period of magnetospheric interaction with especially tangled fields seems difficult to accommodate \cite{fernmet16}. Resonances with a particularly long window are theoretical possible for some EOS; see Sec.~\ref{sec:reswindowdurat}.

\subsubsection{GRB 191221A}
\label{sec:191221A}

The precursor from 191221A is currently the record holder for earliest precursor relative to main event, with a staggering $\sim 20$~s waiting time (though cf. Sec.~\ref{sec:excluded}). No constraints on the spectrum  were placed \cite{Wang20}. This event was coincident with LVK's O3b run though no GWs were detected \cite{grbgws22}, pointing to a non-negligible source redshift. The distinct lack of any X-ray afterglow is also problematic with respect to a stable neutron-star remnant scenario, which would probably be favoured in order to ascribe the long delay to a jet formation timescale \cite{zhang19,burns20}. Premerger models in this direction, assuming a $< 10$~s combined jet production and breakout timescale (see Sec.~\ref{sec:jetform}), are practically limited to low eigenfrequency resonances such as $g$-modes or ocean modes (Sec.~\ref{sec:resfamilies}) unless strong magnetic fields, thermal gradients, or rotation skew the eigenfrequency towards lower values (see Sec.~\ref{sec:spinacc}). A systematic analysis of this event would be worthwhile to carry out because of the above considerations.

\subsubsection{GRB 090510}
\label{sec:090510}

GRB 090510 is special for two reasons: one being that, aside from the unclear case of 081024A, it is the only known merger-driven GRB that showed a \emph{double precursor}. 

\begin{figure}
	\centering
	\includegraphics[width=0.497\textwidth]{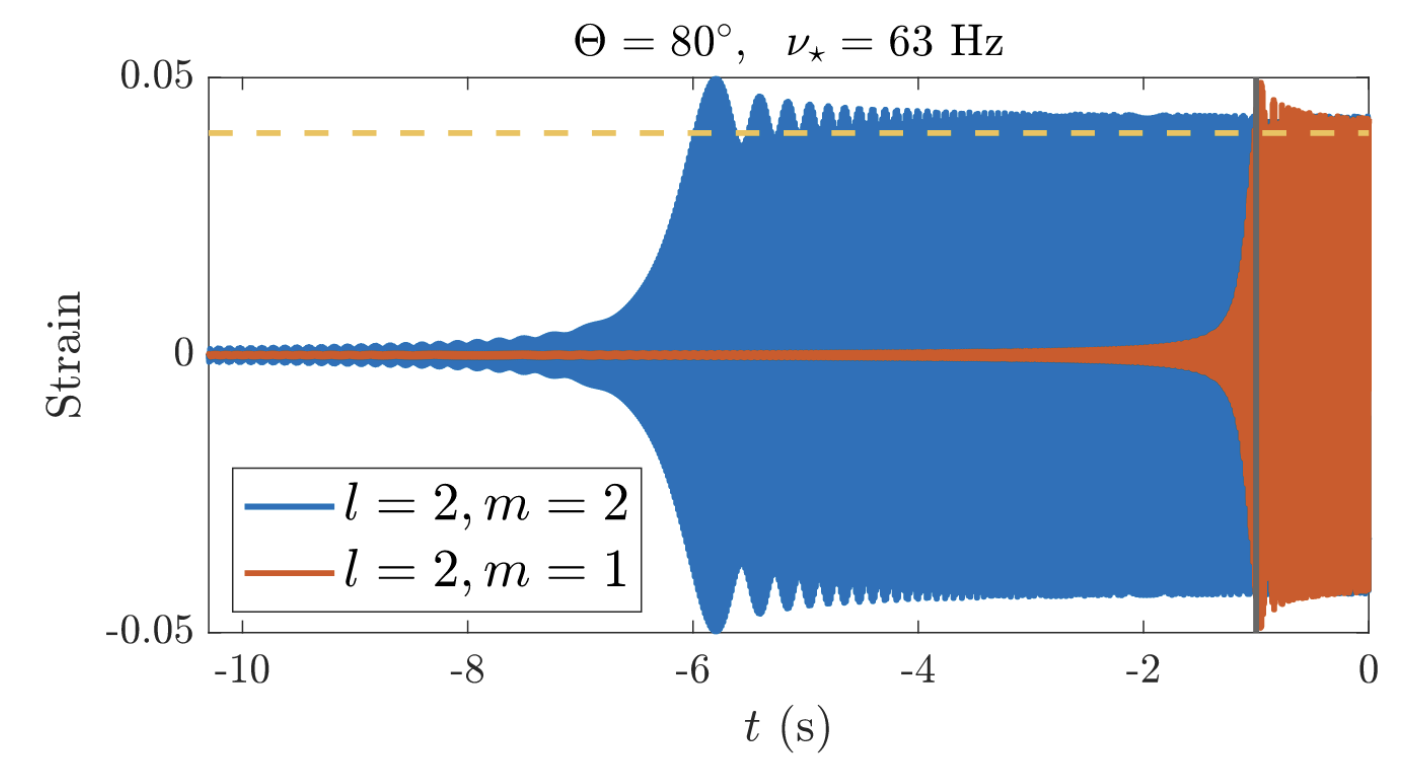}
	\caption{A double resonance of $m=2$ and $m=1$ kink modes in a highly misaligned and rapidly-rotating neutron star. The strain ($\sigma$) can exceed $\sim 0.04$ --- a realistic elastic maximum \cite{bc18,kozh23,b24} --- and thus, theoretically, trigger two episodes of failure separated by a number of seconds. The stratification rises to $\delta \approx 0.021$ for the later episode, due to intervening tidal heating (cf. Sec.~\ref{sec:thermal}), having started from an initial (cold) value of $\delta \approx 0.006$. The binary is symmetric with stars of mass $1.23 M_{\odot}$ and the APR4 EOS. From \citealt{kuan23}.}
	\label{fig:double_prec}
\end{figure}

The two precursors from this event were separated by $\sim 12$~s \cite{troja10}, with the second precursor taking place $\sim 0.5$~s prior to the main event. Several possibilities present themselves, involving (i) two premerger precursors, (ii) two postmerger precursors, or (iii) one of each. The parameter space is large. The first precursor was notably softer ($\sim 30$~keV peak) than the second ($\sim 300$~keV peak), which disfavours the operation of strong \emph{external} magnetic fields at early stages. A possibility put forward by \citealt{kuan23} involves rotation and stellar misalignment. As shown in Figure~\ref{fig:double_prec}, a significant misalignment angle and rapid rotation (though the latter is not strictly necessary, even spins of $\lesssim 10$~Hz are sufficient) can instigate two episodes of crustal failure with $m=2$ and $m=1$ modes (see Sec.~\ref{sec:spinacc}). Although the delay seen in this Figure does not exactly correspond to that seen in 090510, it demonstrates the possibility. Regarding the 090510 double event specifically, parameter-space constraints are given in \cite{kuan23}.

The second remarkable feature of GRB 090510 is that of a rather prominent X-ray afterglow, as discussed by \citealt{rowl13,suvk20p} and others. This afterglow exhibited the `internal plateau' feature thought to be characteristic of a neutron star pumping spindown energy into the forward shock. Importantly, the system showed a steady decline in flux rather than a sharp cutoff, which is often interpreted as the collapse time of the metastable star \cite[see, e.g.][]{ravi14,suvg22}. If indeed a stable neutron star was formed, it is more plausible that the jet formation timescale was comparatively long, and hence that only one precursor was premerger with the second being attributable to a choking jet, possibly because neutron-stars have a harder time forming jets with high Lorentz factors \cite{cio19}. Still, several scenarios are viable.

\subsubsection{GRB 230307A}
\label{sec:230307A}

GRB 230307A is an extremely bright, \emph{long} GRB. The recorded gamma-ray fluence is $\gtrsim 3 \times 10^{-3} \text{ erg cm}^{-2}$ in the 10–1000 keV band \cite{tsang23}, second only to GRB 221009A. Despite its long duration, it is likely associated with a kilonova and heavy element production \cite{lev24}, thus resembling the case of GRB 211211A \cite{tsang23} of a merger-driven long GRB. Taking the best-fit distance of $\approx 291$~Mpc implies an enormous luminosity $L_{\rm prec} \sim 3.6 \times 10^{50} \text{ erg s}^{-1}$. As discussed by \citet{tsang23}, this event is a strong candidate for a magnetar with an external field of $\gtrsim 10^{15}$~G taking place in a merger. {This is not only because of the spectrum and high luminoisty, but because of its long-short nature supporting a magnetar-in-a-merger origin if magnetic barriers are easier to set up (as for GRB 211211A).} If the field was of this magnitude, accounting for magnetic shifting in mode eigenfrequencies would be essential, as would electromotive torques towards the end of the binary evolution (see Sec.~\ref{sec:magacc}). Given the waiting time of only $\approx 0.4$~s however, many ignition scenarios are viable (especially post-merger ones).

\subsection{Connections with post-merger phenomena}
\label{sec:postmerger1}

Insofar as strong magnetic fields are necessary to explain premerger, precursor spectra (Sec.~\ref{sec:spectral}), the remnant environment and structure could be affected. It is fair to say that one of {the} open problems in GRB physics concerns whether or not magnetar remnants are capable of launching a relativistic jet that can drill through the polar baryon pollution \cite[e.g.][]{sarinlasky21}. Most numerical simulations of mergers that leave a magnetar find that (i) a magnetic field of strength approaching equipartition ($\sim 10^{17}$\,G) is needed to a launch a jet (if a jet can be launched at all), and (ii) the magnetic field that is generated is larger if the pre-merger `seed' fields are larger \cite[e.g.][]{c20b}. Point (ii) has been challenged by recent, high-resolution simulations \cite{kiu24}, though it is naturally easier to accept strong remnant fields via flux conservation arguments if there is a magnetar in the merger.

The magnetic field strength of a remnant neutron-star would also influence its spindown energy and GW luminosity. In principle, one could anticipate a subpopulation of remnants with stronger fields if bright precursors are observed therefore \cite{suv24}. Even if irrelevant for precursors, dynamo activity triggered by $f$- or other mode resonances just before merger could also amplify the seed field(s) (see Sec.~\ref{sec:dynamo}).

\section{Precursor flares: theory}
\label{sec:prectheory}
This section of the Review is dedicated to theoretical models for precursor flares. The bulk relates to premerger mechanisms (though see Sec.~\ref{sec:post-merger}), with special focus placed on the resonant failure picture owing to the authors' familiarity. If some stellar oscillation mode comes into resonance with the orbital motion while the stars are inspiralling, significant amounts of tidal energy may be deposited into the mode(s) as described in Sec.~\ref{sec:dyntides}. The resonant amplitude may thus reach the point that the crust yields due to the exerted stresses exceeding the elastic maximum (Sec.~\ref{sec:breaking}). Two of the more promising candidates in this direction are the $g$- and $i$-modes, which appear to lie in a sweet-spot, in the sense that the expected mode frequencies match orbital frequency at times corresponding to many precursor flashes and the overlap integrals are sufficiently large (see Sec.~\ref{sec:resfamilies}). The other main mechanism involves electrodynamic interactions between the binary components (Sec.~\ref{sec:magnetosphericinteraction}) or shocks in a mass-loaded wake \cite{mezr92}; there could be subpopulations of precursors associated with different mechanism.

\subsection{Delay Timescales: postmerger jets}
\label{sec:jetform}

If one is to reliably assign a mechanism to precursor flare ignition, it is important that the physics relating to jet formation and breakout are understood in the context of the waiting time observations described in Sec.~\ref{sec:onset}. 

For better or worse, there is considerable theoretical uncertainty surrounding the orbital/GW frequency at which precursors are launched. This is because the times that are recorded are relative to the main GRB event (see Tab.~\ref{tab:precdata}). However, this will not coincide exactly with the moment of coalescence, which is the relevant quantity as concerns GWs, because there will generically be a delay timescale, $t_{\rm J}$, required in order for the jet to not only form but also break-out (i.e. penetrate through the surrounding baryon pollution) and produce the observed EM radiation (see Fig.~\ref{fig:dia1}). For example, assuming that GWs do indeed propagate at the speed of light, the landmark event GW170817 indicates that the SGRB took $\approx 1.7$~s to launch postmerger. If this estimate were canonical, many of the events listed in Tab.~\ref{tab:precdata} would occur \emph{postmerger}. This is obviously problematic if one wishes to ascribe resonant modes to their excitation mechanism. 

Fortunately --- at least from a premerger perspective --- there is reason to suspect that $t_{\rm J}$ may be considerably shorter than that seen in GRB 170817A in some instances, perhaps even down to a $\sim$~few ms. In this section we closely follow and summarise the compilations of \citealt{zhang19} and \citealt{burns20} in describing various jet formation mechanisms and $t_{\rm J}$ estimates. Where appropriate, some recent numerical simulations or theoretical results concerning remnants are instead quoted to update and/or adjust values presented in Table 1 of \citealt{zhang19}.

We decompose the delay time, $t_{\rm J}$, of a GRB observation relative to merger into three independent terms,
\begin{equation}
 t_{\rm J} = \Delta t_{\rm jet} + \Delta t_{\rm bo} + \Delta t_{\rm GRB},
\label{eq:taujet}
\end{equation}
where $\Delta t_{\rm jet}$ represents how long it takes for the engine to collimate the jet, $\Delta t_{\rm bo}$ is then the extra time for the jet to \emph{break out} from the ambient material ejected before and at merger (and also of the environment generally), with finally $\Delta t_{\rm GRB}$ being an extra propagation timescale associated with the jet to reach some appropriate \emph{energy dissipation radius} where the actual $\gamma$-rays are emitted. In general there will also be a combined cosmological and local (i.e. due to remnant gravity) redshift factor $1+z$ to account for (see also Sec.~\ref{sec:framedragging}), but since this applies to GWs and merger also it is unimportant for the relative timing issue. 

\citealt{zhang19} introduces a second decomposition for $\Delta t_{\rm jet}$, depending on the mechanism responsible for launching. The mechanism is naturally dependent on the progenitors. We write
\begin{equation}
\Delta t_{\rm jet} = \Delta t_{\rm rem} + \Delta t_{\rm acc,B} + \Delta t_{\rm clean},
\end{equation}
where $\Delta t_{\rm rem}$ is a waiting time for the responsible remnant to form (typically zero, unless invoking the collapse of some metastable neutron star), $\Delta t_{\rm acc,B}$ is a further add-on related to the time for either accretion or dynamo activity to actually prime the object for jet launching [i.e. dynamical fall-back timescale (acc) or dynamo timescale for magnetic growth (B)], and lastly we have the time taken for the jet to clean the surroundings (which is connected to the degree of mass loading), $\Delta t_{\rm clean}$. This final quantity is probably the most uncertain, and depends on the Lorentz factor that is achieved, which can be dragged by pollutants surrounding the crash site. For a thermal (neutrino-anti-neutrino annihilation) fireball, one anticipates a short timescale possibly of order only $\sim$~ms. For a magnetically (Poynting-flux) dominated jet, on the other hand, not only will the Lorentz factor generally be smaller \cite{cio18} but the jet might have to reach a turbulence/reconnection radius which is thought to exceed that of the relevant photosphere by a factor $\sim 100$. The main expectations are summarised in Table~\ref{tab:jet_form}.

\begin{table*}
	\centering
	\caption{Expected delays associated with merger-driven GRBs and relative precursor timings, adapted and expanded from \citealt{zhang19}. Abbreviations: NS = Neutron star; SM = Supramassive; HM = hypermassive. Remarks: (*) HMNS collapse time could theoretically be prolonged through thermal \cite{pas12} or magnetic \cite{suvg22} support, extending the upper limit significantly. (**) Result quoted from analysis performed in \citealt{cio19} regarding breakout requirements [see also \citealt{most20} and \citealt{cio18} for detailed discussions]. (\dag) Results anticipated from high-resolution dynamo simulations of \citealt{kiu24}, who found amplification on sub-ms timescales. (\ddag) From simulations of \citealt{ruiz18}, who found that sufficiently low-mass BHNSs result in a HMNS that tends to launch a magnetically-dominated jet. Argued by \citealt{st08} that for canonical neutron-star parameters, SGRB is limited to ``low'' energies of order $\sim 10^{48}$ erg/s because of tidal disruption impact on torus mass. (\#) We assume a fireball-like mechanism for accretion scenarios involving a neutron star, or a Blandford-Znajek mechanism if involving a BH, and a Poynting-flux-dominated jet for dynamo (magnetic) scenarios when estimating $\Delta t_{\rm GRB}$. Though, in principle, any of these mechanisms could have an upper limit of many $\sim$~s owing to theoretical uncertainties; see \citealt{zhang19} and \citealt{burns20}. Black hole topology and modified-gravity terms could also skew these estimates, though such effects are ignored \cite[see, e.g.][]{spiv00,namp20}.}
	\begin{tabular}{cccccccccc}
 \hline
\hline
\label{tab:jet_form}
System & Engine & Mechanism & {} &
$\Delta t_{\rm jet}$ & {} & {$\Delta t_{\rm bo}$} (s) & {$\Delta t_{\rm GRB}$} (s; \#) & $t_{\rm J,min}$ (s) & $t_{\rm J,max}$ (s) \\
\hline
{} & {} & {} & {$\Delta t_{\rm rem}$} (s) &
{$\Delta t_{\rm acc,B}$} (ms) & {$\Delta t_{\rm clean}$} (s) & {} & {} & {} \\
BH-NS & BH & accretion & $\sim$ 0 & $\sim 10$ & $\sim$ 0 & $\gtrsim 10^{-2}$; $\lesssim 10^{-1}$ & $<$~few & $\sim 10^{-2}$ & $\sim$3 \\
BH-NS & HMNS/BH & accretion & $\gtrsim 0.1$; $\lesssim 1$(*) & $\sim 10$ & $\sim$ 0 & $\gtrsim 0.1$; $\lesssim 1$ & $\lesssim$~few(\ddag) & $\lesssim $1  & $>$5(*) \\
BH-NS & HMNS/BH & magnetic &  $\sim$ 0  & $<1$(\dag) & $<$1 & $\gtrsim 10^{-2}$; $\lesssim 1$ & $\lesssim$~few(\ddag) & $\lesssim$1 & $\sim$5 \\
NS-NS & BH & accretion & $\sim$ 0 & $\sim 10$ & $\sim$ 0 & $\gtrsim 10^{-2}$; $\lesssim 10^{-1}$ & $<$~few & $\sim 10^{-2}$ & $\sim$3 \\
NS-NS & HMNS/BH & accretion & $\gtrsim 0.1$; $\lesssim 1$(*) & $\sim 10$ & $\sim$ 0 & $\gtrsim 0.1$; $\lesssim 1$ & $\lesssim$~few & $\gtrsim 10^{-1}$  & $>$5(*) \\
NS-NS & HMNS/BH & magnetic &  $\sim$ 0  & $<1$(\dag) & $\lesssim 1$ & $\gtrsim 10^{-2}$; $\lesssim 1$ & $\sim$~few & $\gtrsim 10^{-2}$ & $>$ 3 \\
NS-NS & SMNS/NS & accretion & $\sim$ 0  & $\sim 10$ & $\lesssim 0.1$ & $\gtrsim 10^{-2}$; $\lesssim 10^{-1}$ & $\lesssim$~few & $\gtrsim 10^{-2}$ & $\gtrsim$ 3 \\
NS-NS & SMNS/NS & magnetic & $\sim$ 0  & $<1$(\dag) &  $\lesssim 0.2$(**)  & $\lesssim 0.2$(**) & $\sim$~few & $\lesssim$1 & $> 3$ \\
\hline
\hline
\end{tabular}
\end{table*}

We close this section by noting that multimessenger events may accompany a collapse event for some of the remnant types listed in Tab.~\ref{tab:jet_form}. Most notably, fast radio bursts (FRBs) have been predicted to be produced following the collapse of a neutron star \cite[see, e.g.][]{fr14,suvg22}. Once an event horizon forms and cloaks the star, the external field lines are no longer anchored and so snap, inciting relativistic, magnetic shocks that accelerate electrons to 
high Lorentz factors, ultimately producing radiation in the $\gtrsim$~GHz band. A coincident detection of a sharp drop in X-ray flux and an FRB would be a ``smoking gun'' for a post-merger, neutron-star collapse; in such a case, at least $\Delta t_{\rm rem}$ could be tightly constrained. See \citealt{lu24} for recent (unsuccessful) search efforts for coincident FRBs and GRBs, though see \citealt{clancy23} who found that FRB 20190425A is coincident with GW190425 at $\sim 2.8 \sigma$. Sadly, no GRB was observed for this event.

\subsection{Magnetospheric interaction and unipolar inductor}
\label{sec:magnetosphericinteraction}

Depending on the persistence (or late-time generation) of strong (crustal) magnetic fields over cosmological timescales, electrodynamic interactions occurring in the final seconds of inspiral could produce fireworks. The mechanisms responsible have been described in detail by \citealt{hl01} and others since. The production of gamma-rays can proceed through essentially three different channels, depending on the relative magnetization and alignment angles of the two stars \cite{piro12,lai12,wang18}. In cases with a black hole, comparable estimates can be found in \citealt{mcw11}.

The simplest, and arguably most probable if a luminous precursor is observed, scenario in the electrodynamic interaction context is that of the unipolar inductor introduced in Sec.~\ref{sec:magacc}. The reason for this is that it is already difficult to explain one magnetar in a merger let alone two, and without strong fields to mediate the interaction the precursor is likely to fizzle out long before it reaches Earth. The other two cases instead apply when the object's have comparable magnetic dipole moments, either aligned or anti-aligned (though in reality of course there will be some angle, $\Theta_{\mu}$). In the aligned case, the field lines will compress at the interaction radius given by $r_{i} \approx a/(1 + q_{\mu}^{1/3})$ with dipole-ratio $q_{\mu} = \mu_{B,d} / \mu_{A,d}$ \cite{wang18}. In the opposite case, we have instead direct reconnection occurring in the interaction zone which leads to explosive event(s) followed by the establishment of a quasi-stable circuit similar to a unipolar inductor \cite[e.g.][]{palen13,Crinquand19}.

\citealt{lai12} argues that the maximum energy that can be relieved relates to $\dot{E}_{\rm EM}$ from equation \eqref{eq:magdiss}. This is controlled by the effective resistance between the two stars through the effective twist parameter $\zeta_{\phi}$, viz.
\begin{equation} \label{eq:dissmax}
\begin{aligned}
    \dot{E}_{\rm EM} &\approx 6.0 \times 10^{43} \zeta_{\phi} \sqrt{\frac{M(1+q)}{1.4M_{\odot}}} \left( \frac{B}{10^{12} \text{ G}} \right)^{2}\\
    &\times \left(\frac{a}{30 \text{ km}}\right)^{-13/2} \left(\frac{R}{13 \text{ km}}\right)^{8} \text{ erg s}^{-1},
    \end{aligned}
\end{equation}
where $B$ here is to be understood as the largest equatorial strength of either star and we have assumed equal stellar radii. The above is difficult to reconcile with precursor luminosities unless $a \sim 3 R$ and $\zeta_{\phi} \gg 1$, though $\dot{E}_{\rm EM}$ can be slightly larger than estimates provided by \citealt{wang18} if $\zeta_{\phi}$ is kept free.

The estimate \eqref{eq:dissmax} is in rough agreement with that found from \emph{ab initio} numerical simulations for large twists. \citealt{mostphil20} found, via special-relativistic simulations, a maximum dissipation of (equation 6 therein)
\begin{equation} \label{eq:mostexpression}
    \dot{E}_{\rm MP20} \approx 4.6 \times 10^{44} \left( \frac{B_{\rm int}}{10^{12} \text{ G}} \right)^{2} \left( \frac{R}{13 \text{ km}} \right)^{3},
\end{equation}
applying just $\sim$~ms before merger for equal mass objects \cite[see also][]{Crinquand19}. Note in particular the relevant $B$ value here is the local value at the interaction radius not that at the stellar surface (though these are approximately equal at merger by definition). Interestingly, the luminosity scaling was found by \citealt{mostphil20} to be $\propto a^{-7/2}$, which is shallower than that of \eqref{eq:dissmax}. Either way, we can match \eqref{eq:mostexpression} with \eqref{eq:dissmax} at merger if $\zeta_{\phi} \lesssim 10$, consistent with the anti-aligned simulations by \citealt{palen13,Crinquand19}. \citealt{mostphil23,carr21} studied neutron-star plus black-hole binaries, finding roughly similar dissipation rates though with the interesting result that the flares present in their simulations were up to $\sim 10^{2}$ times \emph{brighter} than the (relativistic) orbital emission.

An important aspect of the interaction model is that emissions should be generated almost instantaneously. That is, particle production occurs immediately after reconnection and thus the precursor timing should correspond directly to the ignition moment without an additional lag beyond those detailed in Sec.~\ref{sec:jetform}. On the other hand, the \emph{duration} of a precursor born from magnetospheric interaction is likely limited to $\sim 10$~ms based on the relationship between the Poynting flux and chirp length where the separations are low  \cite{fernmet16}. This is somewhat shorter than \emph{all} event durations listed in Tab.~\ref{tab:precdata}, though allowing for a factor $\sim$~few extension would bring many low-waiting-time precursors into view (e.g. 150922A, 100223110 , 100702A, \ldots). Still, \citealt{mz16} argue the detection horizon is limited to $\sim 10$~Mpc for $B \approx 10^{14}$~G unless the twist is large, $\zeta_{\phi} \gg 1$.

Finally, aside from reconnection being generally able to spark a plethora of progenitors (see also Sec.~\ref{sec:multimessenger}), strong magnetic fields in the emission environment could effect photon propagation directly. For example, quantum-mechanical photon splitting could acquire a non-negligible cross section \cite{hard82,medlai10}, thereby impacting on the polarization states and/or Faraday rotation of electromagnetic waves. Such effects, manifesting as vacuum birefringence, have been observed in emissions from the isolated neutron stars RX J1856.5-3754 \cite{mig17} and 4U 0142+61 \cite{tav22} with characteristic field strengths of a few by $10^{13}$~G  \cite{pop17} and $\sim 10^{14}$~G \cite{mcg14}, respectively. As noted by \citealt{wl21}, the next-generation gamma-ray polarimeter POLAR-2 may be able to measure such effects directly and constrain the plasma environment \cite{polar2}.

\subsection{Resonant failure: luminosity and timescales}
\label{sec:resfail}

The resonant failure picture that has been alluded to throughout this Review is covered here and Sec.~\ref{sec:resfamilies}. The model was initially put forward by \citealt{Tsang11} though has since been covered by many authors in different contexts. In a nutshell: a large amount of tidal energy can be rapidly siphoned into modes from the orbit when a resonance is hit depending on the respective overlap integral; see Sec.~\ref{sec:dyntides}. For some modes, the resonant amplitude grows large enough to overstrain the crust (see Sec.~\ref{sec:crust}), releasing magnetoelastic energy that can theoretically fuel a gamma-ray flash. The way in this energy propagates is likely complicated depending on the local field strength, though the recent GR force-free electrodynamic simulations carried out by \citealt{most24} tracked Alfv{\'e}n waves propagating out to the orbital light cylinder with crossing time $t_{\rm cross}^2 \approx a^{3}/G(M_{A} + M_{B})$ where flares were formed self-consistently. For $a \approx 100$~km the light-crossing time is of order ms (see also Sec.~\ref{sec:launching}). Although the mechanism responsible for the overstraining involves resonant pulsations rather than the gradual evolution of the magnetic field itself, the electromagnetic extraction of energy is totally analogous to that put forward for magnetar flares \cite{td95}. 


Based on relationship \eqref{eq:sigmarel} for some mode, one can investigate whether any given mode candidate can be responsible for crustal failure. The mode eigenfrequency relates to the ignition time and GW frequency, which when correcting for jet formation and other factors (Sec.~\ref{sec:jetform}) can be matched to the precursor waiting time. Details of specific mode families are covered in Sec.~\ref{sec:resfamilies}, while below we go into more general aspects relating to luminosity and timescales.

\subsubsection{Energetics}
\label{sec:energetics}

\begin{figure}
	\centering
	\includegraphics[width=0.49\textwidth]{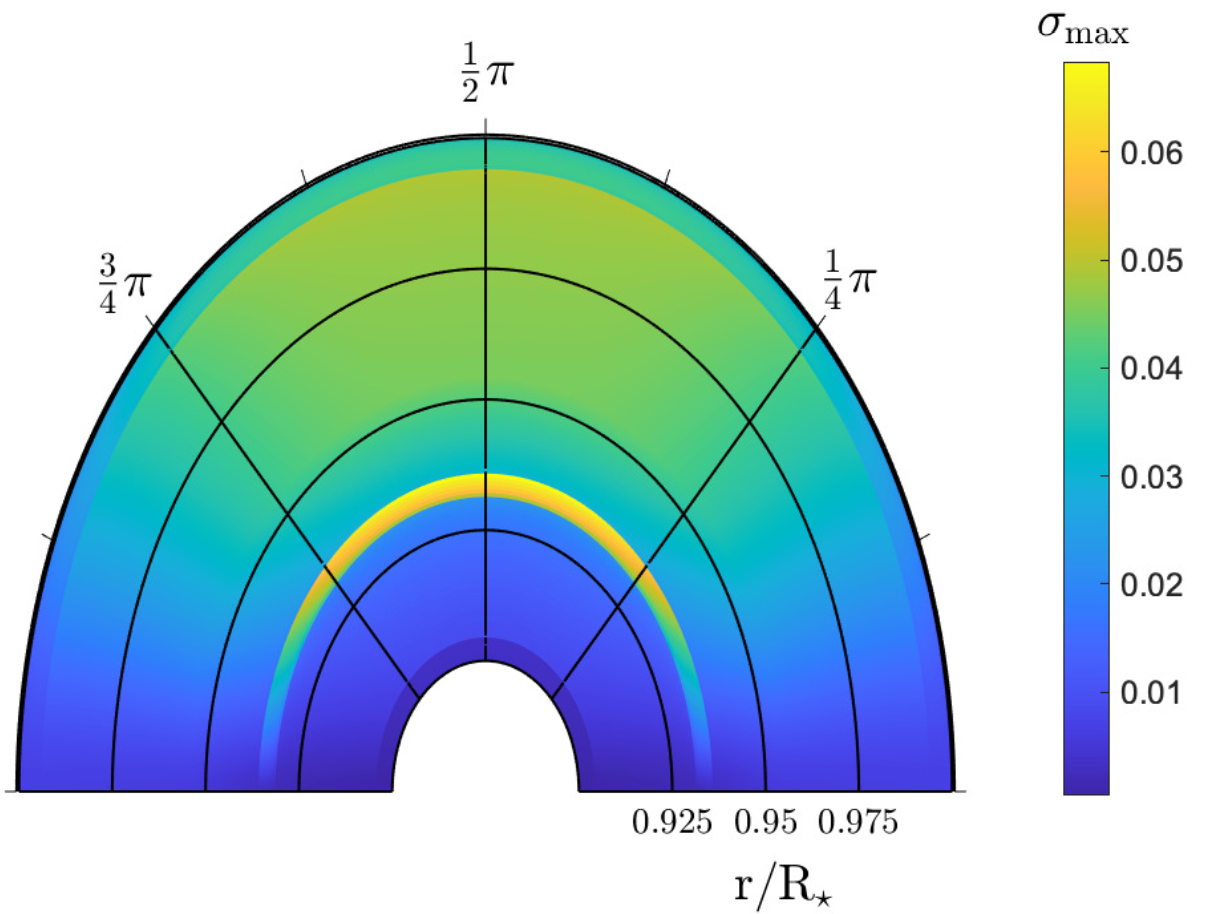}
	\caption{Crustal strain pattern, $\sigma$ (equation~\ref{eq:sigmarel}), in the northern hemisphere as induced by a $g_{2}$-mode for an \emph{unmagnetised star} with the SLy EOS in an equal mass binary with $M = 1.27 M_{\odot}$ and a constant stratification index $\delta = 0.005$. For $\sigma_{\rm max} \lesssim 0.05$ (see Tab.~\ref{tab:sigmamax}) a complicated failure geometry will emerge. From \citealt{Kuan21b}.}
	\label{fig:strainexample}
\end{figure}

The maximum magnetically-extractable luminosity from the crust of a neutron star is given by a Poynting integral, which for a dipole can be estimated through \cite{Tsang11,Tsang13}
\begin{equation} \label{eq:maxlum}
L_{\rm max} \sim 10^{47} \left(\frac{v}{c}\right) \left( \frac {B_{\rm crust}} {10^{13}\, \mbox{G}}\right)^2 \left(\frac{R} {10\, \mbox{km}}\right)^{2} \mbox{~erg~s}^{-1},
\end{equation}
where $v$ is the speed of the mode perturbation. While this value is considerably larger than \eqref{eq:magdiss}, essentially because we have integrated over the whole \emph{surface} rather than just in some small volume within a possibly distant interaction zone, it still requires relatively large fields to explain bright precursors. Such tensions can be alleviated somewhat by beaming (expected if some zone on the surface fails rather than the entire crust), and by noting that there is additional energy available beyond just that of the magnetic field. 

More generally though, the maximum amount of energy during some window that can be liberated within a failed zone reads \cite[e.g.][]{lan15},
\begin{equation} \label{eq:quakeenergy}
	\int dt E_{\text{quake}} \leq \int_{t_{\rm offset}}^{t_{\rm onset}}dt \int_{V_{\text{crack}}(t)} d^{3}x \, u_{\rm tot},
\end{equation}
where the total energy density $u_{\rm tot}$ accounts for all types of energy stored in the region. The above is obtained by integrating the energy stored in the failed zone over a resonance duration $t_{\rm onset} \leq t \leq t_{\rm offset}$ (see Sec.~\ref{sec:reswindowdurat}). From a von Mises perspective, the volume $V$ can be defined through 
\begin{equation}
	V_{\text{crack}}(t)=\{ p\; |\; \sigma(p) \ge \sigma_{\rm max}, \; p\text{ within the crust} \},
\end{equation}
which is just the set where the elastic maximum is exceeded at any given $t$.

The total density, $u_{\rm tot}$, includes a (i) magnetic contribution [which is effectively just a rescaled version of \eqref{eq:maxlum}], the (ii) kinetic energy density of QNMs, $u_{\text{kin}}$, (iii) rotational energy density, $u_{\text{rot}}$, and finally the (iv) tidal energy density, $u_{\text{tid}}$. The three latter contributions are respectively given by
	\begin{equation}
		u_{\text{kin}} = \frac{1}{2} \left( \dot{\boldsymbol{\xi}} \cdot \dot{\boldsymbol{\xi}} \right) \rho
  \end{equation}
  \begin{equation}
		u_{\text{rot}} = \frac{1}{2}\Omega^{2} r^{2}\sin^{2}\theta \rho,
  \end{equation}
and
	\begin{equation}
		u_{\text{tid}} = U \delta\rho, 
	\end{equation}
for tidal field $U$ from equation \eqref{eq:ueqn}. An elastic term may be added also.

Under the approximations that (i) the energy released during a resonance timescale is just the integral of the energy density at the onset of resonance over the cracking area at the offset of resonance, and (ii) $u_{\text{rot}}$ is only associated with uniform rotation and frame dragging is ignorable (cf. Sec.~\ref{sec:framedragging}), these energies were calculated by \citealt{Kuan21b} for some $g$-mode resonances (see Sec.~\ref{sec:gmodes}). The main results are listed in Table 2 therein: for magnetar-level fields strengths, energy outputs can exceed $\sim 10^{46}$ erg each second. (How long such an event may last is covered in Sec.~\ref{sec:reswindowdurat}). This would be sufficient to power most precursors, noting in particular that only some small fraction of the crust actually fails (rather than global events) in those simulations; see Figure~\ref{fig:strainexample} for one such, $g_{2}$-strain pattern. If indeed the precursor was attributable to a $g_{2}$-mode, the isotropic luminosities quoted in Sec.~\ref{sec:spectral} would not be appropriate and in fact the energetics could easily be accommodated through expression \eqref{eq:quakeenergy}.

Aside from beaming considerations, for slow stars the main contributions are from the magnetic field, so that \eqref{eq:maxlum} remains the leading-order piece. However, some $\sim 20$\% or more leeway could be afforded if the resonances are triggered particularly close to merger, relaxing (a little) difficulties associated with magnetic field strengths (see Sec.~\ref{sec:magneticfields}).

\subsubsection{Resonance window duration}
\label{sec:reswindowdurat}

For a given mode with (inertial-frame) eigenfrequency $\omega_{\alpha,i}$,  resonance will be triggered when $\Omega_{\rm orb}$ falls in the interval $[ \frac{1-\varepsilon}{2}\omega_{\alpha,i},\frac{1+\varepsilon}{2}\omega_{\alpha,i} ]$ \cite{Lai94}, where the real part of the eigenfrequency is implied. This frequency interval yields a total duration of 
\begin{equation} \label{eq:precduration}
    t_{\rm prec} = \frac{\varepsilon \omega_{\alpha,i}}{\dot{\Omega}_{\rm orb}} \approx \frac{2 \varepsilon \Omega_{\rm orb}}{\dot{\Omega}_{\rm orb}},
\end{equation}
which can be compared directly to observational durations from Table~\ref{tab:precdata} assuming low redshift \cite[cf. equation 1 in][and Sec.~\ref{sec:framedragging}]{zhang19}. The (dimensionless) detuning parameter $\varepsilon$ can be calculated self-consistently with numerical simulations or the analytic approximation provided by equation (3.10) in \citealt{Lai94}. It is generally expressible as
\begin{equation}\label{eq:resnbh}
	\varepsilon = \chi \sqrt{  \frac{2\pi }{\Omega_{\rm orb}}  \frac{|\dot{a}|}{a}},
\end{equation} 
for some parameter $1 \lesssim \chi \lesssim \text{few}$. Although a value up to $\chi \approx 10$ was found for some $g$-modes by \citealt{Kuan21a}, more realistic values in most cases are $\chi \sim \mathcal{O}(1)$. Either way, the numerical detuning parameter \eqref{eq:resnbh} can be somewhat larger than the analytic estimate made by \citealt{Lai94}. More generally, $\chi$ depends on the EOS, mode family in question, spin, and microphysical (e.g. stratification) assumptions essentially because modes in stars with a specific mass can have the same frequency but different tidal overlap by tuning the aforementioned physics.

To leading order, combining \eqref{eq:adotleadingorder} with the frequency sweeping rate estimated from the Keplerian formula,
\begin{align}
    \frac{\dot{\Omega}_{\rm orb}}{\Oorb}= -\frac{3\dot{a}}{2a},
\end{align} 
we can estimate that at a separation of $a = 170$~km, corresponding roughly to a normal-fluid $g$-mode with $f_{\alpha,i} \approx 88$~Hz and $\approx 3$~seconds before merger, we have $t_{\rm prec} \approx 0.3$~s for stellar parameters such that $\chi=2$. Such durations are more in line with the observational data (Tab.~\ref{tab:precdata}) than the $\sim 10$~ms values anticipated from the magnetospheric interaction picture discussed in Sec.~\ref{sec:magnetosphericinteraction} \cite{fernmet16}.

\subsubsection{Launching timescale}
\label{sec:launching}

One important aspect of the resonant-failure picture is that the flare will not be immediately launched once resonance, which is the quantity deducible from GWs for instance, is met. This couples in with the issues described in \ref{sec:jetform} for deducing precursor-merger waiting times. In particular, as detailed by \citealt{tsang23,kuan23}, there are three timescales to consider. The first of these is the time it takes for overstraining to occur once resonance has been triggered. For the von Mises criterion \eqref{eq:vM}, this failure happens instantaneously once expression \eqref{eq:sigmarel} reaches a significantly large value. In a more realistic stress-strain model (e.g. Zhurkov; see Sec.~\ref{sec:breaking}) there may be a short delay associated with this step; see \citealt{wells}. 

The next timescale pertains to the delay for \emph{failure} once an overstraining has been triggered from resonance. This was estimated by \citealt{Tsang11} to be on the order of $\sim 1$~ms based on elastic-to-tidal energy ratios, though it is sensitive to both the overlap integral and mode frequency (see below equation 10 therein). These authors used a breaking-strain value of $\sigma_{\rm max} \sim 0.1$; all else being equal but using the \citealt{bc18} value $\sigma_{\rm max} \sim 0.04$ would increase this timescale by a factor $\sim (0.1/0.04)^2 \approx 6$ (see Tab.~\ref{tab:sigmamax}). Higher eigenfrequencies or energies would reduce it.

The final timescale of relevance, $t_{\rm emit}$, relates to how long it takes for emissions to be generated following the failure \cite[see][regarding the magnetar flare context]{td93,td95}. \citealt{neil22} argue  $t_{\rm emit}$ can be as long as $\sim 0.1$~s for $B \sim 10^{13}$~G --- see equation (12) therein --- again assuming $\sigma_{\rm max} \sim 0.1$ from \citealt{hk09}. The result scales like \cite{kuan23}
\begin{equation}
t_{\rm emit} \sim \frac{E_{\rm elastic}}{L_{\rm max}} \approx 0.03 \left(\frac{\sigma_{\rm max}}{0.04}\right)^2 \left(\frac {10^{47} \text{ erg/s}}{L_{\rm prec}}\right) \text{ s}.
\end{equation}
For most precursors, $t_{\rm emit}$ will be of order tens of ms. In support of this, \citealt{most24} suggest $t_{\rm emit}$ is related to a light crossing-time that is relatively short unless $a \gg 100$~km.

Altogether there are thus \emph{eight} timescales to keep track of when trying to directly compare a tidal resonance to a precursor timing observation, at least in the resonance picture {for bright \emph{nonthermal precursors}.} These are the five detailed in Sec.~\ref{sec:jetform} relating to jet generation and breakout, and the three described above (or only one in the case of magnetospheric interactions). This highlights the non-trivial nature of the problem, where future numerical simulations of jet break out and radiation transport are likely to prove particularly powerful. {In fact, for thermal or comparatively low luminosity precursors, an additional set of physical timescales become relevant relating to collisions of fireball shells. As described by \citet{neil22}, regions surrounding the emission site may be optically thick to pair production and multiple pair-photon fireballs could be launched which, upon collision, create shocks and synchrotron emissions together with inherent blackbody emissions with luminosities reaching $\sim 10^{48}$~erg/s. The collision times of these shells depends on the Lorentz factors set through the degree of mass loading, with theoretical ranges spanning a few orders of magnitude suggesting a wide range of viable timescales.}

\subsection{Resonant failure: some important families}
\label{sec:resfamilies}

This section details various QNM families which have been invoked to explain precursors in the literature: their physical origin, variation with EOS and microphysical parameters, eigenfrequencies, and overlap integrals.

\subsubsection{i-modes}
\label{sec:imodes}

\begin{figure}
	\centering
	\includegraphics[width=0.497\textwidth]{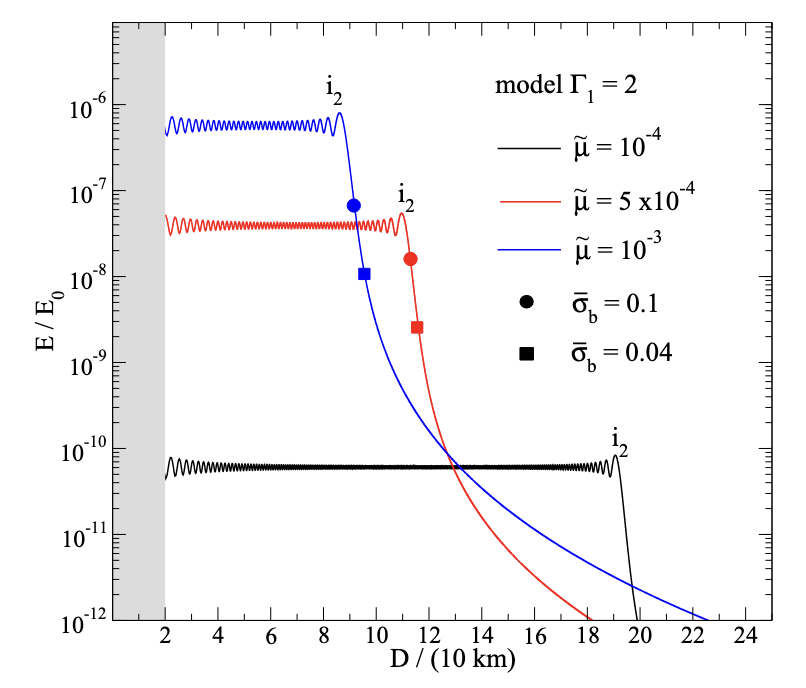}
	\caption{Resonant amplitudes (expressed via energies in units of $E_{0} = G M_{A}^2 /R_{A}$) obtained for the crust-core interface modes $i_{2}$ for a polytropic star in Newtonian gravity ($\Gamma_{1}$ is our $\gamma$), where elastic terms in the crust are self-consistently included for a variety of shear modulii $[\tilde{\mu}$; normalised according to equation~73 in \citealt{Passamonti20}]. Circles (squares) denote distances associated with the moment of crustal failure, if applicable, for a stronger (weaker) breaking strain of 0.1 (0.04) under a von Mises criterion. From \citealt{Passamonti20} with permission.}
	\label{fig:imodes}
\end{figure}

As a neutron star cools, it is expected to undergo a number of localised phase transitions. These can relate to superfluidity and conductivity (Sec.~\ref{sec:superfc}), and the formation of a crust and ocean layers (Sec.~\ref{sec:crust}). The presence of these solid-liquid discontinuities allows for a family of \emph{interface} ($i$-) modes to exist. These were the first family that were considered viable for breaking and arguably remain the strongest candidate \cite{Tsang11,Tsang13,tsang23}, at least for the crust-core ($i_{2})$ rather than crust-ocean ($i_{1}$) variety, though the latter can also induce non-trivial strain and crustal failure (see Sec.~\ref{sec:oceans}). This is because the overlap integrals tend to be large and $i$-mode eigenfrequencies are extremely sensitive to the transition density and stellar macrophysics, allowing for many events to be explained by this one mode in different stars \cite{neil22}. Figure \ref{fig:imodes} shows predictions for the (appropriately normalised) energies deposited into $i$-modes, as found by \citealt{Passamonti20}, who self-consistently included elastic terms into their (Newtonian) equations of motion.

\citet{Passamonti20} found $i$-mode overlaps to be two orders of magnitude weaker than that found by \citet{Tsang11}. Although still plausibly large enough to break the crust (at least for sufficiently large shear modulii; see Fig.~\ref{fig:imodes}), this demonstrates that $i$-modes may not be generically optimal for any given neutron star configuration when incorporating elastic aspects. Another possible origin for this large discrepancy is rooted in the mixed use of Newtonian overlaps/perturbations but with a GR star. Although the use of such a hybrid scheme to solve for stellar spectrum is common in the literature \cite[and was used by][for instance]{Tsang11}, it leads to some non-orthogonality between modes and there can then be ``leakage'' of one mode to others when computing overlap integrals\footnote{In GR, modes are not strictly orthogonal but the inner product between them is much smaller than the extent of this leakage; see also Footnote 6.} [see Appendix A of \citealt{miao24} for a qualitative discussion].

Overall, however, owing to the sensitivity of $i$-modes to crustal microphysics, these modes represent a powerful probe of stellar structure \cite[e.g. for the possible existence of pasta phases;][]{neil22,sot24}. {\citet{neil23} demonstrated that constraints for the nuclear symmetry parameters --- namely the (isovector) symmetry energy parameters $J$, $L$, and $K_{\rm sym}$ --- of a (nucleonic) EOS can be made when combining precursor measurements, assuming $i$-mode triggers, with GWs. The main results are described in figure 3 therein, illustrating the complementary nature of such constraints with those obtained from terrestrial (e.g. PREX-II \cite{reed21}, where the neutron skin thickness of the heaviest lead isotopes were measured) and astrophysical experiments. For example, for a given $i$-mode frequency and ignoring the impacts of rotation or magnetic fields, shifts of $\sim 20$~MeV in uncertainty ranges for $L$ could be anticipated.}

{There are open issues in the modelling of $i$-modes however, such as relates to consistent GR treatments, the choice of shear modulus profile, and the possibility of anisotropic nuclear clusters and pasta. Accurately incorporating these elements is a difficult task owing to both conceptual and numerical issues, and it is unclear how much anticipated constraints on parameter spaces may vary.}

\subsubsection{g-modes}
\label{sec:gmodes}

\begin{figure}
	\centering
\includegraphics[width=0.497\textwidth]{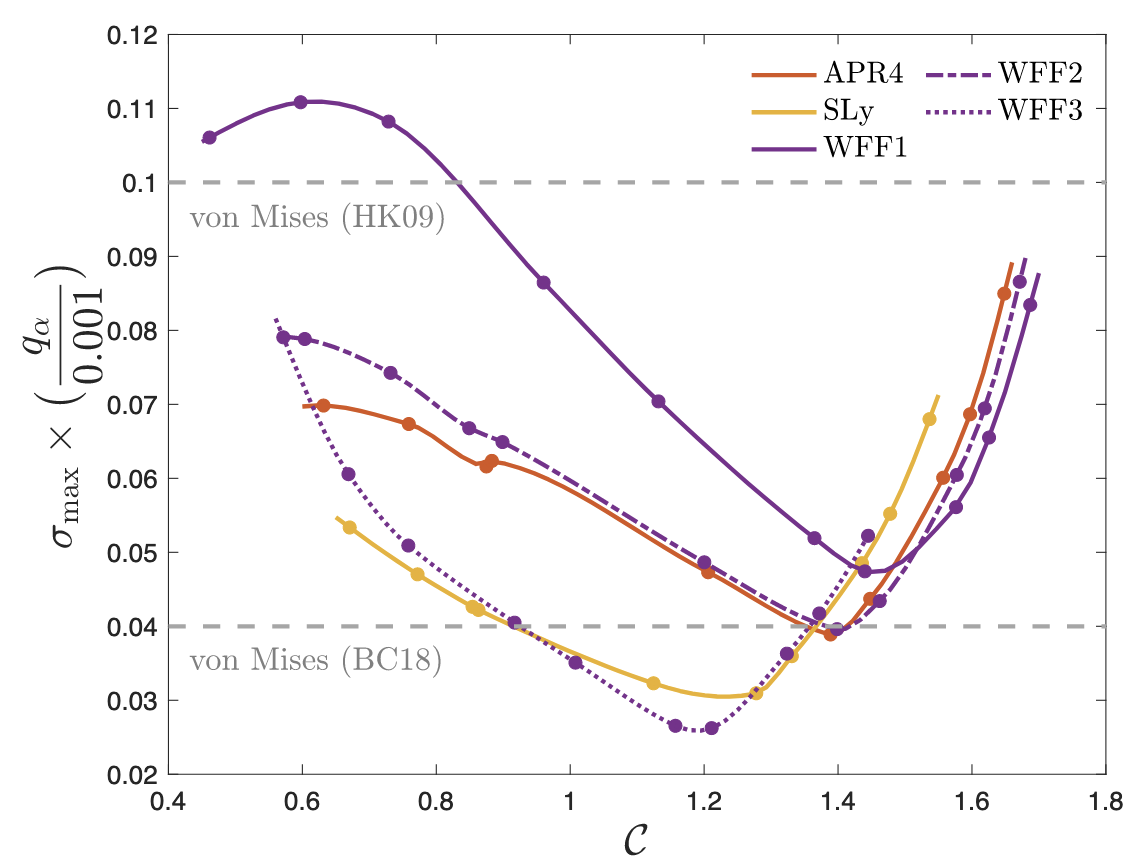}
	\caption{Dependence of the maximally induced stress (equation \ref{eq:sigmarel}) due to $g_1$- modes for various EOS (same for both stars with $q=1$) as functions of the stellar compactness $\mathcal{C} = M_{1.4}/R_{10\text{km}}$. The grey dashed lines represents the von Mises criterion identified by \citealt{bc18} (lower) and \citealt{hk09} (upper). A fixed value of $\delta = 0.005$ is used. From \citealt{Kuan21b}. }
	\label{fig:eosdepend}
\end{figure}

As described throughout, these modes are due to buoyancy resulting from thermal or composition gradients \cite[including discontinuities;][]{Finn87,McDermott90,Sotani01} sourced intrinsically or through the accretion of matter from a companion, crustal movements, or magnetic interactions \cite[e.g.][]{dm24}. For cold stars in a late-stage binary, the frequencies of (normal-fluid) $g$-modes can range from tens to hundreds of Hz, indicating resonances occur $1-10$~s prior to merger when spin is insignificant (see Sec.~\ref{sec:spinacc}). Interestingly, maximum amplitudes and frequency tend to increase with stellar mass for the $g_{1}$-modes, while the amplitudes tend to \emph{decrease} for $g_{2}$-modes; see Table 1 in \citealt{suv24}. On the other hand, $g$-modes in less compact stars have longer growth times, compensating their weaker coupling to the tidal field to some extent. As such, the maximally induced \emph{stress} exhibits a parabola-like shape, at least when the stratification is held fixed, as seen in Figure \ref{fig:eosdepend}. 
The local minima of these curves has been referred to as ``tidal neutrality'', as described in Appendix A of \citealt{kuan22gmode}.

\begin{figure}
	\centering
	\includegraphics[width=0.497\textwidth]{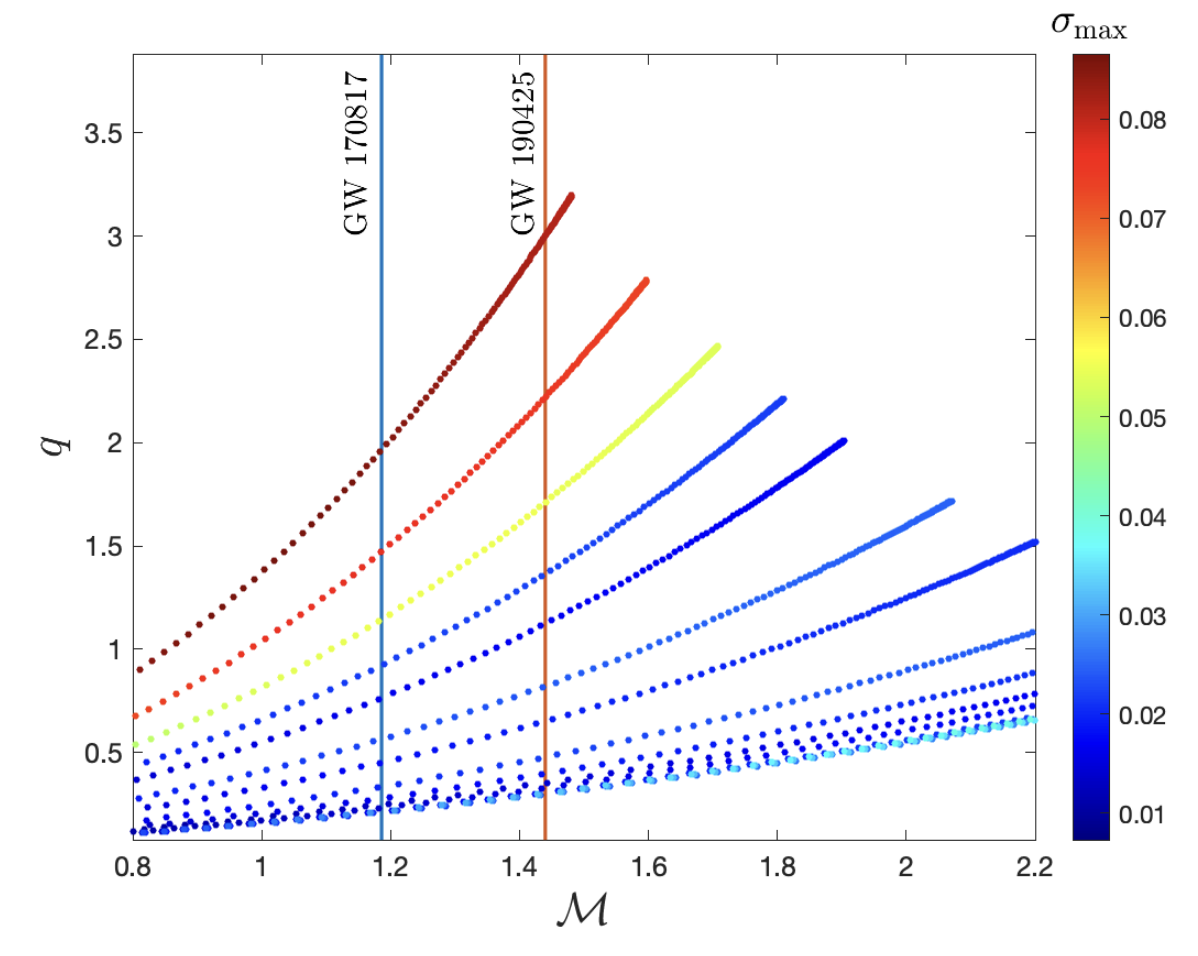}
	\caption{Sequences of maximal strain (redder shades indicating a greater value for $\sigma$) driven by $g_1$-mode resonances for the WFF1 EOS for various chirp masses $\mathcal{M}$ and mass-ratios. The blue vertical line shows the chirp mass of the progenitor of GW 170817, while the red corresponds to GW 190425. From \citealt{Kuan21b}.}
	\label{fig:q-chirp}
\end{figure}

In addition to the intrinsic mode properties, the binary mass-ratio also influences the resonance time-scale through the tidal potential \eqref{eq:ueqn}, and hence the maximal stress also. Fig.~\ref{fig:q-chirp} demonstrates that a larger strain is generally manifest when the companion is heavier (i.e., the $g_1$-mode issues a greater $\sigma$) for a given chirp mass, $\mathcal{M}=M_A q^{3/5}/(1+q)^{1/5}$, of the binary, which can be precisely estimated with GW analysis \cite[e.g.,][]{Cutler94}. This occurs because when $\mathcal{M}$ is fixed a larger $q$ implies a lower mass and pushes the system into the left-half of the compactness range considered in Figure \ref{fig:eosdepend}, noting the use of WFF1 (Sec.~\ref{sec:WFF}). For other EOS, the trend may reverse.

\begin{figure}
	\centering
	\includegraphics[width=0.497\textwidth]{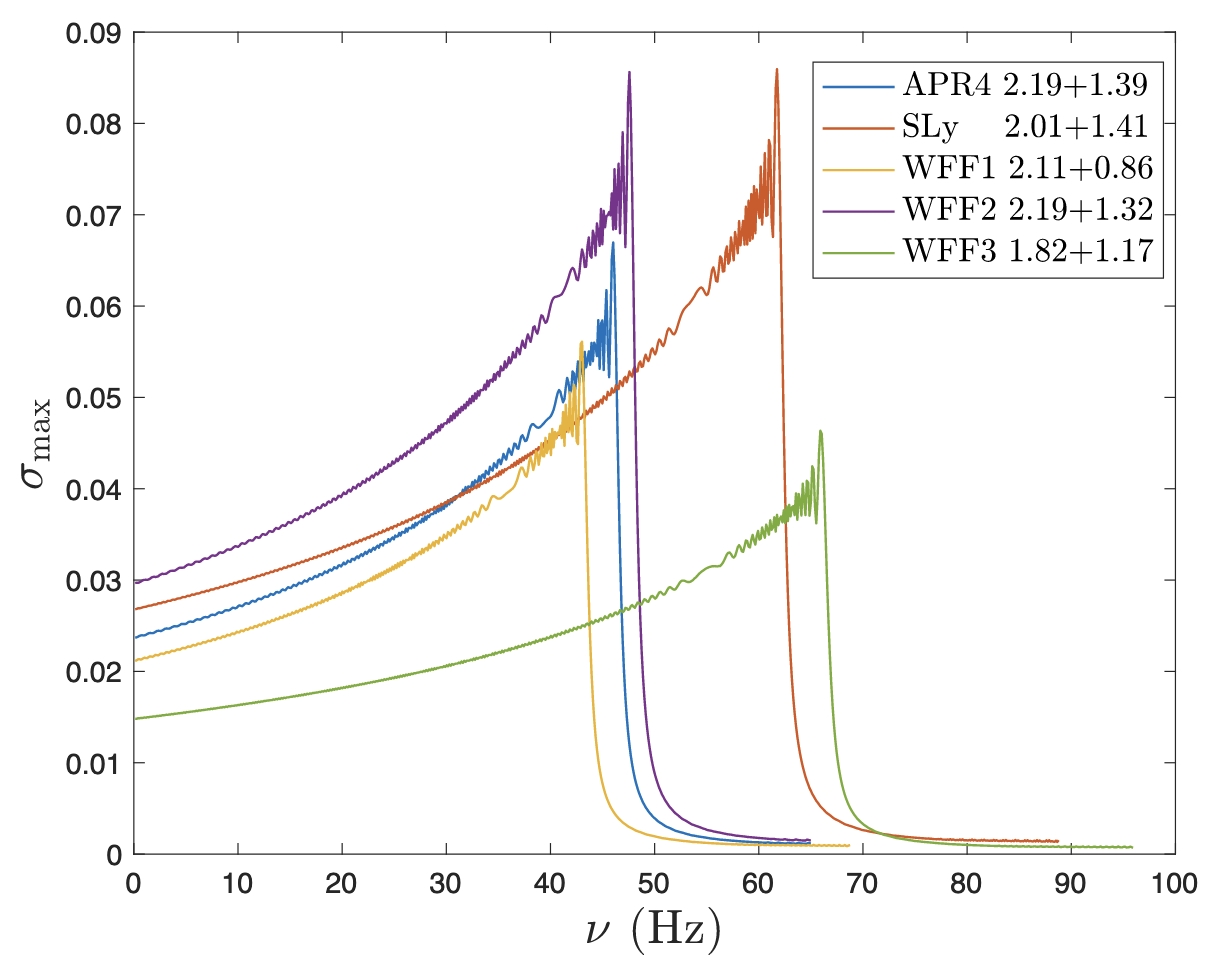}
	\caption{Similar to Fig.~\ref{fig:eosdepend} but as functions of the stellar spin for fixed compactness. Plot legends for the systems define the EOS that is obeyed by the primary and the companion together with individual masses in units of $M_{\odot}$. Note the sharp drops after a certain spin for different modes results because the mode frequency drops lower than the initial tidal pushing frequency used in the numerical computation, and may be analogous to the spectral ``washing out'' described in Sec.~\ref{sec:stratification} once the mode frequencies drop too low. From \citealt{Kuan21b}.}
	\label{fig:spindep}
\end{figure}

As discussed in Sec.~\ref{sec:spinacc}, spin reduces the frequency of retrograde modes. When the stellar angular momentum is in the opposite direction to that of the orbit, the oscillation rate of tidally-susceptible modes (whose phase pattern velocity matches the orbit) is slowed, pushing the resonance time earlier. 
Although perhaps counter-intuitive, \citealt{Lai94} showed that the scaling of the saturation amplitude after resonance is $\propto \omega_{\alpha}^{-5/6}$ under the stationary phase approximation (i.e., when the orbital decay rate is much slower than the mode frequency): an earlier resonance thus generally renders a higher amplitude, leading to the general increase with maximal strain as a function of spin as presented in Figure \ref{fig:spindep} for fixed overlap.
In principle, this increase reaches a peak at the spin such that the mode will be likely smeared out by chemical reaction, leading to an abrupt zeroing of the strain. 
Note, however, such a smearing was not modelled by \citealt{Kuan21b}: the reason for the drop in $\sigma$ in this case is because the mode frequency drops enough that resonance window falls outside of the computational domain at some minimum $\Omega_{\rm Orb}(t=0)$.

Although $g$-modes reside mostly in the core, they are capable of exerting strain at the bottom of crust for two reasons. (i) They penetrate into the crust when $\delta$ is sufficiently large; in fact, \citealt{Passamonti20,Gittins24} demonstrated that the $g_1$-mode eigenfunction can resemble that of the $i$-mode though the frequencies differ \cite[cf.][]{McDermott88}. (ii) Their tangential motion is not necessarily small even when buoyancy is somewhat quenched by the shear modulus. 
Should some crustal activities be triggered by $g$-modes and lead to observables, information about the internal stratification, thermal properties, cooling, phase transitions \cite[including gravitational ones;][]{kuan22}, and heat transport in neutron stars could all be gleaned.

{There are open problems in this direction related to the handling of the microphysics that more sophisticated models hope to treat in future. Even with respect to models described above that do include crustal elasticity explicitly, compositional stratifications at the base of the crust may be driven by neutron drip from nuclear clusters, which is likely to occur at timescales \emph{faster} than that of g-mode oscillations, thereby modifying the mode eigenfunctions there relative to models where perturbed adiabatic indices were put in by hand. Other models described throughout that do use tabulated EOS data for describing the crust and boundary layer composition (see Fig.~\ref{fig:deltafinite}) but which do not include crustal elasticity may similarly overestimate the crustal strain. Efforts to include all such components for $g$-modes are in progress.}

\subsubsection{Ocean modes}
\label{sec:oceans}

Ocean mode oscillations in neutron stars refer rather literally to waves (similar to the surface waves on Earth's oceans) that occur in the thin, penultimate layer of the star (before atmosphere), which is typically a few meters to tens of meters thick \cite{McDermott85,Strohmayer93}. Given the low density of the region, these modes are affected by factors such as the magnetic field, temperature gradients, gravitational forces, and most importantly composition. These modes are important in the context of X-ray bursts, as the thermonuclear reactions that heat up the ocean layer can also instigate drifts that modulate the X-ray emission \cite[see Table 4 and references in][]{watts12}; such drifting can be used to study ocean and atmosphere structure \cite{chamwatts1,chamwatts2,nat24}.

Ocean modes tend to have frequencies of $\sim$~tens of Hz at most, and thus could be excited at early times in a merger (see Fig.~\ref{fig:dia1}). Similar to the $g$-mode case, there is a competition between the excitation window (which is long for low frequencies) and overlap (which tends to be small at low frequencies). This was investigated by \citealt{sul23}, who found largest overlaps for carbon atmospheres with frequencies $\sim 16$~Hz. These authors also found that energy deposits could reach $\lesssim 10^{47}$~erg over the inspiral duration, decreasing by a few orders of magnitude if elasticity or a crust with heavier nuclei (e.g. oxygen or iron) is considered. Atmospheres composed of lighter elements may be expected in a system with a history of accretion; one may therefore expect ``rapid'' rotation to coincide with larger ocean impact (Sec.~\ref{sec:rotation}). Again like the $g$-mode case, there should also be a contest between mode frequencies and the local reaction rate, though the reaction rate in the ocean can be quite different from that in the crust or core. For heavier atmospheres with sub-Hz mode frequencies, space-based GW interferometers may be especially useful since such resonances could occur long before the LVK window and possibly coincide with multimessenger activity \cite{sul23}. Whether such modes could ever instigate crustal failure is unclear, though emissions may still result from particular accelerations via tidal-wave motions \cite{sul24}.

Aside from the crust-core interface, there is also a crust-ocean interface that can be considered (the $i_{1}$ mode). These can potentially grow to large amplitudes; as noted by \citealt{sul24}, while crustal failure may be difficult to achieve through an initial resonance as $i_{1}$ overlaps are not so large \cite[see Figure 9 in][]{Passamonti20}, crustal inhomogeneities may arise from such local failures \cite[see also][]{km22}. This may prime the crust for failures down the line, through either repeated resonances of the same mode, in an eccentric or dynamically-evolving system, or higher-frequency (e.g. $g$-) modes.

\subsubsection{f-modes}
\label{sec:fmodes}

The {\it f-}undamental fluid oscillation mode is a member of the acoustic ($p$-) family, in which the main restoring force is the hydrostatic pressure. This mode has no nodes, and thus couples very strongly to the tidal field which also has no nodes, as noted by \citealt{thorne98} and others before. The linear eigenfrequencies of these modes typically range from 1 kHz to several kHz, depending on the mass, radius, and internal structure of the neutron star. However, several studies have noted an insensitivity to the EOS \cite{1996PhRvL..77.4134A,1998MNRAS.299.1059A}, allowing for universal relations to be written down. Such aspects are powerful in the sense that measuring an $f$-mode frequency would give valuable insight onto the compactness of the system in an EOS-independent way. For a complete list of such relations, including the effects of rotation, we refer the reader to \citealt{2021FrASS...8..166K, 2024PhRvD.109j3033M}.

Although these modes are likely dominant in the tidal dephasing problem (see Sec.~\ref{sec:dyntides}), they are probably not especially relevant for precursors (the topic of this section). This is because their high frequency means that resonances, and hence crustal failure, are unlikely to occur except in cases with rapid rotation \cite{suvkprec19}, high eccentricity \cite{vick18}, or if the stellar compactness is particularly low (e.g. for EOS predicting large radii; see Fig.~\ref{fig:EOS}). Still, if resonant, failure is practically guaranteed as the overlap is of order unity, and thus this mode could explain short-waiting-time events (like in GRBs 150922A, 100223110, ...).

\subsubsection{r-modes}
\label{sec:rmodes}

Inertial (i.e. $r$-) modes are quasi-axial modes, degenerating in the static limit, with relatively weak coupling to the tidal field. However, in systems that are rotating very rapidly ($\nu \gg 100$~Hz), the hybrid character allows the modes to exert non-negligible strain on the crust \cite{Lai06,suvkprec19}, though it is expected that such systems are rare \cite[cf. Sec.~\ref{sec:rotation}][]{zhu18}. They are generically unstable to rotational instabilities \cite{1998ApJ...502..708A,1998ApJ...502..714F,2001IJMPD..10..381A,2016EPJA...52...38K} because their co-rotating frame frequencies are less than $2 \Omega_{s}$ (see equation \ref{eq:inertialframe}) and in this way they may be  important in limiting the maximum rotation of neutron stars through radiation-reaction \cite{1999ApJ...510..846A}. They can be potentially excited in neutron star glitches \cite{2018MNRAS.473.1644A}, the predicted amplitudes of which may be detectable with LVK given $\gtrsim$~months of folding for Galactic sources \cite[e.g. PSR J0537-691;][]{2018ApJ...864..137A}. Mathematical considerations related to tidal and magnetic corrections to the inertial spectrum are considered in \citealt{suvkprec19}, and are similarly not expected to be important unless magnetars are present in the merger (cf. Secs.~\ref{sec:magneticfields} and \ref{sec:spectralmods}). In principle, $r$-modes could also incite dynamo activity prior to merger if reaching a large enough amplitude \cite[][see also Sec.~\ref{sec:dynamo}]{rez01}.

\subsection{Late-stage dynamos?}
\label{sec:dynamo}

\begin{figure}
	\centering
	\includegraphics[width=0.497\textwidth]{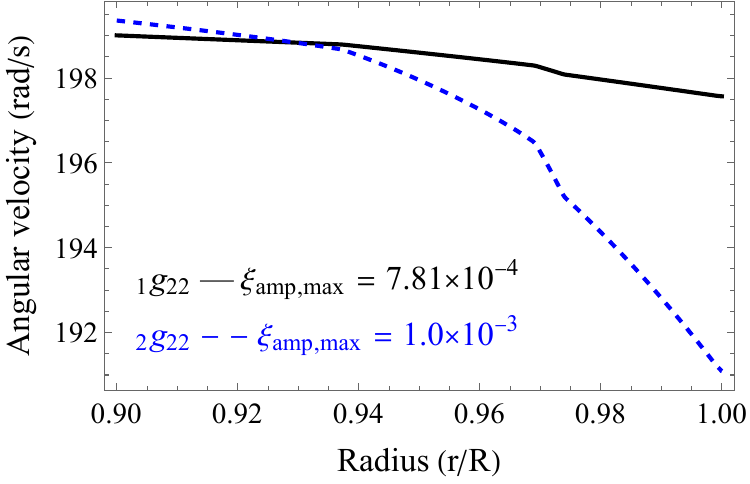}
	\caption{Crustal angular velocities at fixed spherical angles, $\Omega(r)$, for $g_{1}$ (black, solid) and $g_{2}$ (blue, dashed) modes, at a time slice corresponding to the maximum resonant amplitude. The star has an initial angular velocity of $\Omega_{s} = 200$~rad/s with $M = 1.6 M_{\odot}$, and $R = 11.26\,$km with a hybrid APR4 + \citealt{2001A&A...380..151D} (for the crust) EOS. From \citealt{suv24}.
	\label{fig:diffrotex}}
\end{figure}

Tidal resonances, aside from draining energy from the orbit, also add angular momentum to the stellar interior (Sec.~\ref{sec:tidalspin}). The resulting velocity pattern is tied to the mode eigenfunction and generally implies differentially rotating cavities, $\Omega = \Omega(t,\boldsymbol{x})$. One such example, computed self-consistently by \citealt{suv24}, is shown in Figure~\ref{fig:diffrotex} for two different $g$-mode resonances. The snapshots are taken at the moment of peak positive amplitude in the crust. Generally speaking, low-overtone $g$-modes have a monotonic radial profile in the crust, leading to either $\partial_{r} \Omega <0$ or $\partial_{r} \Omega >0$ everywhere there, depending on mode phase.

Differential rotation is typically a necessary (but not sufficient) condition for a dynamo, most notably the magnetorotational instability (MRI): \citealt{bal91} and others since have shown that if the radial angular velocity gradient is sufficiently negative in a cavity, $\partial_{r} \Omega \ll 0$, the magnetic tension can rocket to a large value in an effort to stabilise the system. The exact criterion depends on the microphysics of the cavity itself, such as the chemical diffusivity, stratification gradient, and electric resistivity. As explored by \citealt{suv24}, the magnetic field could be amplified on $\sim$ms timescales in the final $\sim$s of inspiral via the MRI in a mature, recycled star. They found that saturation fields in the crust of a merger participant could exceed $\sim 10^{13}\,$G for certain modes and EOS if the star is spinning sufficiently fast ($\nu \gtrsim 30$~Hz). The reason that some preexisting rotation is necessary is because it influences the growth time of the unstable magnetic (Alfv{\'e}n) modes, in general like $t_{\rm MRI} \propto \Omega_{s}^{-1}$ \cite{bal91,duez06}. This is important because the mode oscillation will flip the sign of $\Omega(\boldsymbol{x})$ every half-period, and thus if the mode growth is too slow (i.e. if $\Omega_{s}$ is too low) then less (or no) amplification is anticipated. The extent can be computed with a dephasing integral.

\begin{figure}
	\centering
	\includegraphics[width=0.497\textwidth]{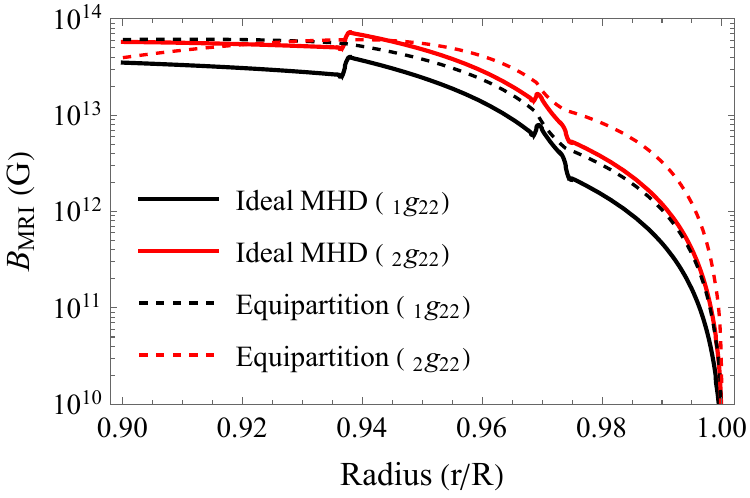}
	\caption{Dynamo amplifications from either a local MHD
analysis (solid curves) or setting equipartition between magnetic and shear energies (dashed curves; upper limits), as triggered by either $g_{1}$- or $g_{2}$-modes in a star with $\Omega_{s,0} = 200$~rad/s. The same macro- and microparameters used in Fig.~\ref{fig:diffrotex} are used. From \citealt{suv24}.
	\label{fig:dynamo1}}
\end{figure}

Fields of strength depicted in Fig.~\ref{fig:dynamo1} would allow for the magnetic energy budget to match the observed luminosities of precursors, possibly resolving the mystery of why precursors are rare: fast stars are needed. \citet{suv24} estimated that the conditions necessary for the MRI are satisfied in $\sim 5\%$ of merger events (see Appendix A therein). It may therefore be that magnetar-level fields do not persist into merger, but rather the field is generated. While appealing, some issues remain with this interpretation however. The main one being that it is difficult to imagine a coherent field being produced; rather one may expect a turbulent, highly-tangled structure to emerge, as is more typical from MRI simulations \cite[e.g.][]{kiu24}. The process of extracting magnetoelastic energy is thus more complicated, and further work is needed to address the topological development of the field. A second issue is that estimates thus far have been carried out in (GR)MHD, but really elastic terms are necessary to adequately describe the crustal system. Whether it is possible for the MRI to activate in an \emph{elastic cavity}, rather than fluid interior, is not obvious. Efforts in this direction are underway. Other dynamos would worth investigating, like those with a chiral origin \cite{deh24} triggered by tide-related processes that force the system out of thermodynamic equilibrium prior to merger \cite{hamm21}.

\subsection{Post-merger models}
\label{sec:post-merger}

If the main GRB episode is produced by (the standard mechanism involving) synchrotron radiation in an internal shock or magnetic dissipation zone, a (thermal) precursor may be expected either as the shock breaks out from the surrounding ejecta or as some (neutrino-anti-neutrino) fireball reaches the photosphere \cite{mr00,wang18,wl21}. Alternatively, it may be that an incipient jet fails to break out (or ``chokes'') which can produce some GRB-like emission prior to the main event, where a renewed jet successfully escapes. In the latter case, a range of spectra are plausible depending on jet nature. A newborn magnetar may also launch a powerful wind independently from the jet, which may interact with the surroundings and produce X- and gamma-ray activity \cite[e.g.][]{dal22}. Overall this Review is however concerned with premerger phenomena, and thus we do not discuss these possibilities further. The interested reader is directed to \citealt{cio18,burns20,sarinlasky21} and references therein.

\section{Multimessenger outlook}
\label{sec:multimessenger}

The previous three sections have been devoted to understanding gravitational and gamma-ray emissions in premerger systems. These can be studied through tidal and magnetic interaction theory, with a careful of the stellar micro- and macrophysics. Here, by contrast, we turn briefly to other multimessenger elements, expected in the radio and X-ray bands, together with neutrino counterparts. We intentionally keep this section rather short: such elements have been described elsewhere \cite[see, for instance,][]{lyu19} and our main focus is on the gamma- and GW aspects. 

What can one learn, broadly speaking, from premerger multimessengers? A rough summary is provided below. 

\begin{enumerate}
    \item[$\boldsymbol{\star}$]{\textbf{Love numbers.}} The effective parameters $\kappa^{T}$, the quadrupolar member of which is defined by \eqref{eq:kappaT}, is directly visible in the gravitational waveform at a leading PN order (see Tab.~\ref{tab:pneffects}). Since this quantity depends on the stellar masses and integrals taken over the internal density, it is clear that EOS information can be gleaned. This is effectively illustrated in Fig.~\ref{fig:EOS}, showing mass-radius contours from GW170817. 
    \item[$\boldsymbol{\star}$]{\textbf{Asteroseismology.}} Dynamical tides also imprint themselves on the waveform. However, since these emerge at finite, non-zero frequencies while the former appear already ``at infinity'', they are generally subleading.  It has been estimated that only in $\sim 1\%$ of neutron-star mergers will one be able to cleanly isolate the impact of dynamical tides with current detectors \cite{gamba23}. However, owing to the discussion provided in Sec.~\ref{sec:dyntides}, there are open questions in this direction which are worth revisiting. For example, the $g$-modes may be comparable contributors to the dephasing if the $f$-mode frequency is very high (cf. Fig.~\ref{fig:mode_evol20}). Strong magnetic fields may also be important, either through modulating the mode frequencies directly or instigating an electromotive spin-up (Sec.~\ref{sec:magacc}), which continuously shifts the spectra. Out-of-equilibrium effects could also distort the spectrum in a complicated, time-dependent way \cite{hamm21}.
    \item[$\boldsymbol{\star}$]{\textbf{Gamma-ray precursors.}} Depending on the ignition mechanism, different kinds of information be may discernible. For premerger precursors, we have argued the resonant failure picture can adequately explain all the observational characteristics (Sec.~\ref{sec:resfail}), though admittedly this is due to the huge range of QNM properties that neutron stars can exhibit. As resonances are obviously tied to the mode spectra, everything above applies here too but in the gamma-ray band. Similarly, since the activity of these modes \emph{in the crust} is the relevant aspect here, microphysical inputs become critical. With the above three (Love number, dynamical tides, and precursors) one may thus learn about both macro- and microphysical elements of the stars taking part in a merger. This can be combined self-consistently with the properties of the post-merger remnant to deduce generative elements of GRBs (e.g. heavier stars will more likely promptly collapse leading to faster jet break-out; see \citealt{zhang19} and Tab.~\ref{tab:jet_form}).
    \item[$\boldsymbol{\star}$]{\textbf{Radio flares.}} Radio activity can be incited premerger from a few different channels. For instance, there may be shock-powered radio emissions through interactions taking place in the accelerating, binary wind left in the wake of the inspiral \cite{mezr92,srid21}. These are likely to be in the form of FRBs or ``giant pulse'' like phenomena in the $\sim$~GHz band. Another possibility discussed by \citet{cooper23} is that acceleration zones may form in regions of interwoven magnetic fields, which could produce  coherent, millisecond bursts in radio frequencies that are theoretically observable out to $\sim$~Gpc distances. Such observations could reveal information about the magnetospheric plasma and radio activation mechanisms, which can be used to deduce information about the general pulsar engine and all that can convey about stellar structure \cite[see, e.g.][]{beskin18}. There is also the possibility of postmerger radio activity associated with a neutron-star collapse, as described in Sec.~\ref{sec:jetform}. The Square Kilometer Array (SKA) will go a long way towards detection prospects.
    \item[$\boldsymbol{\star}$]{\textbf{X-ray flares.}} We have earlier described how the emission mechanism for GRB precursors resembles that of giant flares from magnetars in terms of energy extraction and propagation. Given that X-ray activity from magnetars is commonplace (and arguably a defining feature), the same applies for premerger stars with strong fields. X-ray emission mechanisms in a premerger system are described by \citealt{belo21}, the key ingredient being the nonlinear development of magnetospheric Alfv{\'e}n waves \cite[see also][]{most24}. Such observations may reveal crucial information about the local magnetic field strength and radiation transport physics, from which information about neutron-star evolutionary pathways can be deduced. X-ray activity may also be related to magnetic instabilities arising by late stage dynamo activity or magnetic reconfiguration more generally (see Sec.~\ref{sec:dynamo} and also \citealt{mast15b,suv23} for instance). The planned ECLAIRs telescope will help enable searches for X-rays out to cosmological distances.
    \item[$\boldsymbol{\star}$]{\textbf{Neutrinos.}} Neutrinos could be emitted from a premerger system through at least two distinct means. One involves tidal heating. As described in Sec.~\ref{sec:thermal}, rapid episodes of heating could lead to chemical imbalance and the production of neutrinos which could be theoretically observed for a close merger. Such observations would provide important information on the microphysical heat capacity. The second mechanism is indirectly via the production of cosmic rays. As discussed by \citet{cop19}, it is thought that GRBs could be sources of ultra-high-energy cosmic rays. In this case, the interaction of high-energy protons in relativistic wind wakes (as above) near the source could produce neutrinos carrying a non-negligible fraction of the proton energy. 
     Ice-Cube has reported detection of some \emph{cosmic neutrinos} \cite{IceCube:2013low,IceCube:2015gsk,IceCube:2020wum}, while the sources are only identified in a few cases; next-generation neutrino observatories \cite[like IceCube-Gen2;][]{IceCube-Gen2:2020qha} will synergise with other instruments to better constrain the sources \cite[e.g.][]{mura19}.
     In addition, such observations could be used to place constraints on binary neutron-star merger abundances and provide tests of the standard model \cite{anch14}. For instance, neutrinos are thought to carry magnetic moments \cite{Fujikawa:1980yx,Shrock:1982sc} through which they interact with the ambient magnetic field that depends on whether they are of a Dirac or Majorana nature \cite{giun15}. Thus, the neutrino flux from a strongly magnetised environment, such as the engine of GRBs \cite{waxman95,viet95}, may encode neutrino's nature \cite[e.g.][]{brdar24}. An excellent review of future neutrino observatories, and instruments capable of follow-up, are described in Tables 1 and 2 of \citet{ok22}.
\end{enumerate}

In addition to the above, there are the broadband kilonovae \cite{pac98,val17}. With the upcoming Roman Space Telescope they may be detectable out to redshifts $z \sim 1$ \cite{chase22}, which is considerably further than the GWs from neutron-star mergers. Even without GWs, more kilonova observations would be valuable to disentangle the types of GRBs that can be associated with mergers; see Sec.~\ref{sec:longvsshort}.
\subsection{A brief look at future possibilities}
\label{sec:outlook}

One of the key ways in which multimessenger astrophysics will be propelled forward in the future comes from next-generation GW interferometers. Although there are a number of proposed technologies, some of the most notable are (i) the Einstein Telescope \cite[ET;][]{Maggiore:2019uih,punt10,Hild:2010id,Branchesi:2023mws}. The sensitivity band of ET is similar to that of the LVK network, though notably deeply with considerable improvement in the $\lesssim 10$~Hz range. The reason is that the gravitational gradient (aka ``Newtonian''), seismic, and thermal noises are less of a problem for ET, as it will be built underground and include cryogenic technologies to reduce thermal vibrations \cite{punt10}. This latter band is relevant to capture the onset of low-frequency oscillations and track the orbit for longer periods of time. (ii) Cosmic Explorer \cite[CE;][]{LIGOScientific:2016wof,Reitze:2019iox,Evans:2021gyd}. This detector is similar to ET in terms of sensitivity band, though is planned to go even deeper. CE may detect \emph{thousands} of neutron stars in merger \cite{ce23}, and thus lead to a true era where statistical methods can nail down the neutron-star EOS. (iii) The Laser Interferometer Space Antenna (LISA). LISA is a planned space-based detector that could capture the very early aspects of inspiral while the orbit is at $\gtrsim$~mHz frequencies \cite{lisa23}. Wide binary elements could be useful for ocean resonances \cite{sul23,sul24}, low frequency $g$-modes (assuming these are not washed out by reaction rates; Sec.~\ref{sec:stratification}) and the testing of spacetime structure generally \cite[see, e.g.][]{dest20,dest21}.

With CE and ET, many of the high-amplitude QNMs of a merger remnant may be resolvable out to cosmological distances. Aside from providing tests of GR directly \cite[see, e.g.][]{kk99,2009CQGra..26p3001B,2011RvMP...83..793K,sv21}, this could be connected with GRB observations to deduce remnant nature and its impact on jet structure; see \citealt{chirenti23}, who reported the discovery of $\sim$~kHz modulations in the short GRBs 910711 and 931101B. Identifying oscillation frequencies in both the remnant and the premerger stars (via dynamical tides) would prove a powerful probe of fundamental physics. In a similar vein, GWs from a resonant failure itself may be detectable with a sufficiently sensitive telescope for a Galactic binary merger \cite{zink11}. This would allow for the precursor-merger waiting time to be cleanly isolated, removing uncertainties relating to jet-break out (see Sec.~\ref{sec:jetform}).

Innovations on the electromagnetic side are no less promising. For instance, with the development of the SKA and next-generation X-ray observatories the types of premerger precursors could increase significantly. Although we have limited our use of the word \emph{precursor} to specifically mean the gamma-ray flashes occurring before the main event (see Tab.~\ref{tab:precdata}), it is hoped that such terminology will prove ambiguous in the future. Tables 1 and 2 of \citealt{ok22} provide a summary of next-generation telescopes that could coincidentally detect signals associated with neutrino bursts originating from a binary merger. One intriguing possibility comes from the planned POLAR-2 detector, which should able to measure polarization and similar effects related to the propagation of gamma-rays \cite[a survey of other next-generation gamma-ray facilities are described in Section 2 of][]{polar2}. This could be used to probe the local magnetic field strength in the region where the rays were produced. This would allow one to test, for instance, the extent to which non-thermality is tied to magnetars in mergers (see also Sec.~\ref{sec:magacc}).

\section{Conclusions}
\label{sec:conclusions}

This Review is dedicated to the study of premerger phenomena in compact binaries involving at least one neutron star. Although there are a number of reviews of binary mergers on the market, we feel that none have attempted to put the theory of dynamical tides on a consistent footing together with the relevant GW and electromagnetic elements. Fig.~\ref{fig:dia1} provides an overall picture of that which we consider.

In compact binaries without pure black holes, tidal forces become large in the late stages of inspiral. These strong tides distort the stars geometrically through the equilibrium tide and subsequently initiate large-amplitude fluid motions through dynamical tides (see Fig.~\ref{fig:tidesfig}). Resonant pulsations can drain significant amounts of energy from the orbit, thereby allowing for avenues to test the highly-coveted core EOS of neutron stars by examining \emph{dephasing} predictions relative to cases where tides are ``switched off'' (see Fig.~\ref{fig:teob_A}). A review of tidal theory, with a particular emphasis on GW phenomena, was given in Sec.~\ref{sec:inspiralbasics}, building on the earlier two sections which introduced neutron-star macro- and microphysical elements. Although views on the detectability of dynamical tides in binaries varies in the literature, it is hoped that our Review gives call for optimism, noting that there are open problems on both the theoretical side (e.g. modelling out-of-equilibrium elements related to tidal heating) and modelling sides (e.g. accounting for time-dependent spins from tidal or electromotive torques) which, when resolved, may improve the outlook.

High-energy phenomena in late-stage inspirals are not limited to the GW sector, as the observation of early \emph{precursors} establishes (see Sec.~\ref{sec:precursors}). Observational characteristics of these first-round gamma-ray flashes are highly-varied in terms of their spectra, energetics, and waiting times (see Tab.~\ref{tab:precdata}). It is likely therefore that a single, theoretical model cannot account for all of them. The two more popular models considered thus far involve either electrodynamic interactions (Sec.~\ref{sec:magnetosphericinteraction}) or, as above, resonant tides (Sec.~\ref{sec:resfail}). Even within these two classes there are varied predictions. For example: which mode may be responsible for initiating crustal failure that leads to precursor emission? Several families have been proposed in the literature, as covered in Sec.~\ref{sec:resfamilies}. Aside from the gamma-ray flashes, a number of other multimessenger elements have been predicted, ranging from radio-band pulsations to neutrino floods. These are described in Sec.~\ref{sec:multimessenger}. 

{What type of constraints may the multimessenger channels described herein place on neutron-star magnetic fields, for instance? In no particular order, we have: 
\begin{itemize}
\item[$\boldsymbol{\star}$]{Magnetospheric interactions will accelerate the inspiral and lead to electromotive spinup (quantified in Sec.~\ref{sec:magacc}), both of which may be visible in GWs either directly (dephasing) or indirectly (shifting mode frequencies as a function of time). Flares, radio pulsations, and/or FRBs may also be triggered by reconnection; such observations can place constraints on the large-scale, far-field dipole moments of old stars.}
\item[$\boldsymbol{\star}$]{Strong internal fields ($B \gtrsim 10^{15}$~G) could, in principle, reveal themselves by shifting the stellar QNM frequencies or tidal Love numbers (Sec.~\ref{sec:internalfields}). This could also be used to statistically constrain the presence or absence of superconducting states, as the superconductor tension force may also shift these properties even if $B$ itself is partially expelled.}
\item[$\boldsymbol{\star}$]{Near-surface fields could be constrained by bright, nonthermal precursors if the magnetic field dominates the energy transport process (e.g. the maximum luminosity is set by the near-surface field; see Sec.~\ref{sec:energetics}).}
\item[$\boldsymbol{\star}$]{Remnant properties may be tied to the strength of the fields of the inspiralling constituents. Although very recent, high-resolution simulations suggest the magnetisation of the remnant may be set independently by dynamo activity \cite{kiu24}, disk material may still be highly magnetised \cite{kiu14} which could prolong subsequent GRB activity by setting up magnetic barriers that sporadically halt accretion. This channel could constrain internal fields and disc dynamics more generally.}
\item[$\boldsymbol{\star}$]{Field longevity is a critical question in the precursor scenario. How is it possible that magnetars take place in mergers when the anticipated decay timescales are orders of magnitude lower than a typical inspiral time? If the answer is that it is impossible, we appear to left with only two options, being either (i) bright, nonthermal precursors must be \emph{post-merger} phenomena (Sec.~\ref{sec:post-merger}), or (ii) the field is generated by dynamo activity in the seconds leading up to coalescence (Sec.~\ref{sec:dynamo}). Either of these would lead to constraints on jet formation and propagation together with the microphysics of the crust where the dynamo would operate. If such fields can persist into late times, this may call into question assumptions about field evolution generally \cite[e.g. whether one can ignore meridional circulation when evolving the induction equation;][]{suvg23b}.}
\end{itemize}}

We hope to leave the reader with the impression that premerger phenomena has the capacity to educate us on fundamental physics related to the low-temperature, high-baryon-density sector of quantum chromodynamics, GR, quantum electrodynamics, and more to a level that rivals that of postmerger phenomena. 

\section*{Acknowledgements}
We are grateful to Armen Sedrakian, Jan Steinhoff, and Sebastiano Bernuzzi for comments. We thank the referees and editor for comments which improved the quality of this manuscript. AGS acknowledges support provided by the Conselleria d'Educaci{\'o}, Cultura, Universitats i Ocupaci{\'o} de la Generalitat Valenciana through Prometeo Project CIPROM/2022/13. This project has received funding from the European Union's Horizon-MSCA-2022 research and innovation programme ``EinsteinWaves'' under grant agreement  No 101131233.


\appendix

\end{document}